\documentclass{aa}

\usepackage{natbib,twoopt}
\usepackage[breaklinks=true]{hyperref} %% to avoid \citeads line fills
\bibpunct{(}{)}{;}{a}{}{,}             %% natbib format for A&A and ApJ
\makeatletter
  \newcommandtwoopt{\citeads}[3][][]{\href{http://adsabs.harvard.edu/abs/#3}%
    {\def\hyper@linkstart##1##2{}%
     \let\hyper@linkend\@empty\citealp[#1][#2]{#3}}}
  \newcommandtwoopt{\citepads}[3][][]{\href{http://adsabs.harvard.edu/abs/#3}%
    {\def\hyper@linkstart##1##2{}%
     \let\hyper@linkend\@empty\citep[#1][#2]{#3}}}
  \newcommandtwoopt{\citetads}[3][][]{\href{http://adsabs.harvard.edu/abs/#3}%
    {\def\hyper@linkstart##1##2{}%
     \let\hyper@linkend\@empty\citet[#1][#2]{#3}}}
  \newcommandtwoopt{\citeyearads}[3][][]%
    {\href{http://adsabs.harvard.edu/abs/#3}
    {\def\hyper@linkstart##1##2{}%
     \let\hyper@linkend\@empty\citeyear[#1][#2]{#3}}}
\makeatother

\usepackage{array} %options pour les tableaux
\usepackage{tabularx}
\usepackage{multirow}
\usepackage{graphicx}
\usepackage[version=3]{mhchem}
\usepackage{txfonts}
\usepackage{color}
\usepackage{ulem}

\begin{document}

\title{A reduced chemical scheme for modelling warm to hot hydrogen-dominated atmospheres}
\author{O. Venot\inst{1}, R. Bounaceur\inst{2}, M. Dobrijevic\inst{3}, E. H\'ebrard\inst{4},  T. Cavali\'e\inst{3,5}, P. Tremblin\inst{6}, B. Drummond\inst{4},  B. Charnay\inst{5}
}

\institute{Laboratoire Interuniversitaire des Syst\`{e}mes Atmosph\'{e}riques, UMR CNRS 7583, Universit\'{e}s Paris Est Cr\'eteil (UPEC) et Paris Diderot (UPD), Cr\'{e}teil, France\label{LISA}\\
\email{olivia.venot@lisa.u-pec.fr}
\and Laboratoire R\'{e}actions et G\'{e}nie des Proc\'{e}d\'{e}s, LRGP UMP 7274 CNRS, Universit\'{e} de Lorraine, 1 rue Grandville, BP 20401, F-54001 Nancy, France
\and Laboratoire d'Astrophysique de Bordeaux, Univ. Bordeaux, CNRS, B18N, all\'ee Geoffroy Saint-Hilaire, Pessac 33615, France
\and Astrophysics Group, University of Exeter, EX4 4QL, Exeter, UK
\and LESIA, Observatoire de Paris, Universit\'e PSL, CNRS, Sorbonne Universit\'e, Univ. Paris Diderot, Sorbonne Paris Cit\'e, 5 place Jules Janssen, 92195 Meudon, France
\and Maison de la Simulation, CEA, CNRS, Univ. Paris-Sud, UVSQ, Universit\'e Paris-Saclay, 91191 Gif-sur-Yvette, France
}

\titlerunning{A reduced chemical scheme}
\authorrunning{Venot et al.}
\date{Received <date> /
Accepted <date>}

\abstract{
%Context
Three dimensional models that account for chemistry are useful tools to predict the chemical composition of (exo)planet and brown dwarf atmospheres and interpret observations of future telescopes, such as JWST and ARIEL. Recent Juno observations of the NH$_3$ tropospheric distribution in Jupiter \citep{Bolton2017} also indicate that 3D chemical modelling may be necessary to constrain the deep composition of the giant planets of the Solar System. However, due to the high computational cost of chemistry calculations, 3D chemical modelling has so far been limited.}
{%Aims
Our goal is to develop a reduced chemical scheme from the full chemical scheme of \citet{Venot2012} able to reproduce accurately the vertical profiles of the observable species (H$_2$O, CH$_4$, CO, CO$_2$, NH$_3$, and HCN). This reduced scheme should have a size compatible with three dimensional models and be usable across a large parameter space (e.g. temperature, pressure, elemental abundance). The absence of C$_2$H$_2$ from our reduced chemical scheme prevents its use to study hot C-rich atmospheres.}
{%Methods
We use a mechanism-processing utility designed for use with Chemkin-Pro to reduce a full detailed mechanism. ANSYS Chemkin-Pro Reaction Workbench allows the reduction of a reaction mechanism for a given list of target species and a specified level of accuracy. We take a warm giant exoplanet with solar abundances, GJ 436b, as a template to perform the scheme reduction.
To assess the validity of our reduced scheme, we take the uncertainties on the reaction rates into account in Monte-Carlo runs with the full scheme, and compare the resulting vertical profiles with the reduced scheme. We explore the range of validity of the reduced scheme even further by applying our new reduced scheme to GJ 436b's atmosphere with different elemental abundances, to three other exoplanet atmospheres (GJ 1214b, HD 209458b, HD 189733b), a brown dwarf atmosphere (SD~1110), and to the troposphere of two giant planets of the Solar System (Uranus and Neptune).}
{%Results
For all cases except one, the abundances predicted by the reduced scheme remain within the error bars of the model with the full scheme. Expectedly, we found important differences that cannot be neglected only for the C-rich hot atmosphere. The reduced chemical scheme allows more rapid runs than the full scheme it derived from ($\sim$30 times faster).}
{%Conclusion
We have developed a reduced scheme containing 30 species and 181 reversible reactions. This scheme has a large range of validity and can be used to study all kind of warm atmospheres, except hot C-rich ones, which contains a high amount of C$_2$H$_2$. It can be used in 1D models, for fast computations, but also in 3D models for hot giant (exo)planet and brown dwarf atmospheres.}

\maketitle

\section{Introduction}

The next generation of space telescopes, such as the \textit{James Webb Space Telescope} (JWST) and the \textit{Atmospheric Remote-Sensing Infrared Exoplanet Large-survey} (ARIEL) will certainly revolutionise our knowledge and understanding of exoplanetary worlds. The high sensitivity of these future telescopes will permit the determination of horizontal and vertical variations of temperature and chemical composition in the atmospheres of hot and warm giant planets (e.g. \citealt{Bean2018, Tinetti2018, Venot2018ARIEL} and \citealt{VenotWASP43b}). To fully benefit from these high-resolution data and accurately interpret those spectra, three-dimensional (3D) models that include a detailed description of both physical and chemical processes are needed. Such models have not been developed yet, because of the very high computational time they require. Indeed, full chemical networks used to model Hot Jupiters atmospheres \citep[e.g.][]{moses2011,Venot2012} are very large, e.g. $\sim$2000 reactions including a hundred species. Resolving the corresponding large system of non-linear and highly coupled differential equations, in addition to the physical processes described in General Circulation Models (GCMs) would result in unreasonable computational times. To tackle this problem, most GCMs currently assume atmosphere at thermochemical equilibrium. Even if correct for very hot environments, this assumption might be wrong in cooler atmospheres and/or atmospheres subject to very strong transport.

Two attempts have been undertaken so far to couple dynamics and disequilibrium chemistry. First, \cite{CS2006} have studied carbon chemistry in the atmosphere of HD~209458b thanks to a three-dimensional model. In this model, the repartition of carbon into carbon monoxide (CO) is determined by relaxing the mole fraction of CO to its equilibrium value and that into methane (CH$_4$) is obtained through mass balance. This simple approach has shown that, at all latitudes, the mixing ratios of CO and CH$_4$ are homogenised with longitude in the 1--1000 mbar pressure region. The authors argue that vertical mixing is more efficient than horizontal mixing in the atmosphere of HD~209458b. This method was later used by other groups. The same chemical relaxation was applied to the atmospheres of HD 209459b \citep{Drummond2018} and HD 189733b \citep{Drummond2018b} using a different GCM, with the chemistry consistently coupled with the radiative transfer. They found also vertical mixing to be the dominant disequilibrium process over a large pressure range, albeit horizontal mixing has a non-negligible effect at low pressures and especially for CH$_4$. Also, \cite{Mendonca2018} used another relaxation scheme \citep{Tsai2018} to study the disequilibrium chemistry of CO, CO$_2$, H$_2$O, and CH$_4$ in WASP-43b atmosphere. Contrary to the previous studies of \cite{CS2006} and \cite{Drummond2018,Drummond2018b}, they found that horizontal quenching is the dominant process. This dissimilarity was attributed to differences in dynamical timescales, due to different model parameters (clear/cloudy atmosphere, surface gravity, rotation rate). As shown by \cite{mayne2017}, variations in numerical settings, such as numerical drag or boundaries conditions, could also be responsible for these different results.\\
An alternative method to study disequilibrium chemistry in Hot Jupiters has been undertaken by \cite{agu2012, agu2014}. They have developed  pseudo-2D photo-thermochemical models of HD~209458b and HD~189733b with full chemical kinetics. With this method, the 1D atmospheric column rotates along the equator to mimic horizontal mixing. The chemical composition is determined at steady-state by integrating a full chemical kinetics scheme \citep{Venot2012}. As \cite{CS2006}, they found an homogenisation of the chemical composition with longitude, but their conclusion on the relative importance of horizontal and vertical quenching phenomenon differs. \cite{agu2012, agu2014} have determined that vertical mixing and horizontal mixing are both at play in Hot Jupiter atmospheres.
These various results obtained with different methods show the complexity of atmospheric modelling and the difficulty to draw a general behaviour. Finally, the best way to determine 3D maps of the atmospheric composition of exoplanets is to include a realistic and detailed description of chemical kinetics in 3D dynamical models.

While hot giant exoplanet atmospheric studies would undoubtedly benefit from coupled chemical and dynamical modelling in 3D, it seems to be more and more the case for Solar System giant planet interiors as well. Until now, only 1D models have been applied to the deep tropospheres of these planets to try to unveil their primordial composition \citep[e.g.][]{Visscher2010, mousis2014, Cavalie2014, cavalie2017}. However, recent Juno/MWR observations of the meridional distribution of NH$_3$ in Jupiter's troposphere \citep{Bolton2017} surprisingly show that NH$_3$ is not well-mixed below its condensation level. This may then also be the case for other minor species, and 3D chemical and dynamical modelling of Jupiter's troposphere may be required when attempting to use upper tropospheric abundance measurements to derive deep elemental composition. This is probably also valid for the other Solar System giant planets, in which the temporal and meridional variability in the appearance of convective storms and clouds \citep[e.g.][]{Fischer2011, dePater2015, Irwin2016} may be the observable outcome of deep tropospheric inhomogeneities.

In this view, we have developed a reduced chemical scheme from the full chemical network published by \cite{Venot2012}, called C$_0$-C$_2$ scheme. We have chosen as a template the atmosphere of the warm Neptune GJ~436b, where thermochemistry and disequilibrium chemistry are both at play, and with solar abundances. This reduced scheme has a size comparable to the ones already included in GCMs developed for terrestrial planets/satellites of the Solar System, that is to say 40 species and 284 reactions for Titan \citep{lebonnois2001}, 12 species and 42 reactions for Mars \citep{Lefevre2004}, 34 species and 121 reactions for Venus \citep{stolzenbach2014, stolzenbach2016, gilli2017}. Therefore, we presume that our reduced chemical scheme, containing 30 species en 181 reversible reactions (i.e. 362 reactions in total), should be implementable in 3D models. We explain in Sect.\ref{sec:method} the method we have used to develop the chemical scheme and to determine its level of accuracy. Then, we present in Sect.~\ref{sec:results} the reduced chemical scheme we have obtained and compare the atmospheric composition of GJ 436b obtained with this scheme with that obtained with the full original scheme. In Sect.~\ref{sec:validity}, we study the range of validity of this scheme, i.e. the range of atmospheric conditions for which the chemical composition obtained with the reduced scheme is similar (or not) to the one obtained with the full chemical scheme for a set of species of interest, i.e. the observable ones. We present our conclusions in Sect.~\ref{sec:concl}.

\section{Method}\label{sec:method}

\subsection{Reduction of the chemical scheme}
\begin{figure}[b]
\centering
\includegraphics[angle=0,width=\columnwidth]{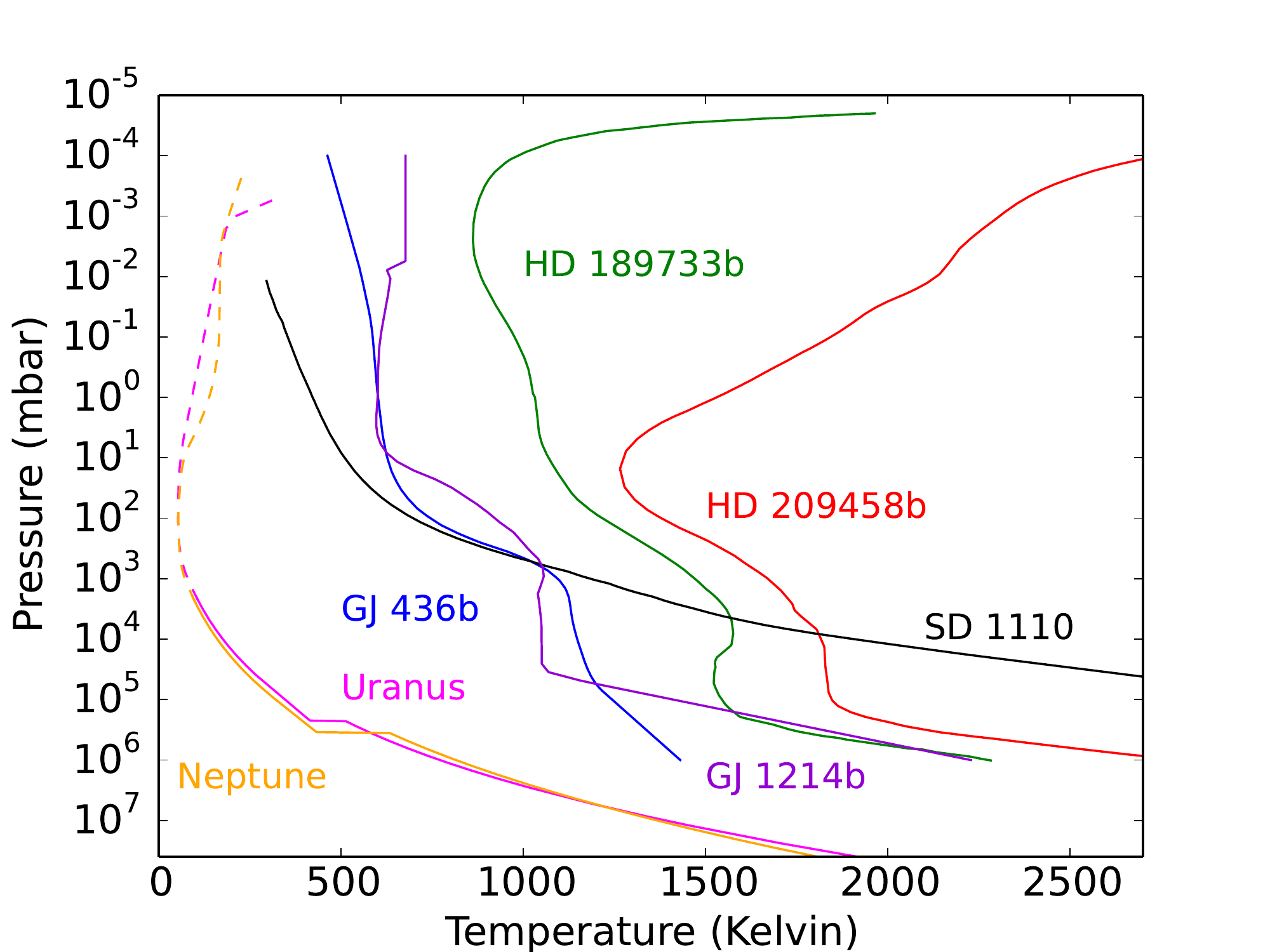}
\caption{Thermal profiles used in this study, as labelled on the figure. Only the deeper part of Uranus and Neptune (represented in solid lines) have been modelled. The upper part (dashed lines) is shown for information.}
\label{fig:PT}
\end{figure}

We have used the kinetic model of \citet{Venot2012} to perform the chemical scheme reduction presented in this paper. The initial full chemical scheme contains only neutral chemical reactions and we have excluded photodissociations. There are several methods that exist to reduce a chemical scheme, like the one proposed by \citet{Dobrijevic2011} for Solar System Giant Planet stratospheres. However such methods are not adapted to cases with reversible reactions. To reduce our chemical scheme, we have used the package ANSYS Chemkin-Pro Reaction Workbench \citeyearpar{chemkin}.
\begin{figure*}[t]
\centering
\includegraphics[angle=0,width=0.9\columnwidth]{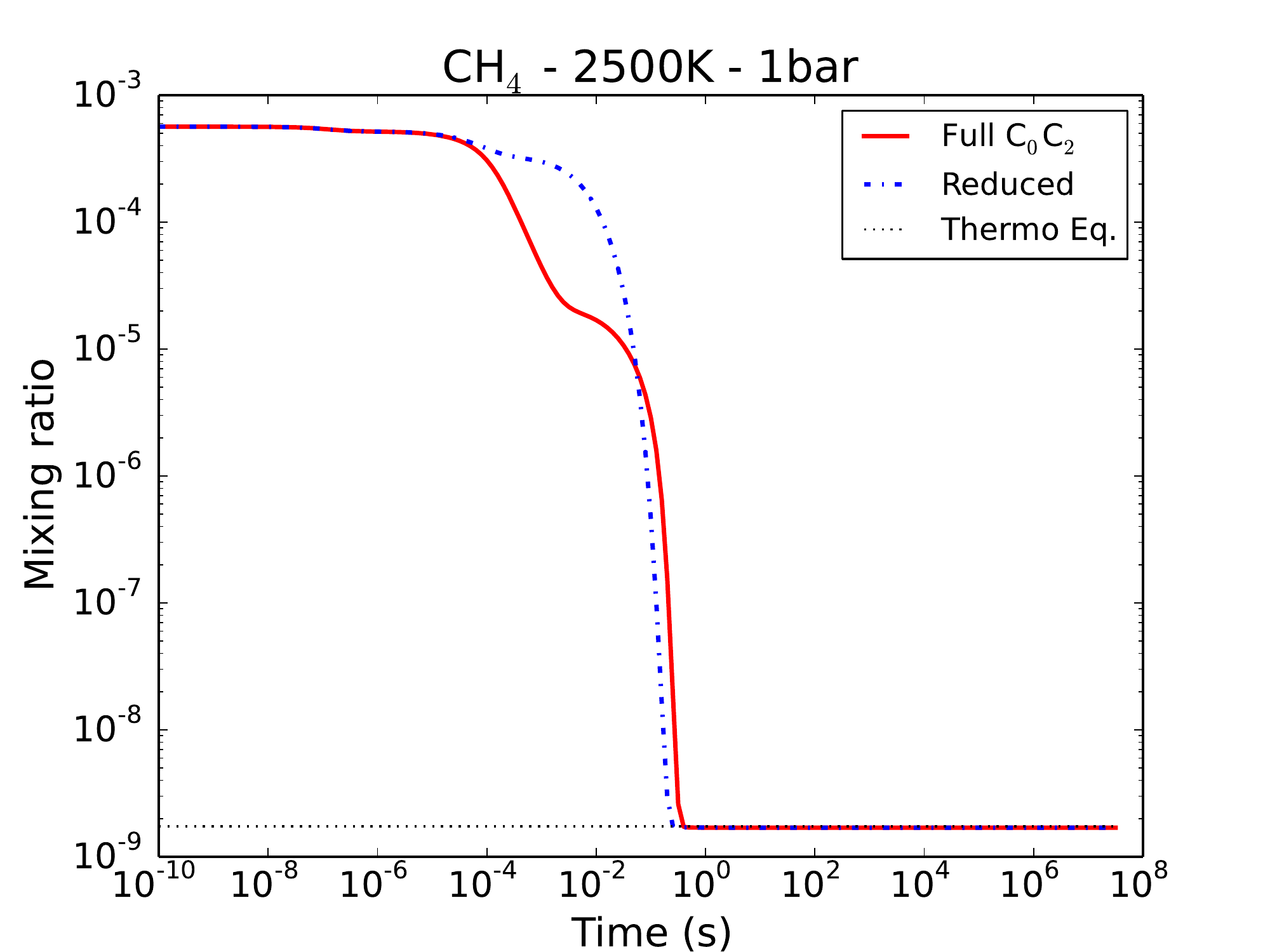}
\includegraphics[angle=0,width=0.9\columnwidth]{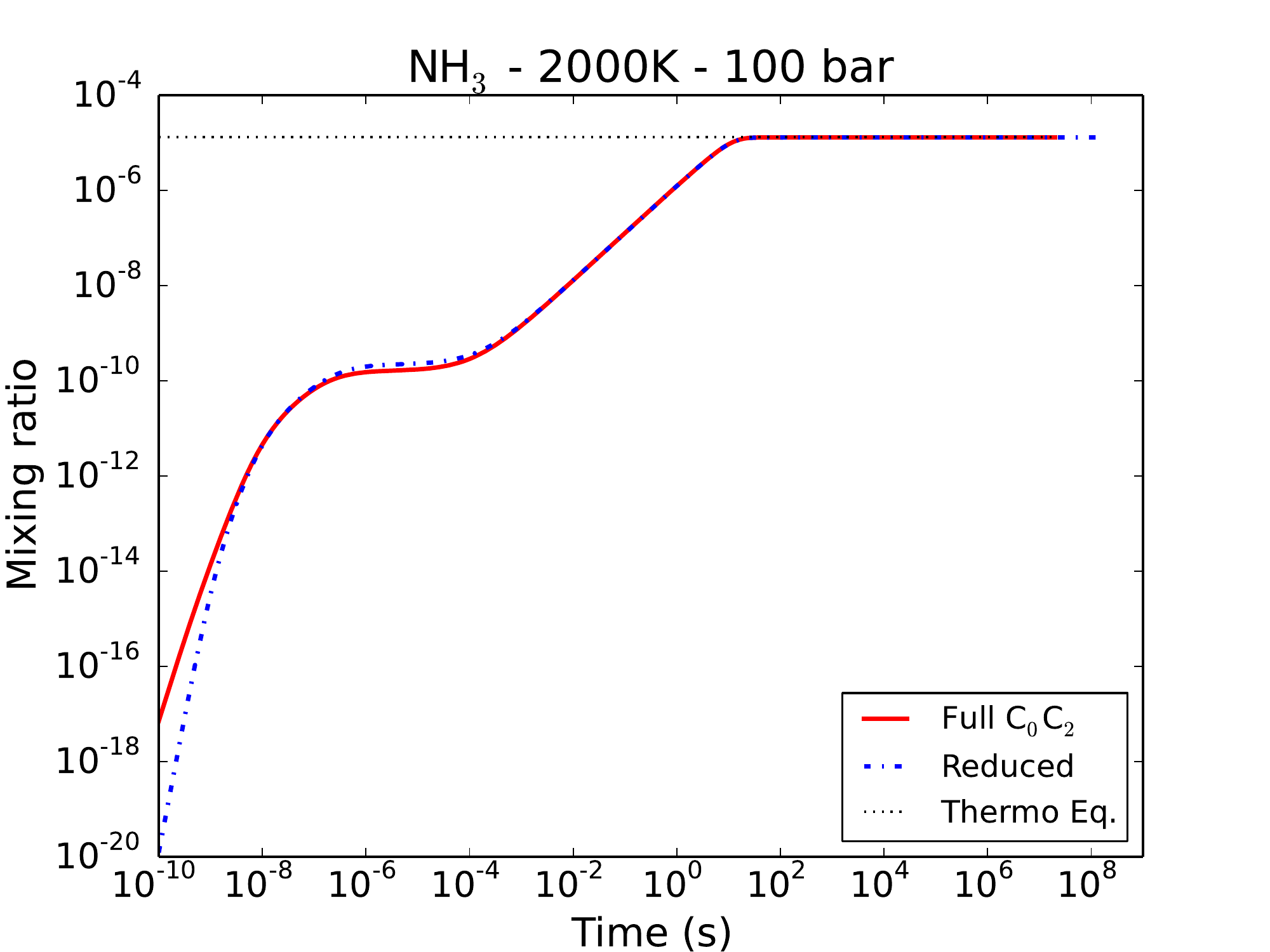}
\includegraphics[angle=0,width=0.9\columnwidth]{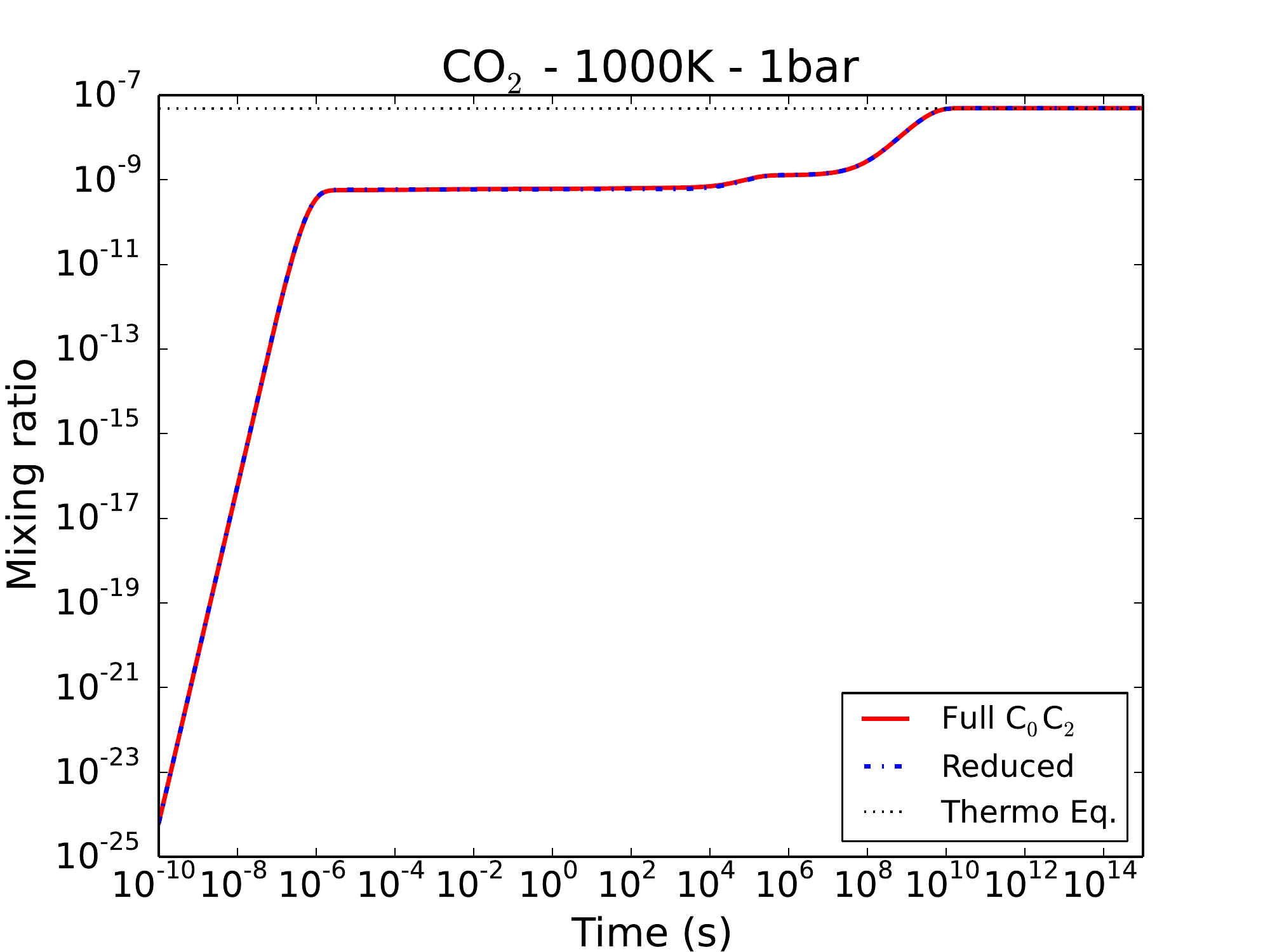}
\includegraphics[angle=0,width=0.9\columnwidth]{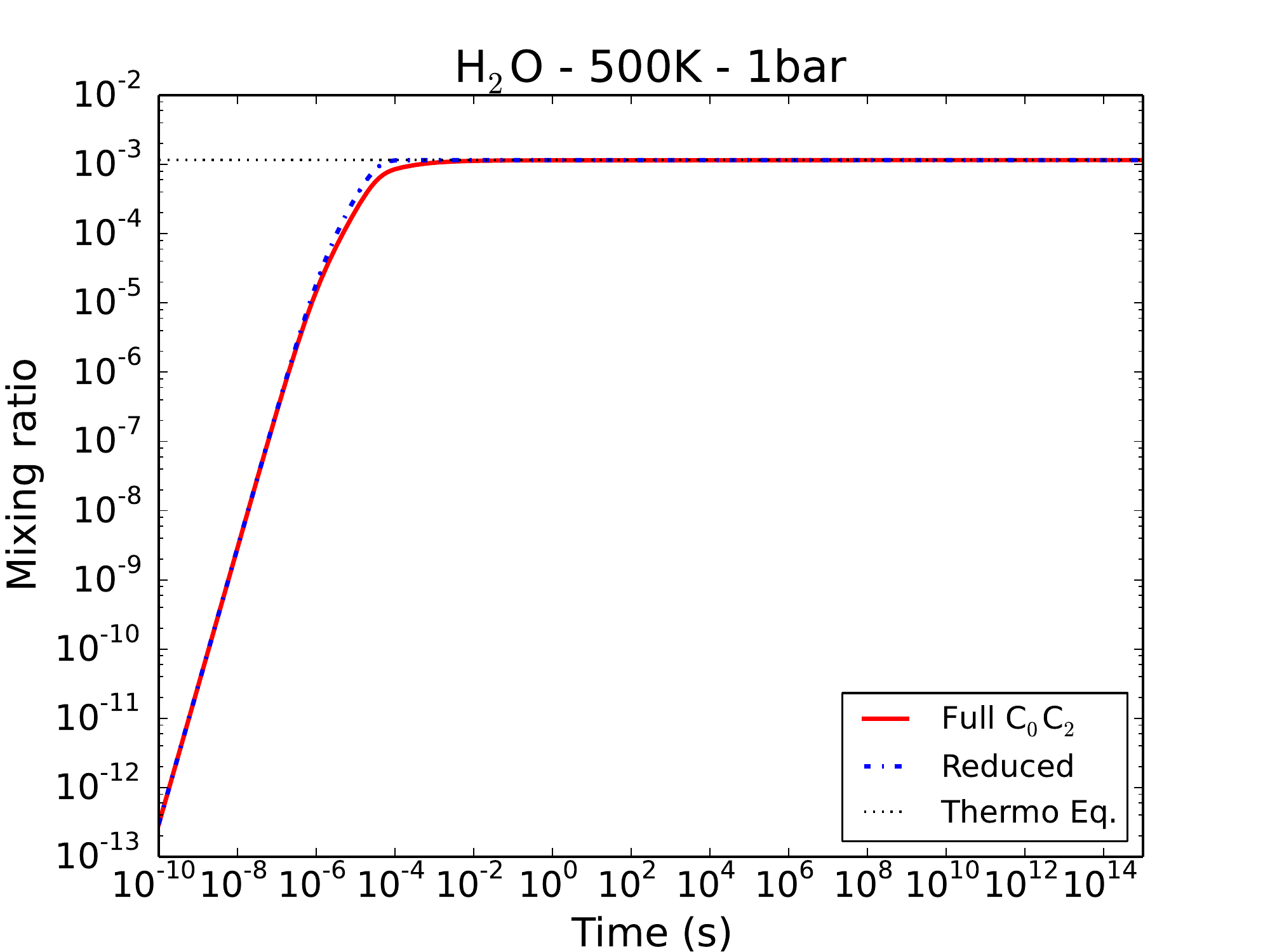}
\caption{Temporal evolution of abundances for CH$_4$, NH$_3$, CO$_2$, and H$_2$O at different temperature and pressure conditions, as labeled on each figure. Abundances are calculated with the full chemical scheme (red solid lines) and with the reduced scheme (blue dot-dashed lines). These abundances are compared with the thermochemical equilibrium value (black dotted lines). Initial conditions are solar abundances spread into H$_2$, He, CH$_4$, O($^3$P), and N$_2$.}
\label{fig:0D}
\end{figure*}
We have used the method DRGEP (Directed Relation Graph with Error Propagation), followed by a Sensitivity Analysis (SA). Initially, the Directed Relation Graph (DRG) method has been developed by \cite{lu2005} and extended to the DRGEP method by \cite{pepiot2008} followed by \cite{liang2009}.
The mathematical explanation of these methods is beyond the scope of this paper, but the detailed algorithms of these mechanism reduction methods were described in \cite{lebedev2013}. To summarize, from a list of species of interest (target species) defined by the user, in the DRGEP method, the significance of all the non-target species for description of the target species is determined by analysing the rates of transformations between them. Each non-target species is thus characterised by an importance index regarding to each target species. The maximum of these values (amount equals to the number of target species) define the overall importance of each non-target species in the chemical scheme. According to the level of accuracy/completeness desired by the user, all species with an overall importance index above the threshold value are included in the reduced scheme. In the SA method, the maximum value of the concentration sensitivity to each reaction is calculated and compared to a threshold value. The reactions with all sensitivity coefficients smaller than the threshold value are excluded from the mechanism.

We can mention that these approaches are commonly used in the field of combustion. Recently, \cite{qiu2016} have used these approaches to transform an n-decane/$\alpha$-methylnaphthalene/polycyclic aromatic hydrocarbon (PAH) kinetic mechanism involving 108 species and 846 reactions into a reduced model including 56 species and 236 reactions. The newly developed mechanism was validated by experimental data in fundamental reactors, including mole fraction of key species in the ethylene premixed flame and jet stirred reactor, as well as ignition delay time of pure and mixed fuel. 
In our study, these methods identify and select the key reactions that govern the chemical composition for a given thermal profile and given elemental abundances. We have first employed the DRGEP methodology to identify and eliminate unimportant species and associated reactions. We then implemented a reaction sensitivity analysis module to eliminate less important reactions. The use of DRGEP before the sensitivity analysis is required owing to the fact that SA is very time-consuming.

We have applied these two methods to our nominal atmospheric structure of GJ 436b: a thermal profile of GJ 436b (Fig.~\ref{fig:PT}) determined with the code \texttt{ATMO} \citep{tremblin:2015aa}. For the opacities, we used the parameters of the benchmark study of \cite{Baudino2017} and assumed a composition at thermochemical equilibrium with solar elemental abundances (but with a reduction of 20\% of oxygen due to sequestration). In our kinetic model, we used the same elemental abundances and assumed a vertical mixing $K_{zz}$ = 10$^8$cm$^2$s$^{-1}$. With the reduced scheme, we aim at being able to reproduce the abundances of a series of species of interest :  H$_2$O, CH$_4$, CO, CO$_2$, NH$_3$, and HCN. These species are both the most abundant ones and all have strong infrared spectral signature. Note that we could add acetylene (C$_2$H$_2$) to this list, as this species is potentially important for C-rich atmospheres \cite[e.g.][]{Moses2013CO,Venot2015}. We chose instead to focus on species that have already been observed in exoplanets, which is not the case of acetylene currently, albeit it is present in Jupiter's atmosphere \citep{nixon2010}.  As we will see in Sect.~\ref{sec:validity}, the absence of this species in our reduced scheme limits the range of validity of the latter to atmospheres in which C$_2$H$_2$ is not very abundant: atmospheres with both a high C/O ratio (larger than 1) and a high temperature (T $\gtrsim$1000 K) will not be perfectly represented with the reduced chemical scheme; C-rich atmospheres with a lower temperature can be studied with the reduced scheme as acetylene is not a major component of these atmospheres \citep{Venot2015}.
The methodology used in (DRGEP+SA) stops the reduction process when the error on targeted parameters goes beyond the user-specified tolerance level during removal of species from the master mechanism. For the set of species cited above, we have chosen a value of relative tolerance of 10\%, globally representative of the uncertainty on the predicted abundance determined by our uncertainty propagation study (see Sect.~\ref{sec:MC}). Our goal is to generate a reduced mechanism able to reproduce the prediction of concentrations given by the original mechanism, within the error bars. We want to obtain a reduced mechanism as small as possible and valid on a very large range of operating conditions, allowing to study various kind of atmospheres (Fig.~\ref{fig:PT}).
The original mechanism includes 957 reversible reactions, involving 105 neutral species (molecule or radical). The reduced mechanism includes 181 reversible reactions involving 30 different species: H$_2$, H, H$_2$O, CO, N$_2$, CH$_4$, CH$_3$, NH$_3$, NH$_2$, HCN, OH, CO$_2$, H$_2$CO, HCO, $^3$CH$_2$, $^1$CH$_2$, O($^3$P), CH$_3$OH, CH$_2$OH, CH$_3$O, H$_2$CN, HNCO, HOCN, CN, NCO, NH, NNH, N$_2$H$_2$, N$_2$H$_3$, and He (the latter only being involved as a third-body).

In terms of temperature and pressure, the reduced chemical scheme has the same range of validity of our full chemical scheme (i.e. [300-2500] K and [0.01--100] bar). The temporal evolution of abundances with time are very similar with both chemical schemes and they are both able to reproduce thermochemical equilibrium (Fig.~\ref{fig:0D}). 
Finally, we obtained a reduction factor of about 3 for the number of species and about 5 for the number of reactions, which leaves hope for a possible coupling of such mechanism in a 3D model. For a given atmosphere, using the reduced scheme instead of the full chemical scheme in our 1D kinetic model permit $\sim$30$\times$ faster runs. This reduced scheme is available in the KIDA Database \citep{KIDA2012}\footnote{\href{http://kida.obs.u-bordeaux1.fr/}{kida.obs.u-bordeaux1.fr}}.

\subsection{Uncertainty propagation model}

\begin{figure*}[t]
\centering
\includegraphics[angle=0,width=0.9\columnwidth]{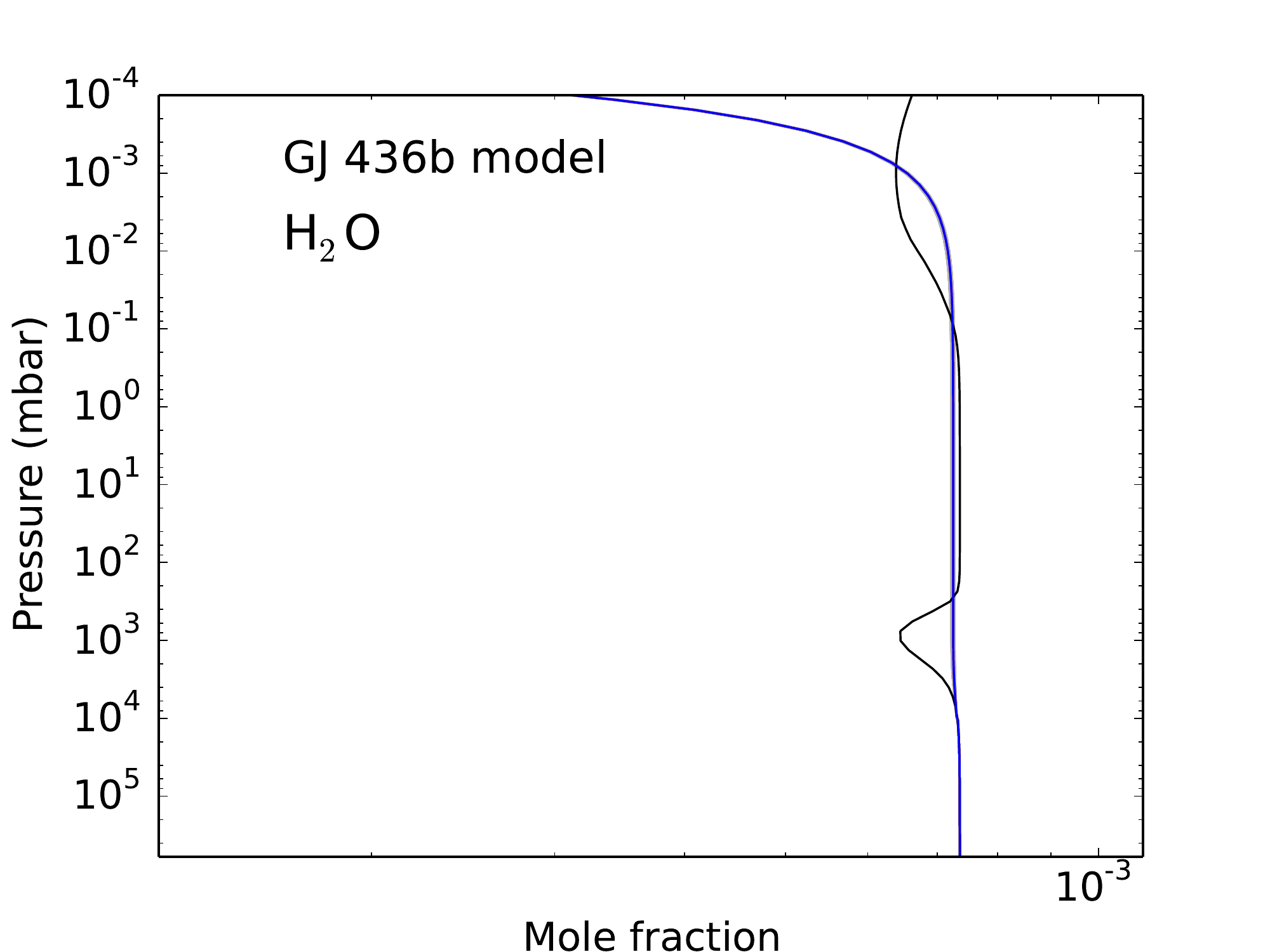}
\includegraphics[angle=0,width=0.9\columnwidth]{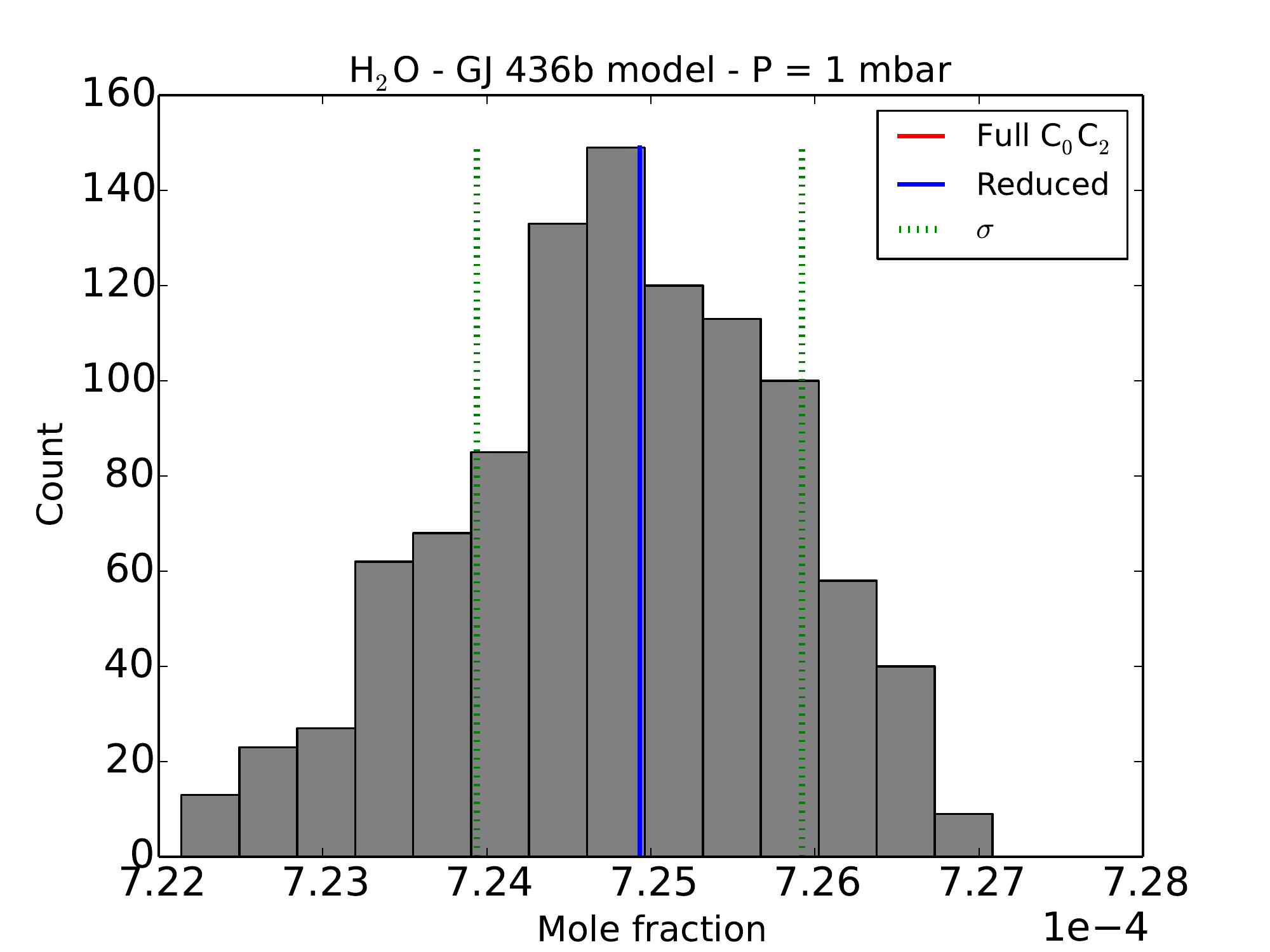}
\includegraphics[angle=0,width=0.9\columnwidth]{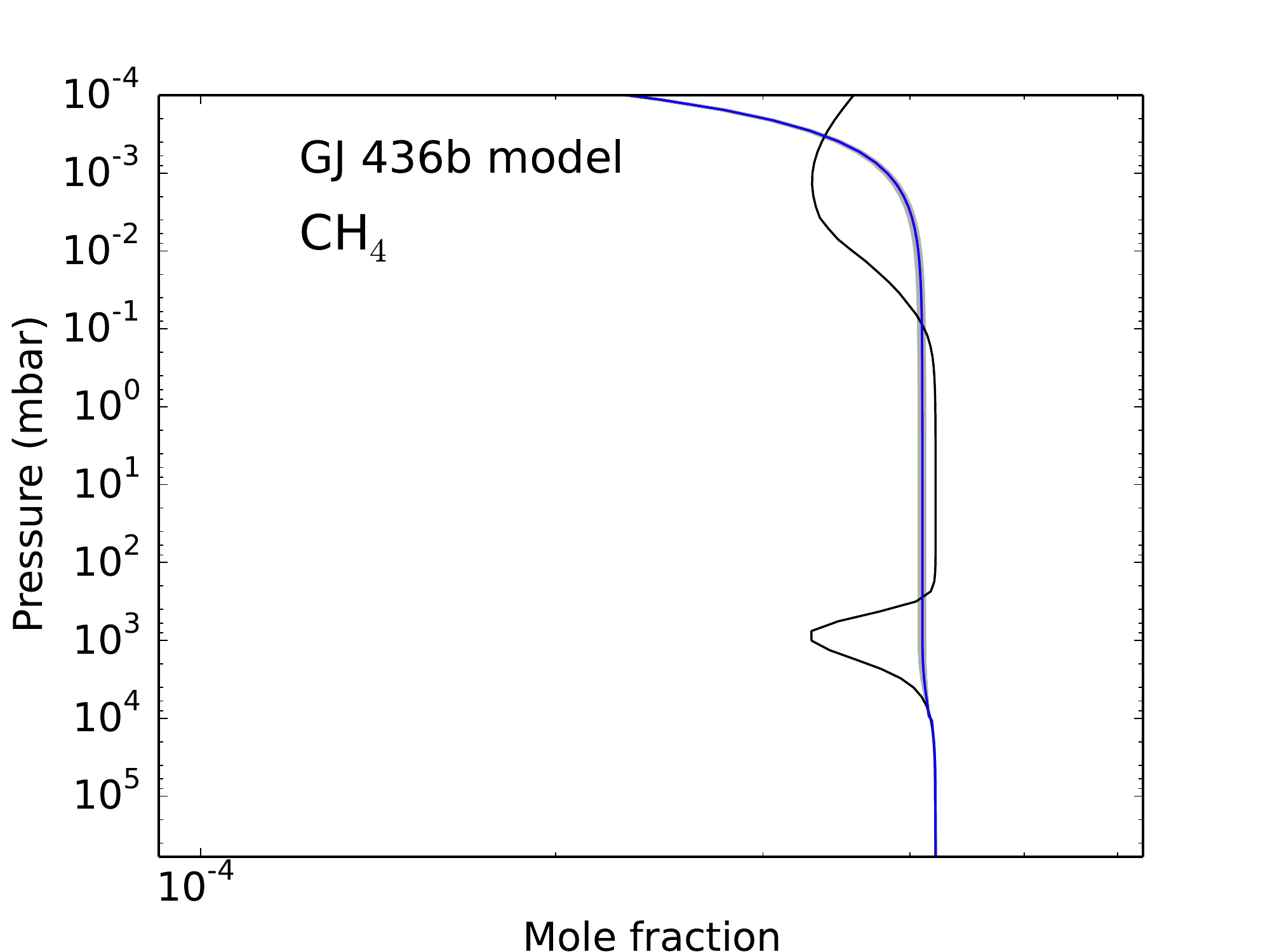}
\includegraphics[angle=0,width=0.9\columnwidth]{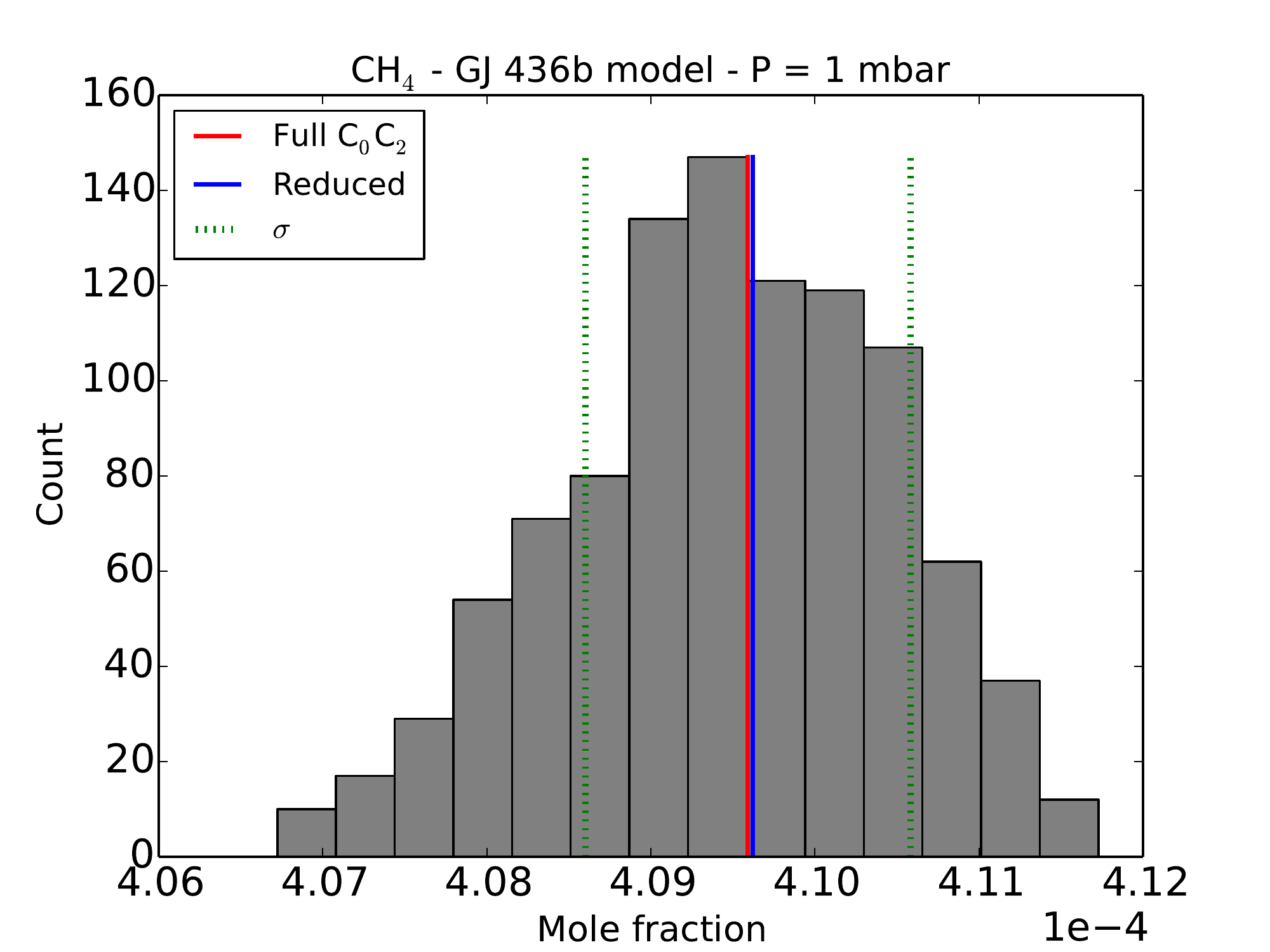}
\includegraphics[angle=0,width=0.9\columnwidth]{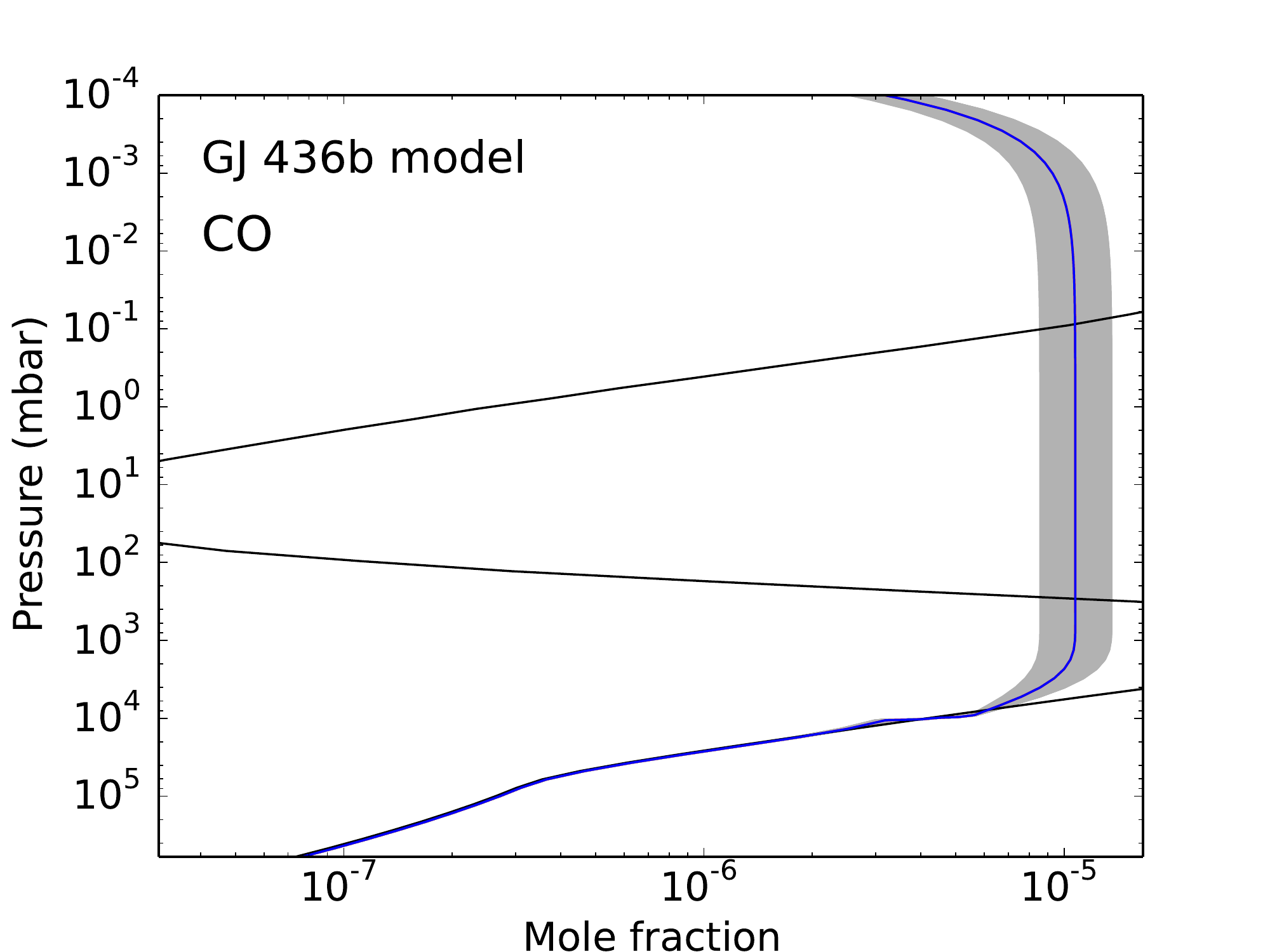}
\includegraphics[angle=0,width=0.9\columnwidth]{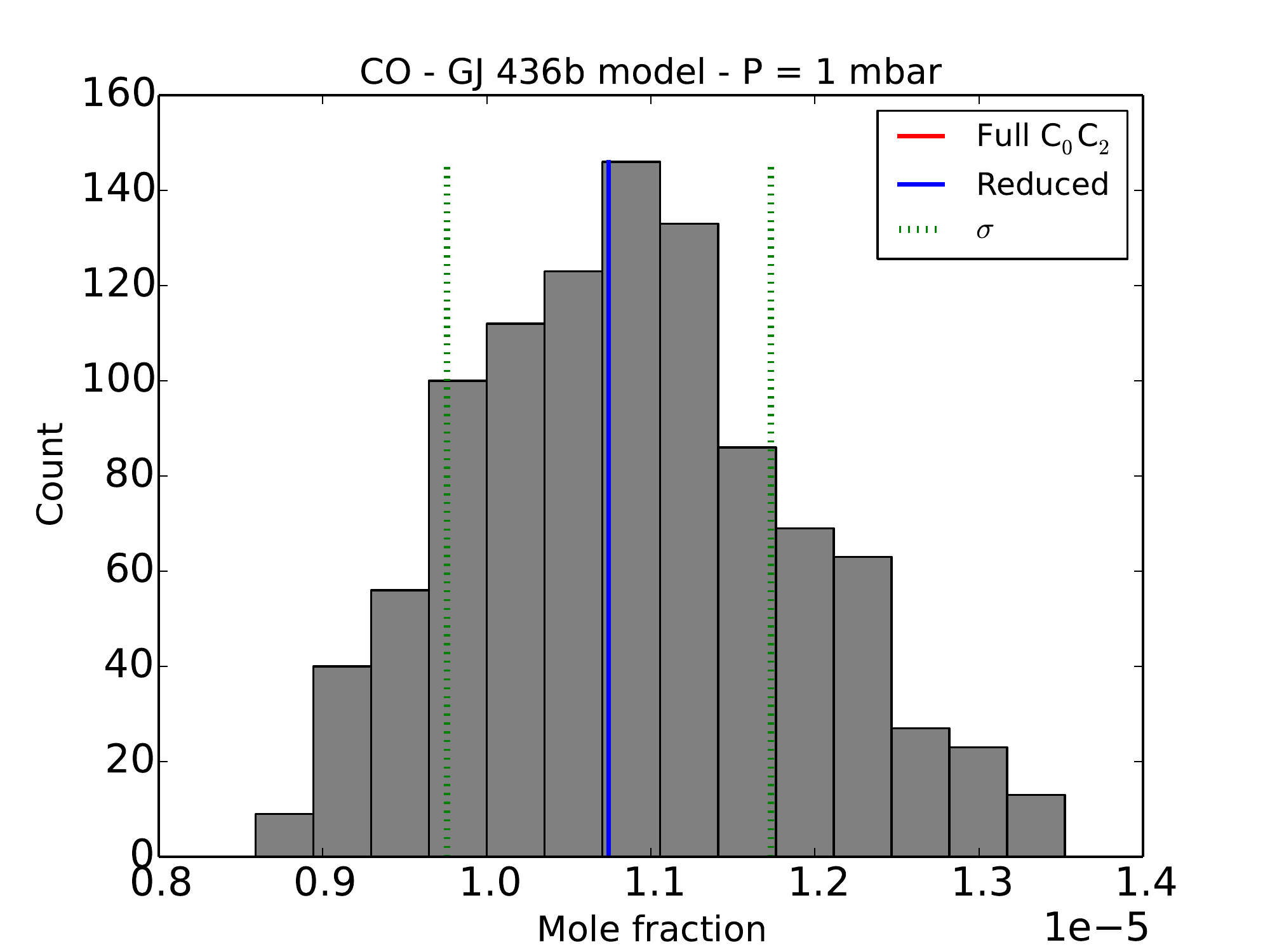}
\caption{Left: Vertical abundance profiles determined after 1000 runs with the full chemical scheme (grey lines) for H$_2$O, CH$_4$, and CO. The nominal results of the full chemical scheme (red lines),  thermochemical equilibrium (black lines), and results obtained with the reduced scheme (blue lines) are also shown. Right: Corresponding distribution of abundances at 1 mbar represented with grey bars, with nominal abundances of the full scheme (red lines), abundances obtained with the reduced scheme (blue lines), and the 1-$\sigma$ intervals around the nominal abundances (green dotted lines). Blue and red lines are very close on each plot.}
\label{fig:MC_abundances_nominal}
\end{figure*}
\begin{figure*}[t]
\centering
\includegraphics[angle=0,width=0.9\columnwidth]{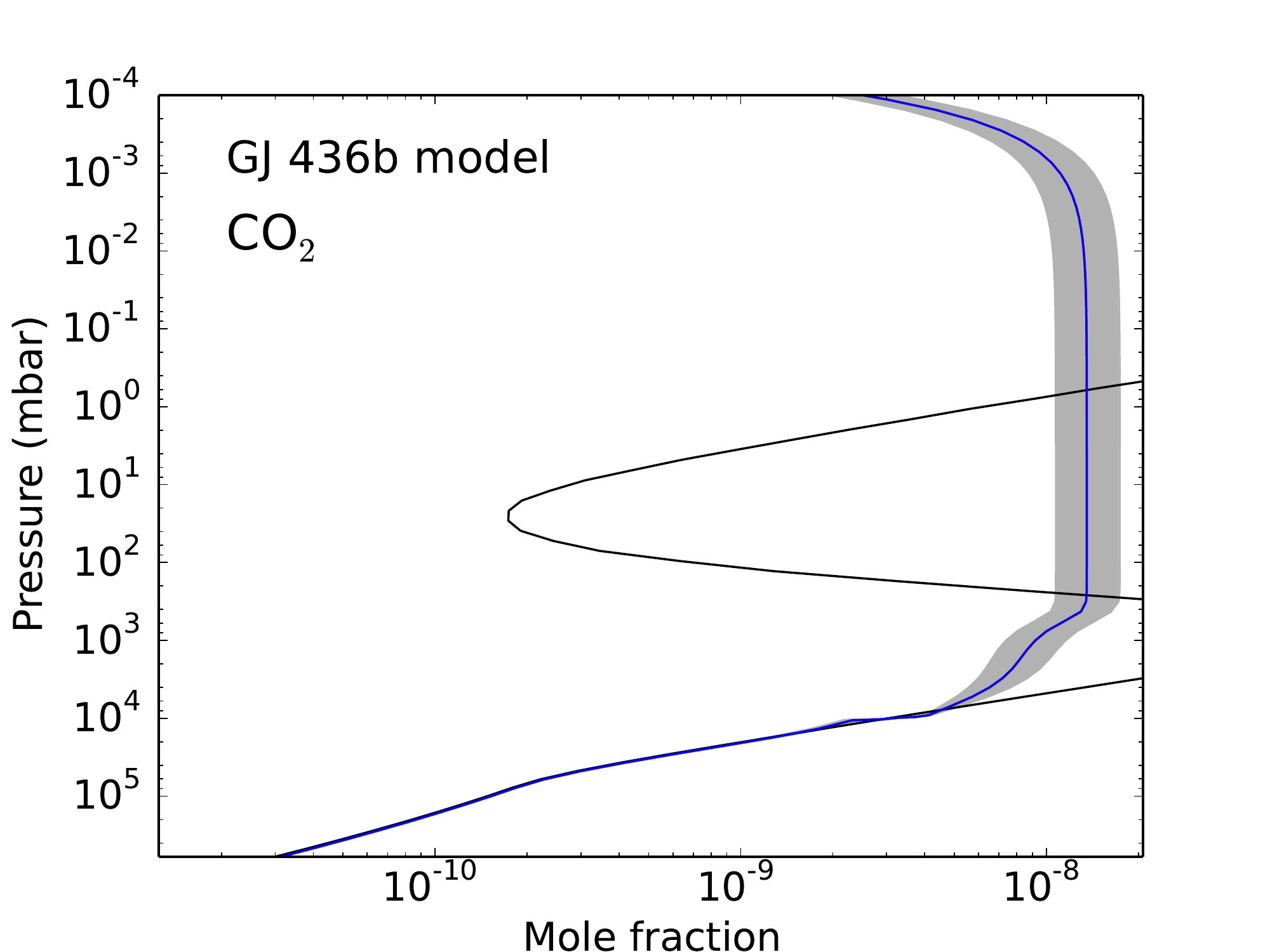}
\includegraphics[angle=0,width=0.9\columnwidth]{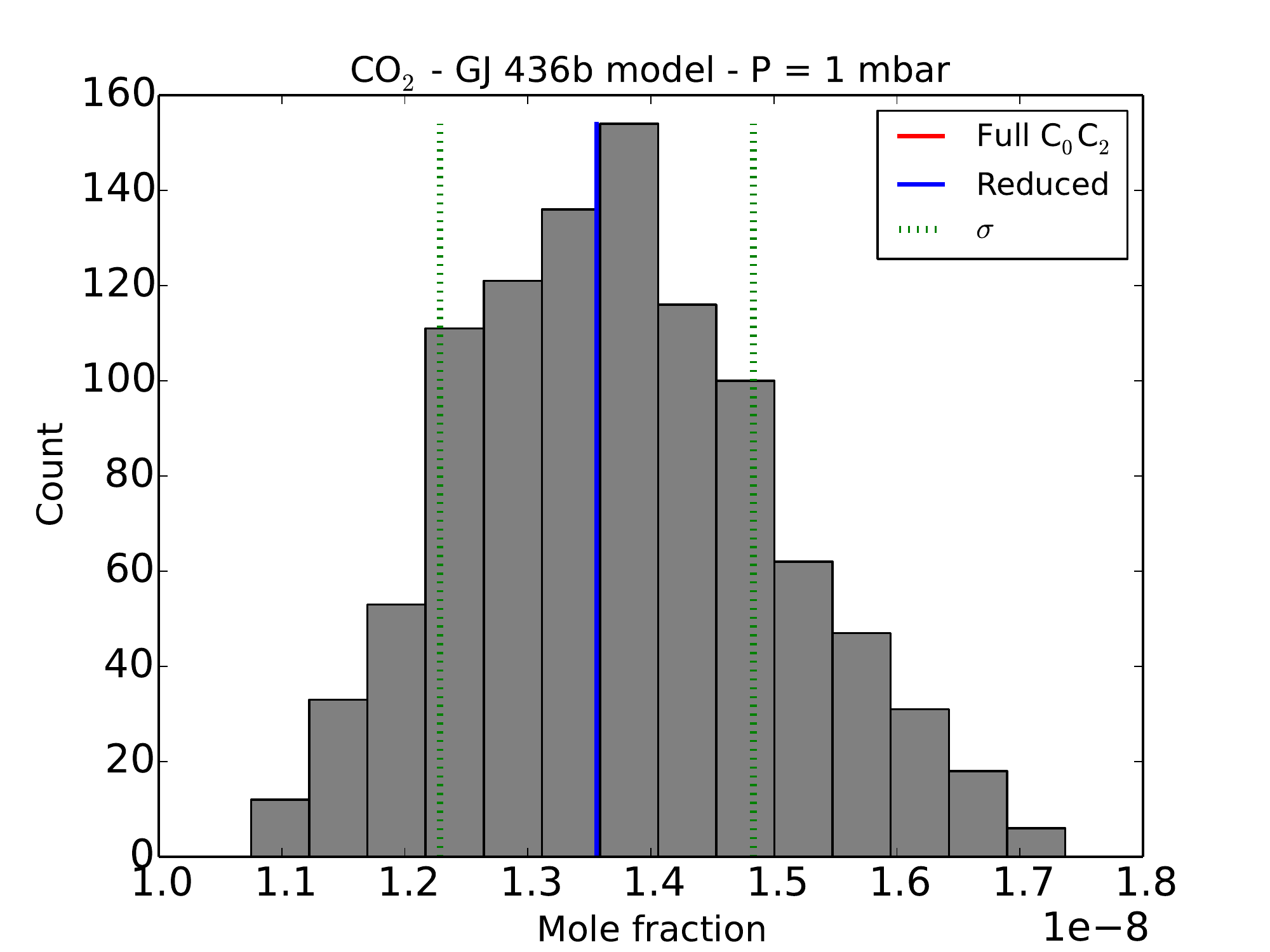}
\includegraphics[angle=0,width=0.9\columnwidth]{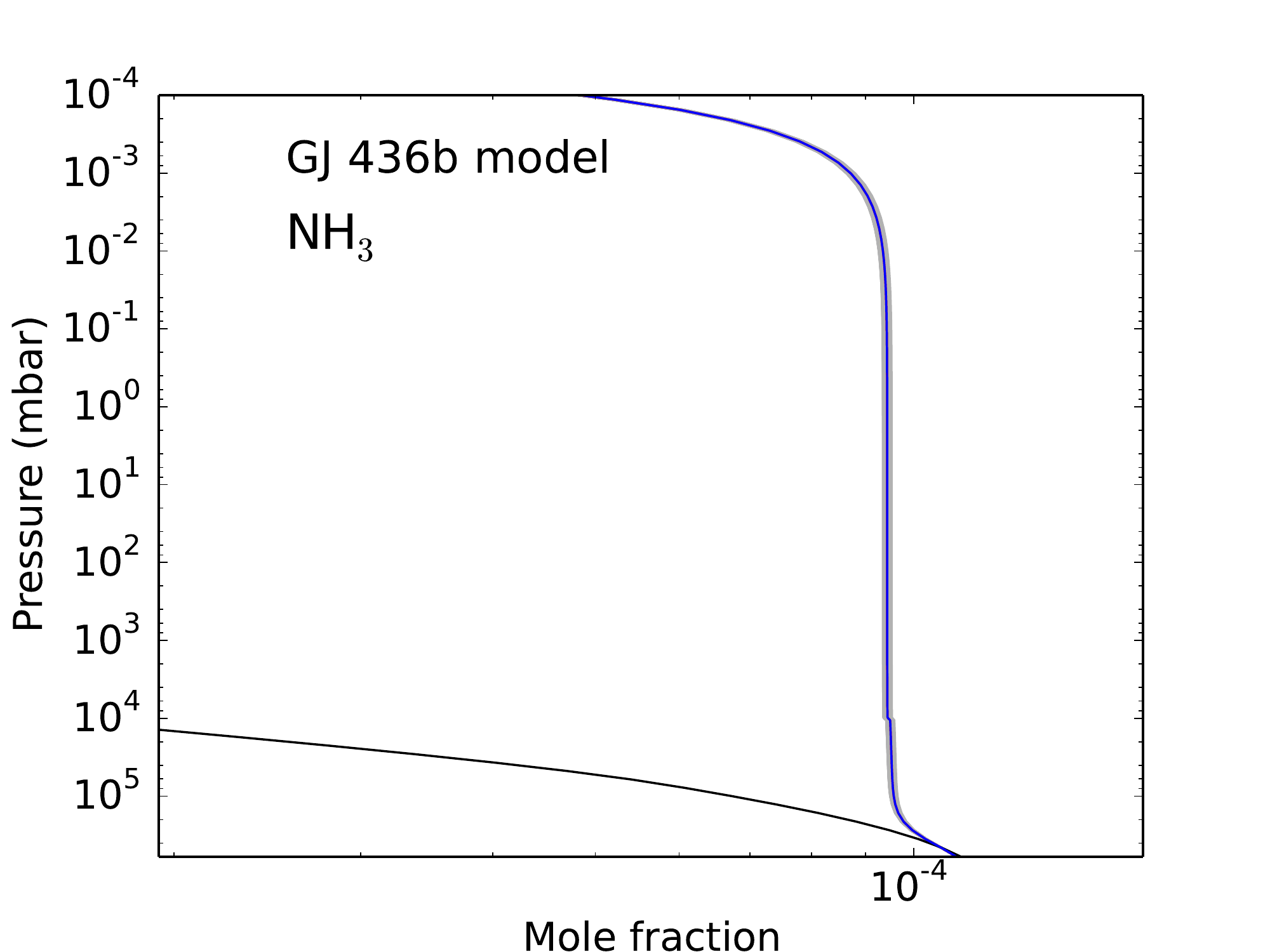}
\includegraphics[angle=0,width=0.9\columnwidth]{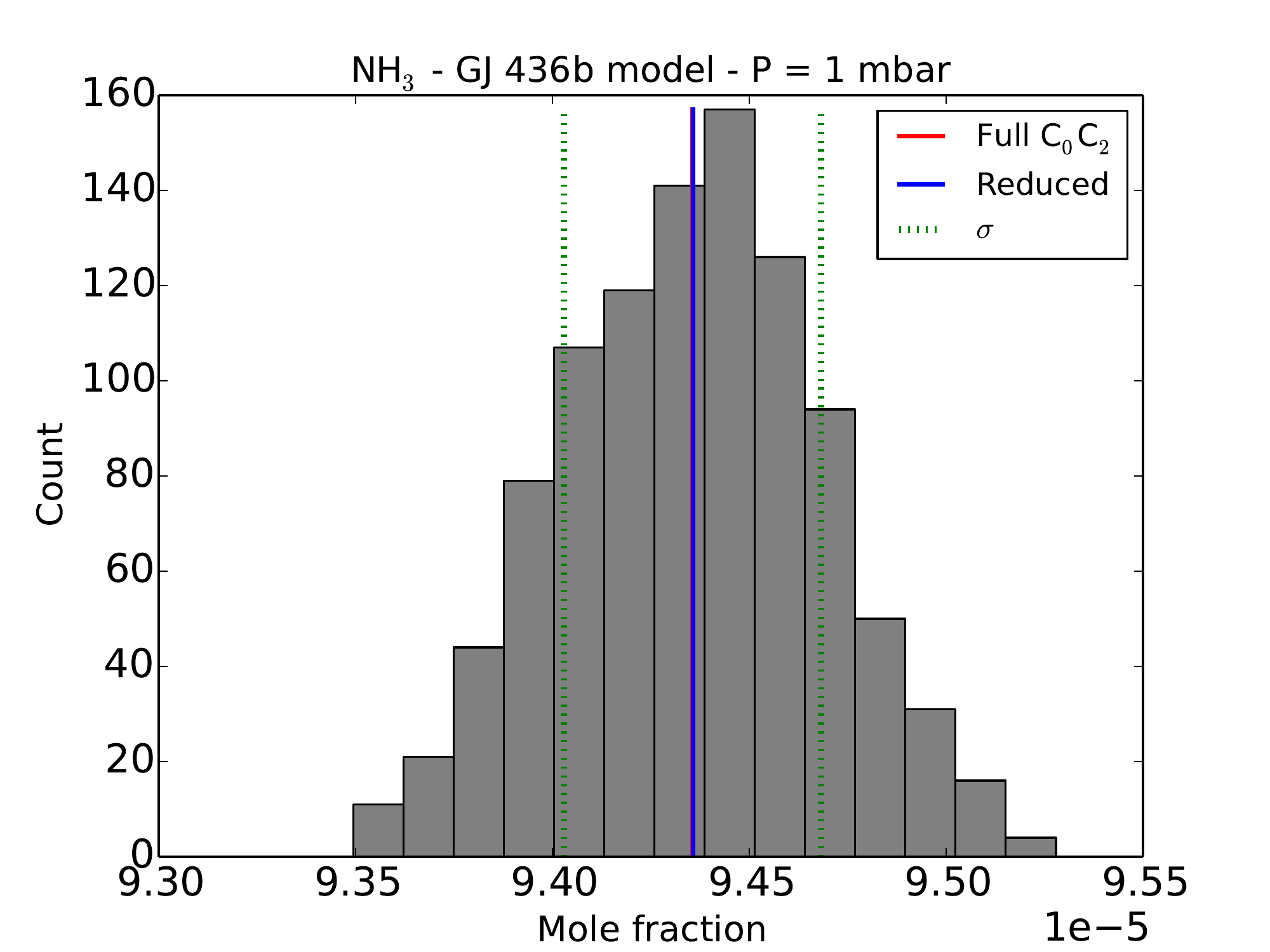}
\includegraphics[angle=0,width=0.9\columnwidth]{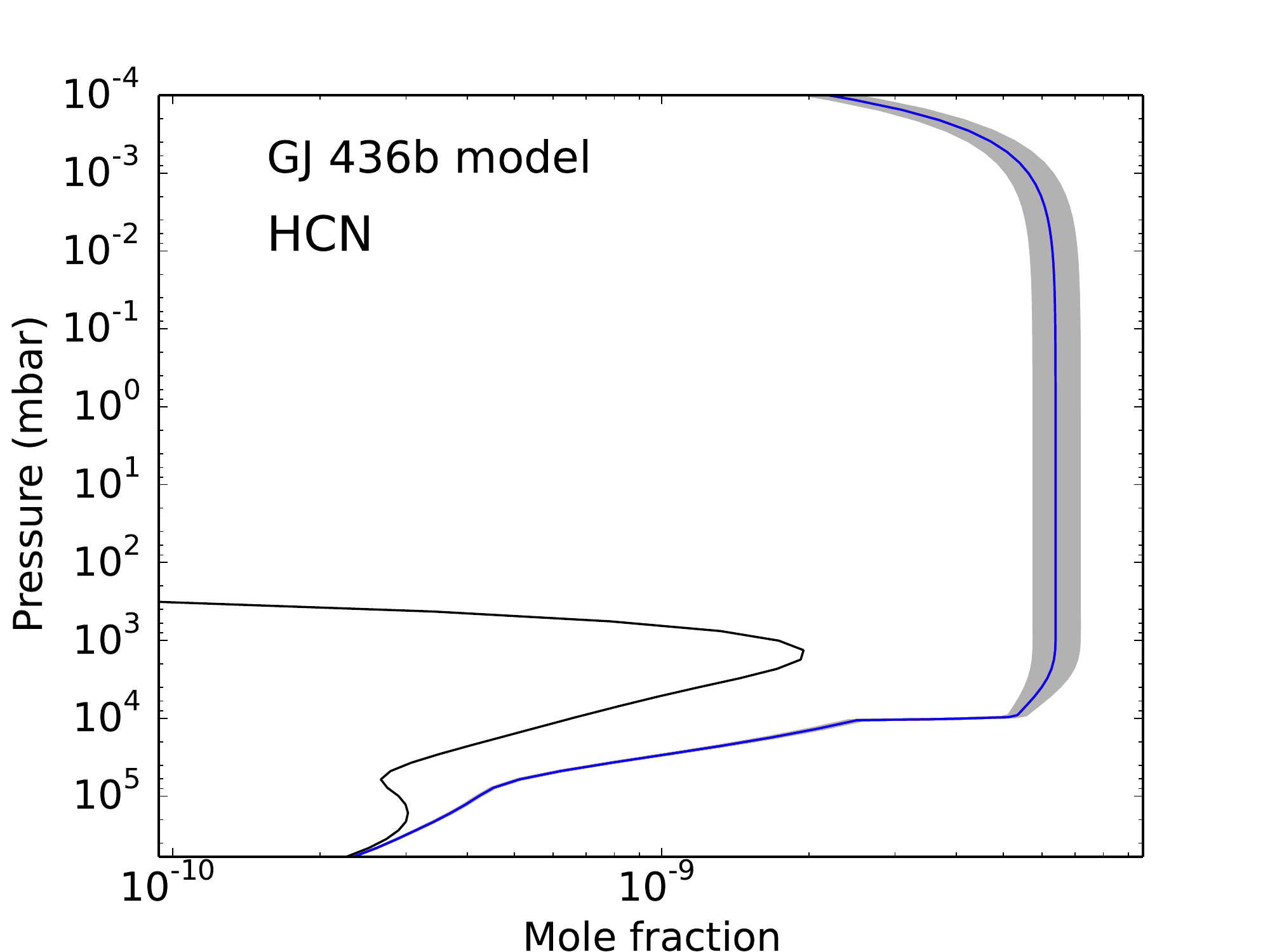}
\includegraphics[angle=0,width=0.9\columnwidth]{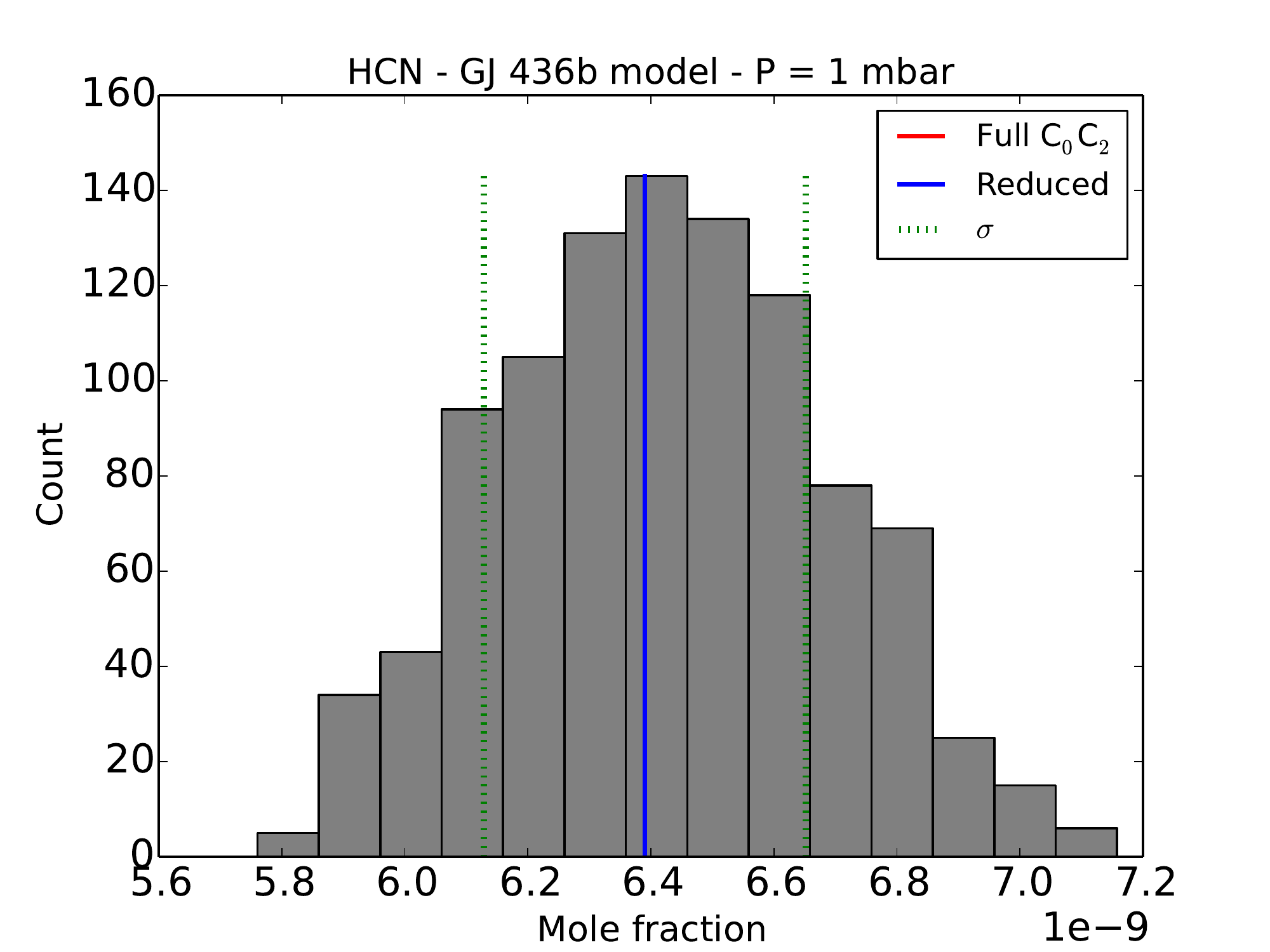}
\caption{Same as Fig.\ref{fig:MC_abundances_nominal} for CO$_2$, NH$_3$, and HCN.}
\label{fig:MC_abundances_nominal2}
\end{figure*}
\begin{figure*}[!htb]
\centering
\includegraphics[angle=0,width=0.9\columnwidth]{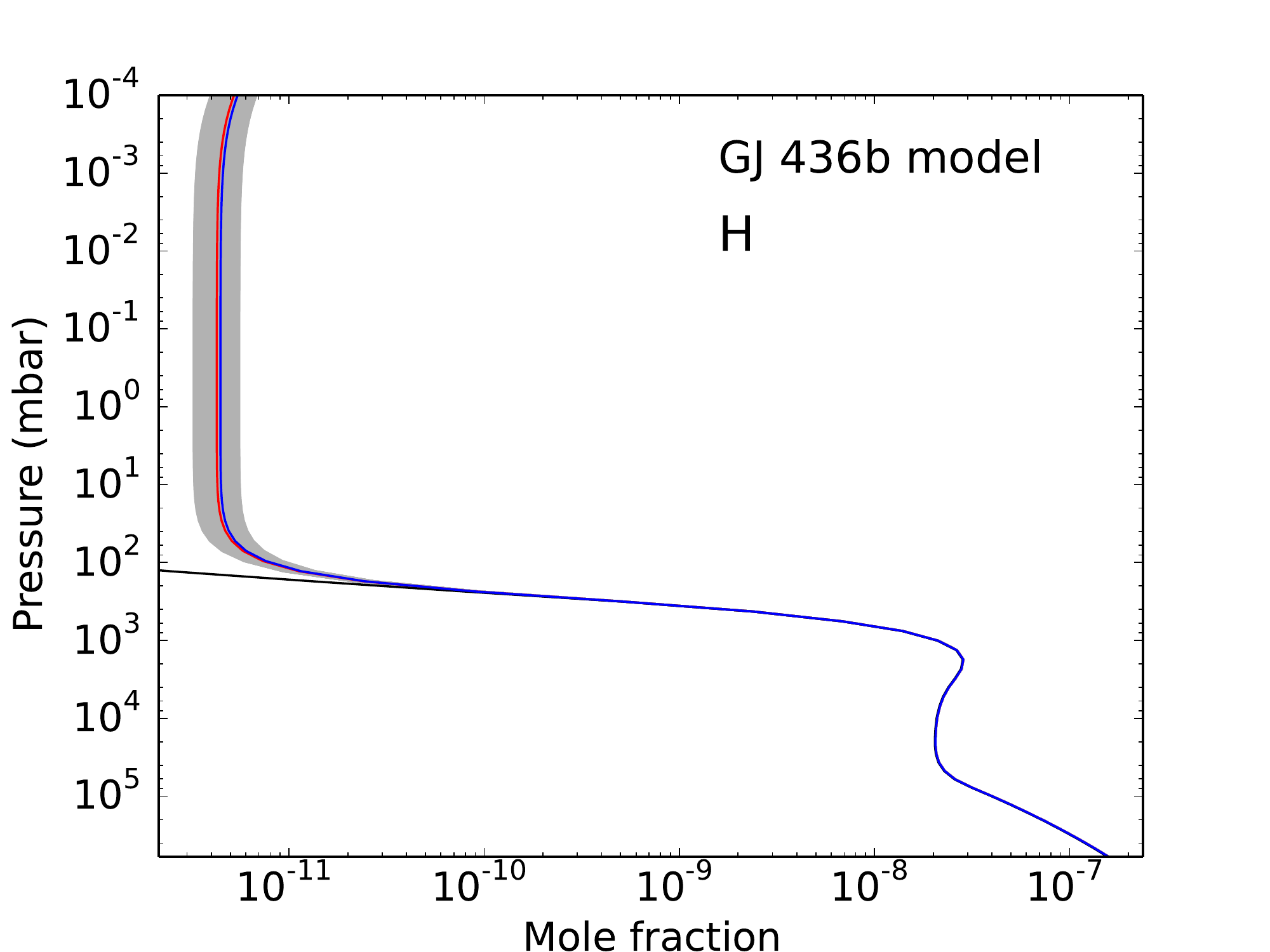}
\includegraphics[angle=0,width=0.9\columnwidth]{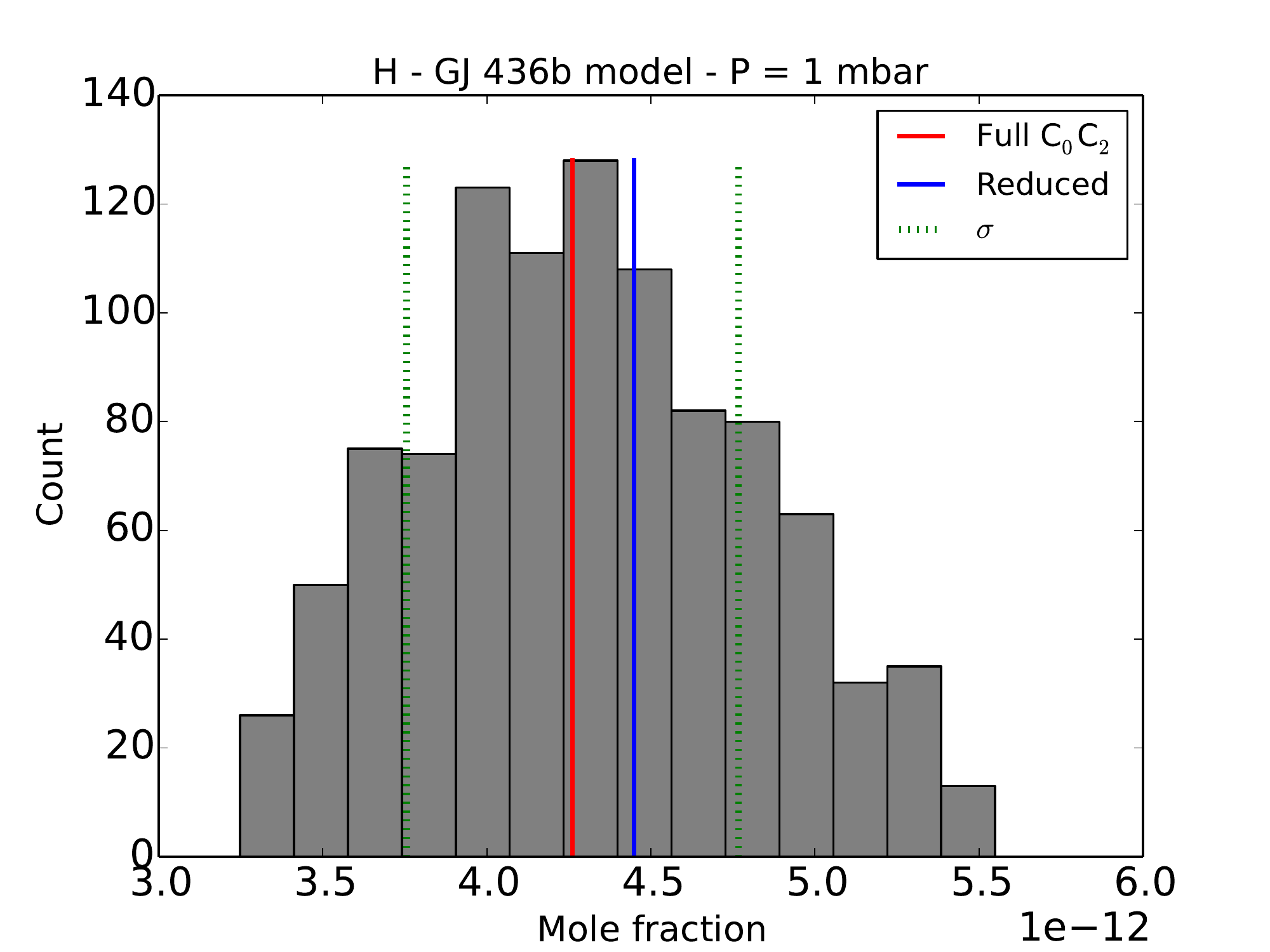}
\includegraphics[angle=0,width=0.9\columnwidth]{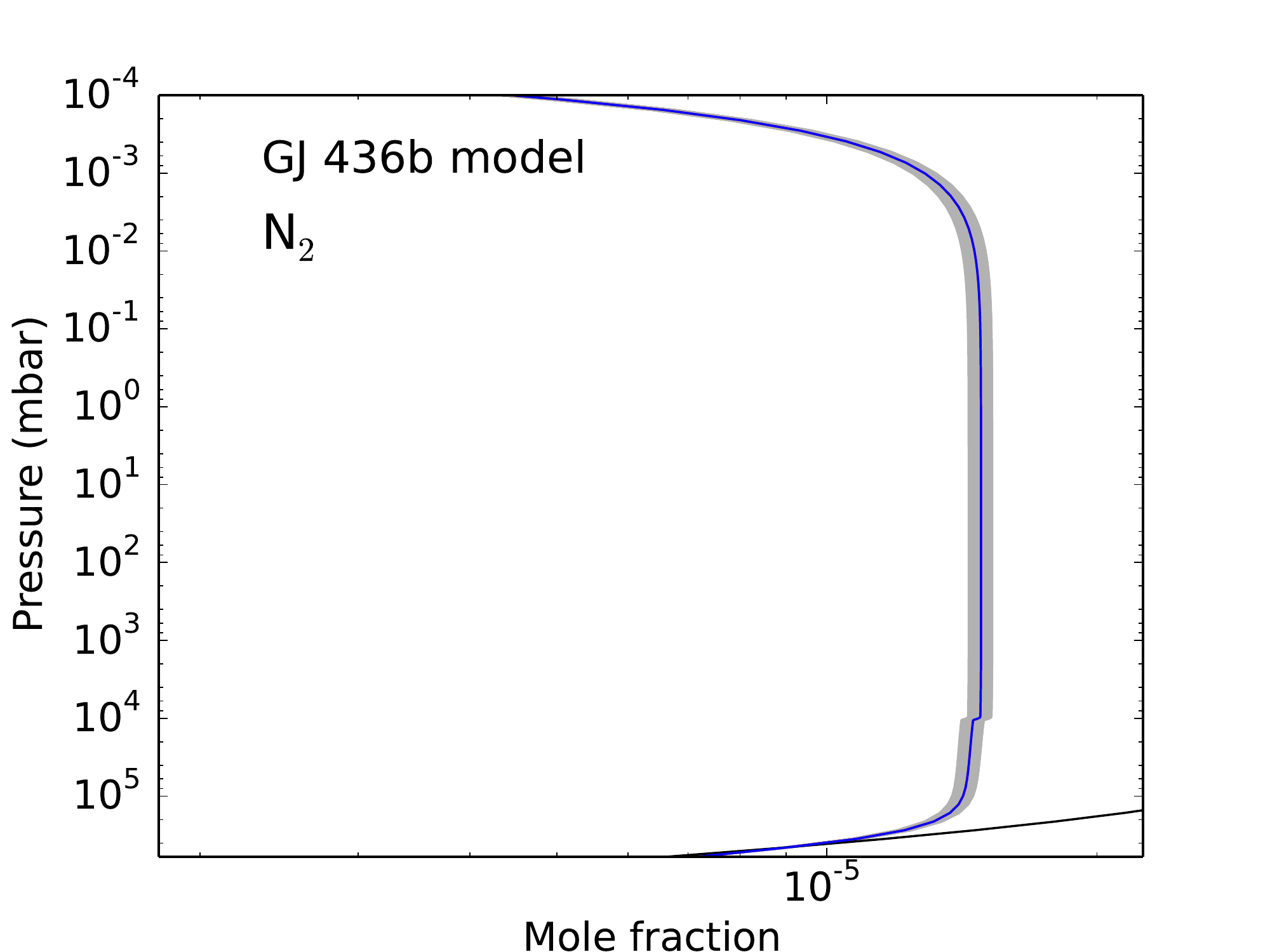}
\includegraphics[angle=0,width=0.9\columnwidth]{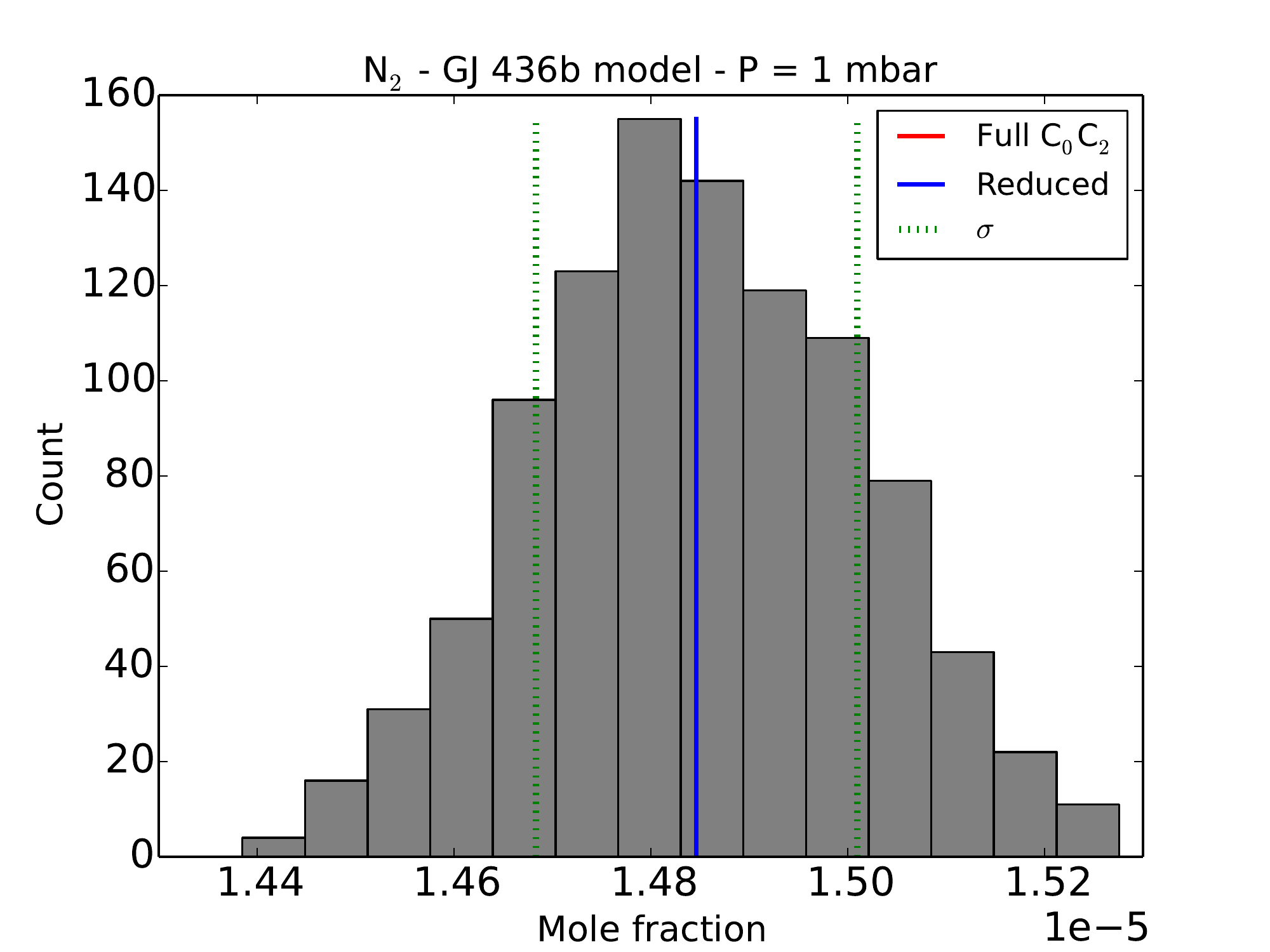}
\includegraphics[angle=0,width=0.9\columnwidth]{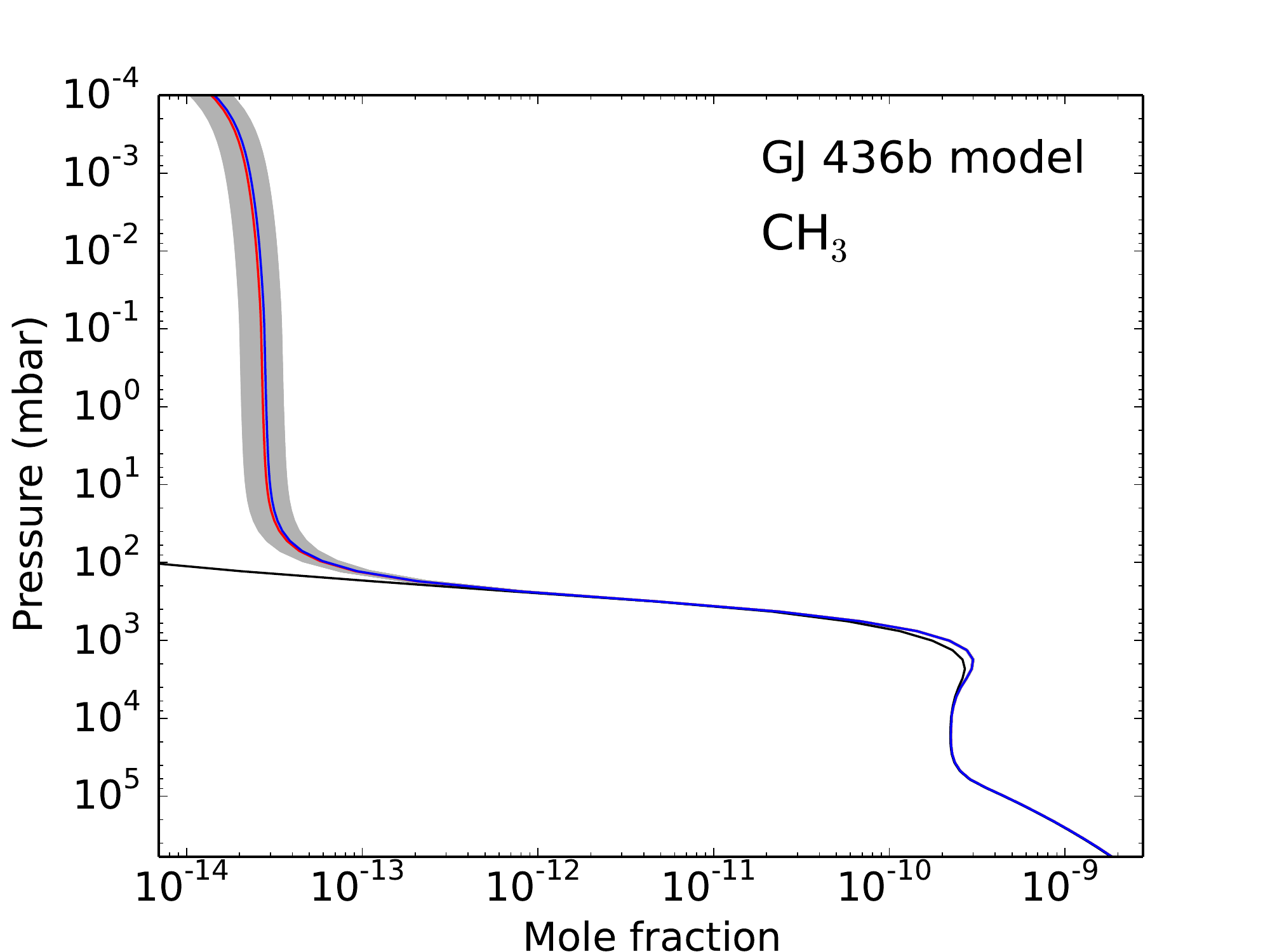}
\includegraphics[angle=0,width=0.9\columnwidth]{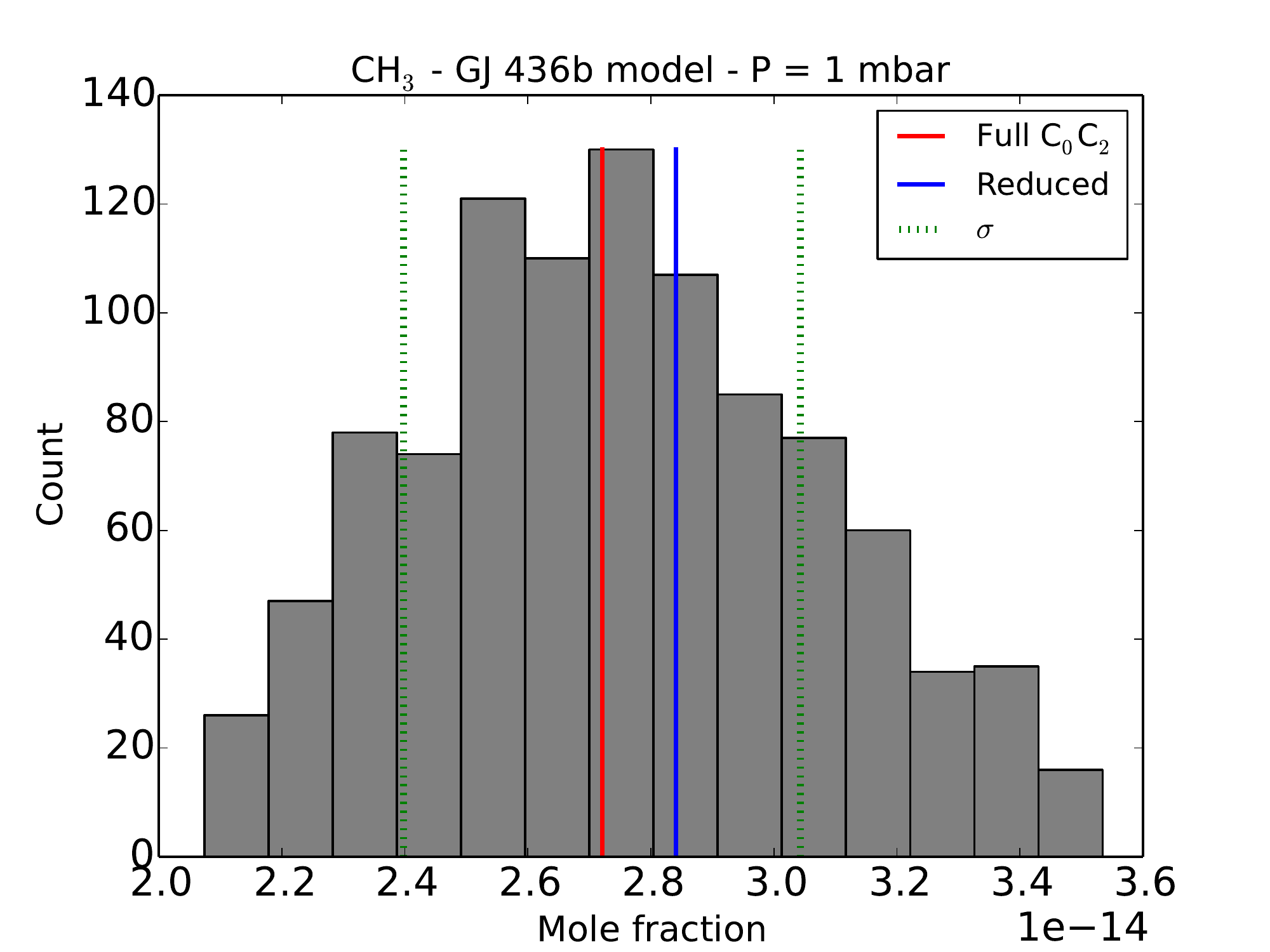}
\caption{Same as Fig.\ref{fig:MC_abundances_nominal} for H, N$_2$, and CH$_3$. Note that these three species are not in the list of species for which the reduced scheme has been designed.}
\label{fig:MC_abundances_nominal3}
\end{figure*}
To go further than a simple comparison of the vertical abundances given by the kinetic model when using the full and the reduced chemical schemes, we performed an uncertainty propagation study with the full scheme to have an objective reference. Indeed, the precision of the kinetic model is related to the uncertainties in the rate coefficients of the chemical scheme. Uncertainties of all reactions combine to each other and propagate non-linearly in the kinetic model through the system of differential equations, resulting from the continuity equations \citep[for details on continuity equations, see e.g.][]{moses2011, Venot2012, Drummond2016}. To study this uncertainty propagation, we use the method described in \cite{dob1998, dob2003}, applied to a chemical scheme whose reactions are reversible. The rate constant of each forward chemical reaction was pseudo-randomly estimated according to its own temperature-dependent uncertainty factors $F(T)$ following the data evaluation and the numerical method described in  \cite{hebrard2015}. The corresponding reverse rate constant was determined using the reaction equilibrium constant. This way, thermochemical equilibrium is maintained in the deep layers of the atmosphere, or in absence of disequilibrium processes. The sets of perturbed rates are used to run the kinetic model and determine `alternative' atmospheric compositions for our model of GJ 436b. They represent the uncertainty on the predicted abundances. We have performed 1000 Monte-Carlo simulations allowing us to have statistically significant results. All simulations are run for $t$=10$^9$s, which is the time needed to reach steady-state for the atmosphere of GJ 436b. The uncertainty factors for chemical reactions included in our initial C$_0$-C$_2$ scheme have been published in \cite{hebrard2015}. Note that we do not consider any uncertainty that is potentially associated with the NASA polynomial coefficients \citep{mcbride1993nasa}, used to calculate the equilibrium constant. To our knowledge, no uncertainty on these coefficients have been evaluated so far \citep{goldsmith2012}. Such ambitious work of bibliographic census and analysis would be necessary for a complete evaluation of the uncertainty on the predicted abundances. This is beyond the scope of this study that aims at developing a reduced scheme reproducing the equilibrium state determined with a more complete one and nominal values of NASA polynomial coefficients.

\section{Results}\label{sec:results}

\subsection{GJ 436b model}\label{sec:GJ436b}

As can be seen on Figs.~\ref{fig:MC_abundances_nominal}, \ref{fig:MC_abundances_nominal2}, and \ref{fig:MC_abundances_nominal3}, the agreement between the atmospheric compositions obtained with the two chemical schemes is very good. Table \ref{tab:variation} gathers, for each selected species, the maximum difference of abundances obtained using the two chemical networks and at which pressure level this value is reached. We focus our study on the [0.1--1000] mbar region only, which is the region probed by infrared observations.
The species that presents the largest difference is CO$_2$, with 0.02\% at 300 mbar. The other species present differences always lower than this value.

\begin{table}[!h]
\caption{For the nominal model, maximum ($\Delta max$) differences of abundances (in \%) for each species for which the reduced scheme has been built. The pressure level (@level in mbar) where the maximum variation is reached is indicated within parentheses. These values are calculated within the [0.1--1000] mbar region only, which is probed by infrared observations.} \label{tab:variation}
\centering
\begin{tabular}{l|l}
\hline \hline
Species & $\Delta max$ \\
\hline
% OK - actualisé red8
H$_2$O  & 2$\times$10$^{-4}$ (@6$\times$10$^{2}$)\\
 \hline
CH$_4$ & 7$\times$10$^{-3}$ (@8$\times$10$^{2}$)\\
  \hline
CO &   1$\times$10$^{-2}$ (@1$\times$10$^{-1}$)\\       
 \hline
CO$_2$   &  2$\times$10$^{-2}$ (@3$\times$10$^{2}$)\\  
 \hline
NH$_3$  &  2$\times$10$^{-3}$ (@1$\times$10$^{-1}$)\\
 \hline
HCN   &   2$\times$10$^{-3}$ (@8$\times$10$^{2}$)\\
\hline
\end{tabular}
\end{table}

\subsection{Uncertainty propagation model}\label{sec:MC}
The 1000 Monte-Carlo runs performed for the atmospheric model of GJ 436b allow us to determine the distribution of abundance profiles of each species. It can be seen on Figs. \ref{fig:MC_abundances_nominal}, \ref{fig:MC_abundances_nominal2}, and \ref{fig:MC_abundances_nominal3} that these distributions are highly species- and altitude-dependent. The plots on the left column of these figures represent the vertical abundances profiles, whereas the plots on the right column represent the histogram of abundances at 1 mbar. We chose this pressure level because it is located in the quenched zone and the abundance is representative of the [0.1--1000] mbar region probed by infrared observations. 
For all species, thermochemical equilibrium is maintained in the deep atmosphere. The quenching area (proper to each species) remains the same but the exact quenching pressure level is slightly modified, leading to variations above this point, at lower pressures. For H$_2$O, CH$_4$, and NH$_3$, the width of the distribution of abundance is very small, whereas this width is more important for species such as CO, CO$_2$, and HCN. It is interesting to note that some important species, although not in the list of species that must be well described by the reduced scheme, are however well reproduced by the reduced scheme, e.g. H, N$_2$, CH$_3$.

We plot on each histogram the standard deviation ($\sigma$) of the distribution. For each species of the reduced scheme, the reduced chemical scheme gives an abundance well within the 1-$\sigma$ interval around the nominal abundance obtained with the full C$_0$-C$_2$ chemical scheme. 

\section{Range of validity of the reduced chemical scheme}\label{sec:validity}

The reduced chemical scheme has been designed originally for the atmosphere of GJ 436b, with solar elemental abundances. However, it turns out that this chemical scheme has a very large range of validity. We have used it to model GJ 436b with different atmospheric compositions in terms of metallicities and C/O ratios (Sect. \ref{sec:GJ436b_various}). We have also modelled planets that have been studied with the C$_0$-C$_2$ scheme in the past: in Sect. \ref{sec:HJ} we show results for the hot Jupiters HD 209458b and HD 189733b \citep{Venot2012}. In Sect. \ref{sec:UN}, we show the results for the deep atmosphere of Uranus and Neptune \citep{cavalie2017} and in Sect. \ref{sec:BD}, we present the results obtained for SD 1110-like brown dwarf, modelled for the first time with the C$_0$-C$_2$ scheme. Finally, we explore a range where the reduced scheme is not valid anymore, by modelling HD~209458b with high C/O ratios (3 and 6 times solar) in Sect.~\ref{sec:HD209_Crich}. All the thermal profiles are presented in Fig.~\ref{fig:PT}. It can be viewed that the range of temperatures thus scanned is large: $\sim$[300--2500] K. For each case, the kinetic code is run until steady-state of the atmosphere is reached (i.e. 10$^7$, 10$^8$, or 10$^9$s depending on the planet) and chemical composition obtained with the two schemes are compared.

\subsection{GJ 436b with different elemental abundances}\label{sec:GJ436b_various}
\begin{figure*}[ht]
\centering
\includegraphics[angle=0,width=\columnwidth]{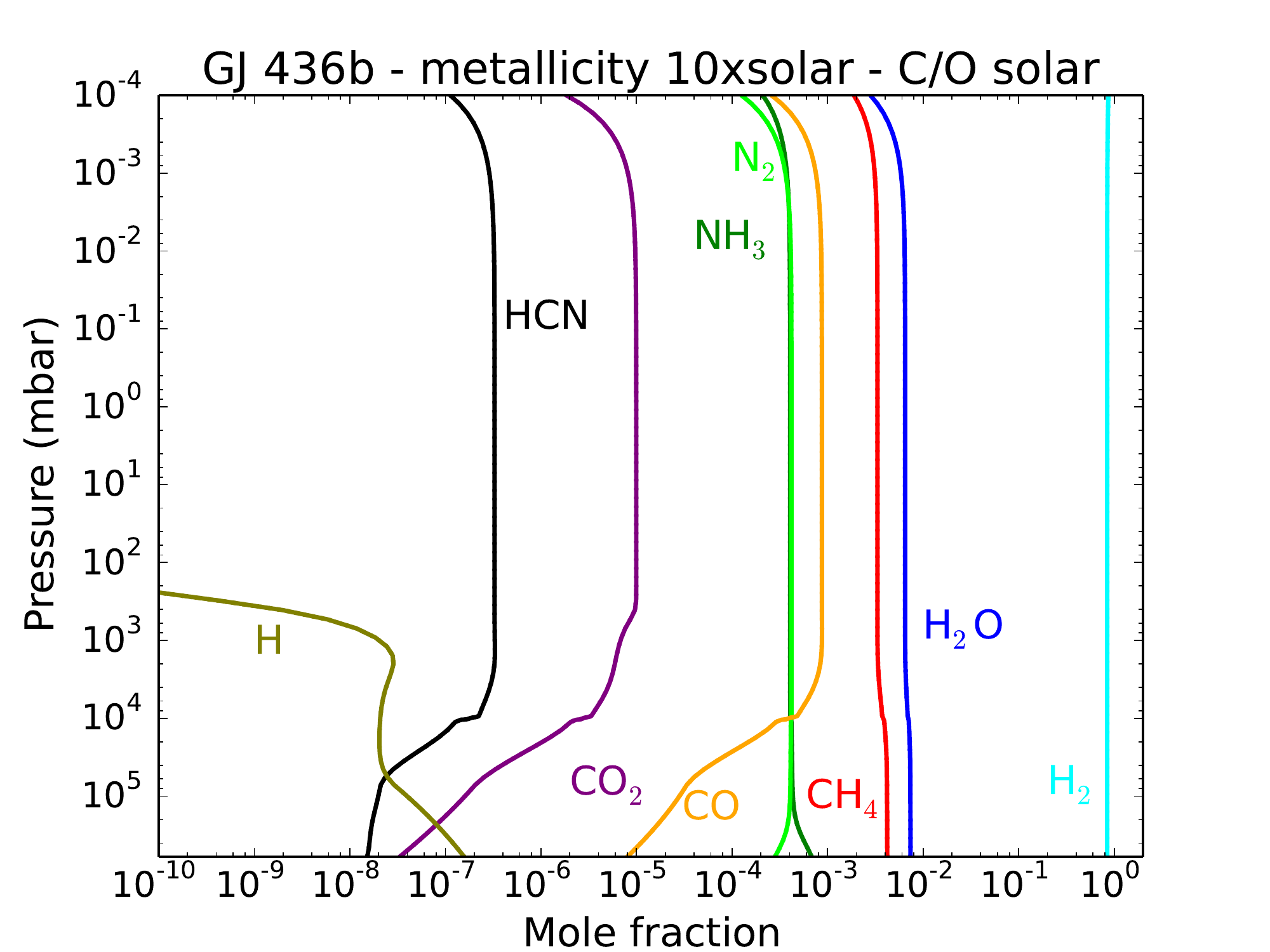}
\includegraphics[angle=0,width=\columnwidth]{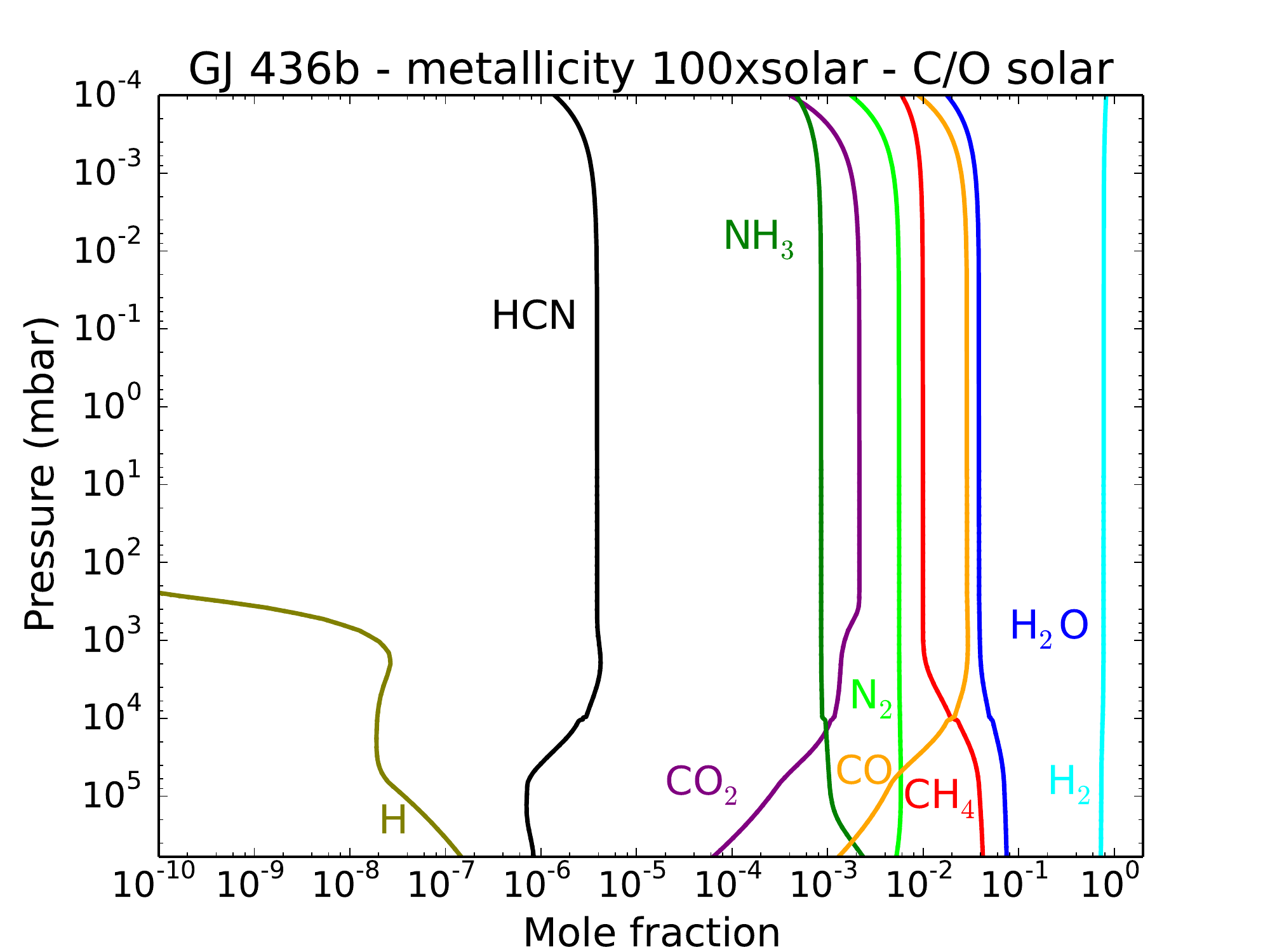}
\includegraphics[angle=0,width=\columnwidth]{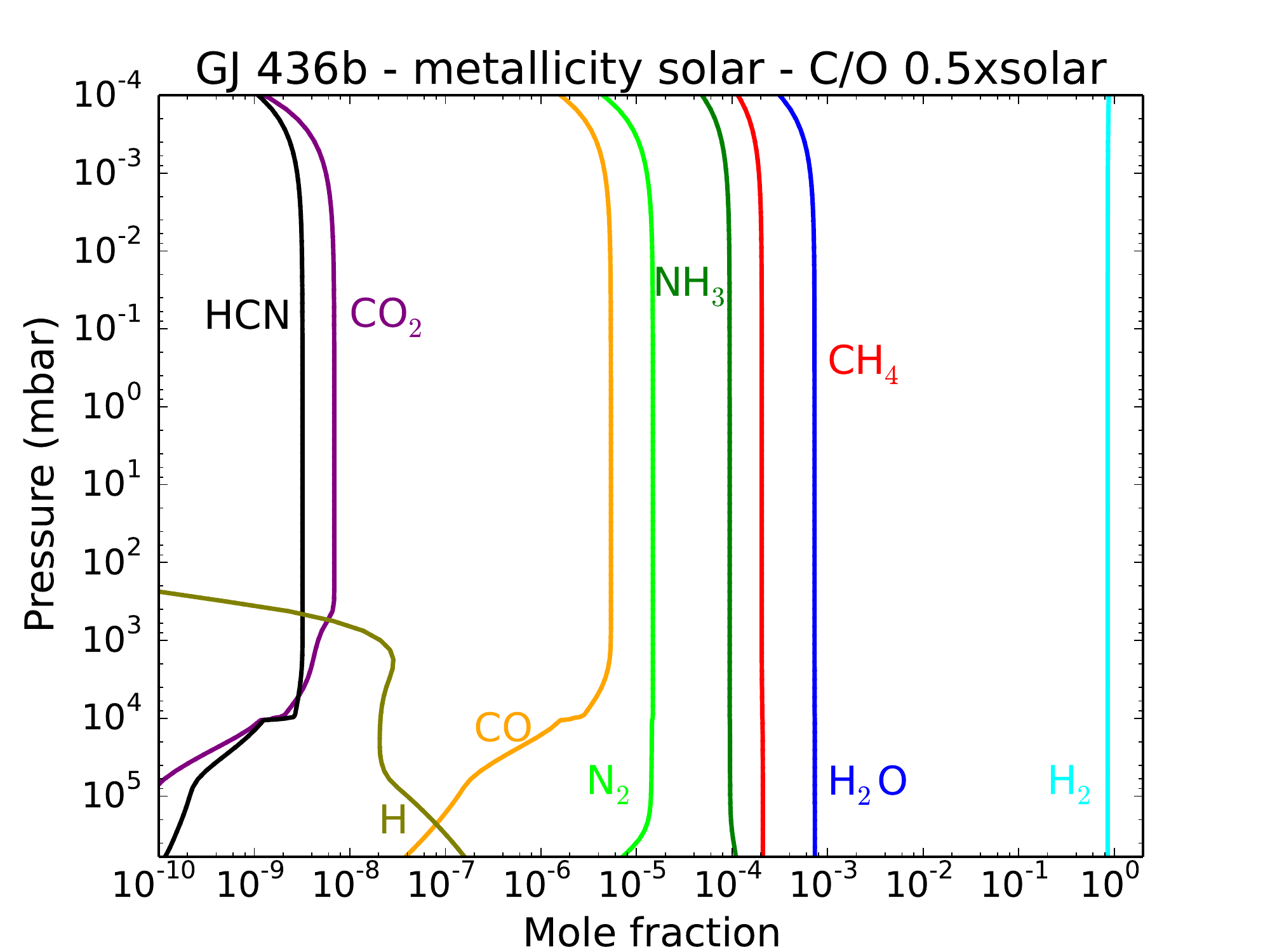}
\includegraphics[angle=0,width=\columnwidth]{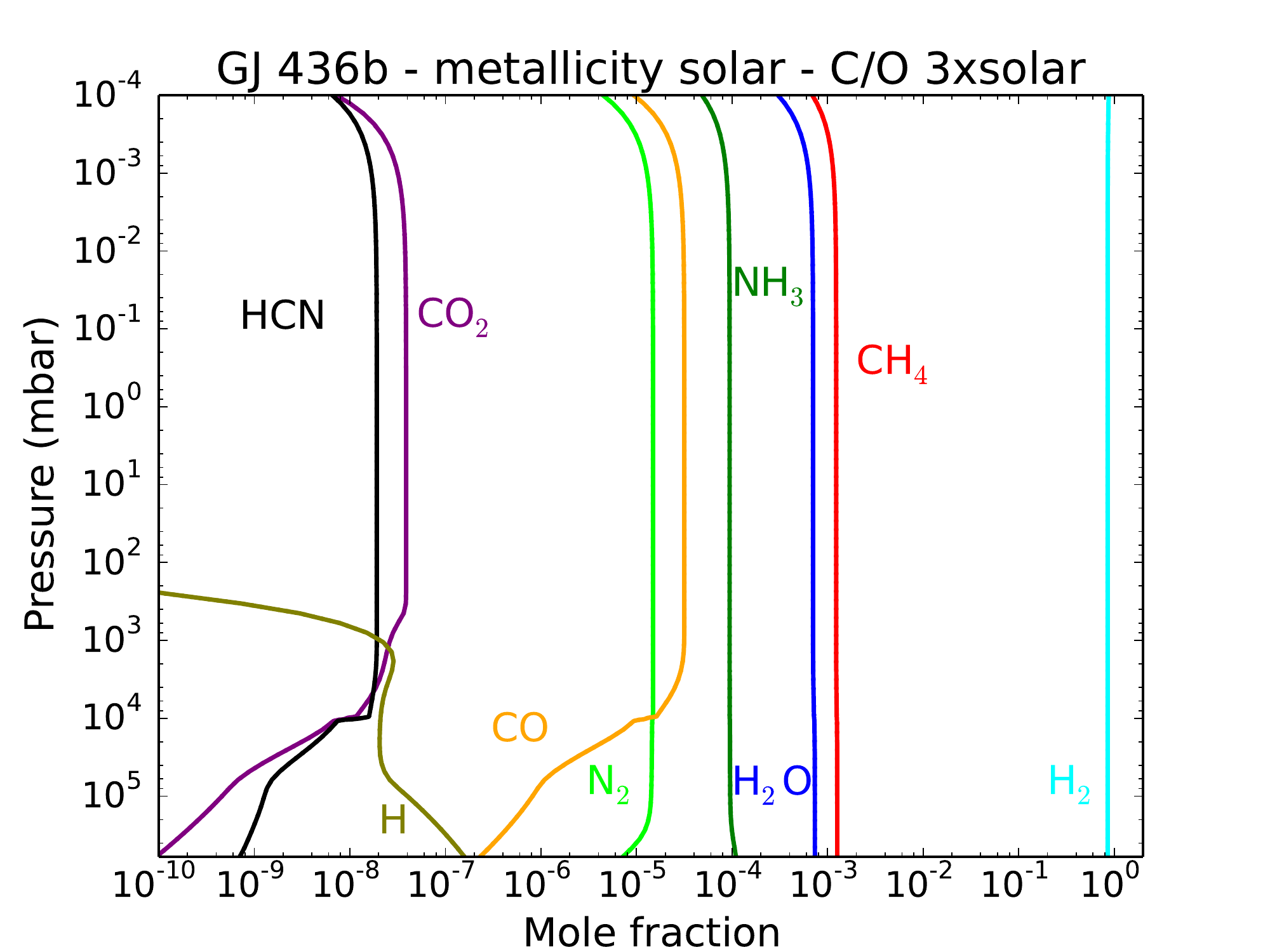}
\caption{Atmospheric chemical compositions of GJ 436b assuming different enrichment relative to the solar elemental abundances \citep{lodders2010}: on the top, C/O ratio solar and metallicities of 10 $\times$ (left) and 100 $\times$ (right) the solar value. On the bottom, solar metallicity and C/O ratios of 0.5 $\times$ (left) and 3 $\times$ (right) the solar value. Results obtained with the full scheme of \cite{Venot2012} (solid lines) and the reduced scheme (dotted lines) are superimposed and cannot be distinguished from one another.}
\label{fig:GJ436b_various}
\end{figure*}
We modelled GJ 436b with the same thermal profile than in Sect.~\ref{sec:GJ436b}, but with different elemental abundances. On one hand, keeping the C/O/N ratios solar, we increase the metallicity by a factor of 10 and 100. In these cases, as can be seen on Fig. \ref{fig:GJ436b_various}, the abundances of all species, except H, globally increase compared to the nominal case (Figs. \ref{fig:MC_abundances_nominal}, \ref{fig:MC_abundances_nominal2}, \ref{fig:MC_abundances_nominal3}). Note that an increase of the metallicity has an effect on the thermal profile \citep[e.g.][]{charnay2015}, that we didn't take into account here as our goal is only to compare the outputs of the two chemical schemes.
On the other hand, keeping O/H and N/H ratios equal to the solar values, we modify only the C/H ratio by a factor of 0.5 and 3 relatively to the solar value. One can observe on Fig. \ref{fig:GJ436b_various} that a decrease of the C/H ratio, leads to a decrease of the abundance of CH$_4$, CO, CO$_2$, and HCN. Inversely, an increase of C/H ratio leads to an increase of the abundance of these carbonaceous species. Methane becomes more abundant than water in this case of carbon-rich atmosphere. These evolutions of abundances with the metallicity and the elemental C/O ratio have been shown also by e.g. \citealt{Moses2013,Venot2014, Venot2015, Tsai2017}. 

The reduced chemical scheme describes very well the atmospheric composition of the four aforementioned models. The results obtained with the full chemical scheme and the reduced one are not distinguishable on log-log plots (Fig. \ref{fig:GJ436b_various}). On Table \ref{tab:variation_GJ436b_various}, we report the maximum and mean differences of abundances between the two chemical schemes for these four models, within the pressure range of interest [0.1--1000] mbar. For each species, we observe that the maximal departures are reached for the high metallicity case. For this case, the maximal difference is 0.6\% and is reached by CH$_4$ at 600 mbar. For the other cases and species, deviations are always lower than this value.

We have also tested much higher metallicities for the atmosphere of GJ 436b: 500, 1000, and 10 000 $\times$ solar metallicities (not shown here). For all of these models, we have found a very good agreement between the reduced and the full chemical scheme, with maximal difference of CH$_4$ between the two schemes of 1, 2, and 11 \% for the cases 500, 1000, and 10 000 $\times$ solar metallicities, respectively (in the pressure range [0.1--1000] mbar). The other species present variations less than 0.5\%.
\begin{table*}[h]
\caption{For GJ 436b models, with various elemental abundances, maximum differences of abundances (in \%) for each species for which the reduced scheme is designed. The pressure level (@level in mbar) where the maximum variation is reached is indicated in parenthesis. These values are calculated within the [0.1--1000] mbar region only, which is probed by infrared observations.} \label{tab:variation_GJ436b_various}
\centering
\begin{tabular}{l|l|l|l|l}
\hline \hline
% OK - actualise avec Red8
Species &  Metallicity = 10 & Metallicity = 100 & C/O =0.5$\odot$ & C/O = 3$\odot$\\
\hline
H$_2$O  &   2$\times$10$^{-2}$ (@1$\times$10$^{-1}$) & 2$\times$10$^{-1}$ (@7) &  4$\times$10$^{-4}$ (@1$\times$10$^{3}$)&  3$\times$10$^{-3}$ (@5$\times$10$^{-1}$)\\
 \hline
CH$_4$ &   6$\times$10$^{-2}$ (@9$\times$10$^{2}$)    & 6$\times$10$^{-1}$ (@6$\times$10$^{2}$)  & 2$\times$10$^{-2}$ (@1$\times$10$^{3}$) & 6$\times$10$^{-3}$ (@7)\\
  \hline
CO            &   1$\times$10$^{-1}$ (@5$\times$10$^{2}$)  &  2$\times$10$^{-1}$ (@5$\times$10$^{2}$)   &  8$\times$10$^{-3}$ (@1$\times$10$^{3}$)  &  7$\times$10$^{-2}$ (@6$\times$10$^{2}$)\\
 \hline
CO$_2$   &    2$\times$10$^{-1}$ (@1$\times$10$^{-1}$)     &  1$\times$10$^{-1}$ (@7)   & 2$\times$10$^{-2}$ (@2$\times$10$^{1}$)   & 9$\times$10$^{-2}$ (@1$\times$10$^{2}$)\\
 \hline
NH$_3$   &    1$\times$10$^{-3}$ (@1$\times$10$^{-1}$)   & 2$\times$10$^{-2}$ (@6$\times$10$^{2}$) &  2$\times$10$^{-3}$ (@1$\times$10$^{-1}$)   &  1$\times$10$^{-3}$ (@8$\times$10$^{2}$)\\
 \hline
HCN       &  1$\times$10$^{-1}$ (@1$\times$10$^{-1}$) & 3$\times$10$^{-1}$ (@4$\times$10$^{2}$) & 1$\times$10$^{-2}$ (@1$\times$10$^{3}$)  &  3$\times$10$^{-2}$ (@1$\times$10$^{-1}$)\\ 
\hline
\end{tabular}
\end{table*}

\subsection{Another warm Neptune: GJ 1214b}\label{sec:GJ1214}

\begin{figure}[!htb]
\centering
\includegraphics[angle=0,width=\columnwidth]{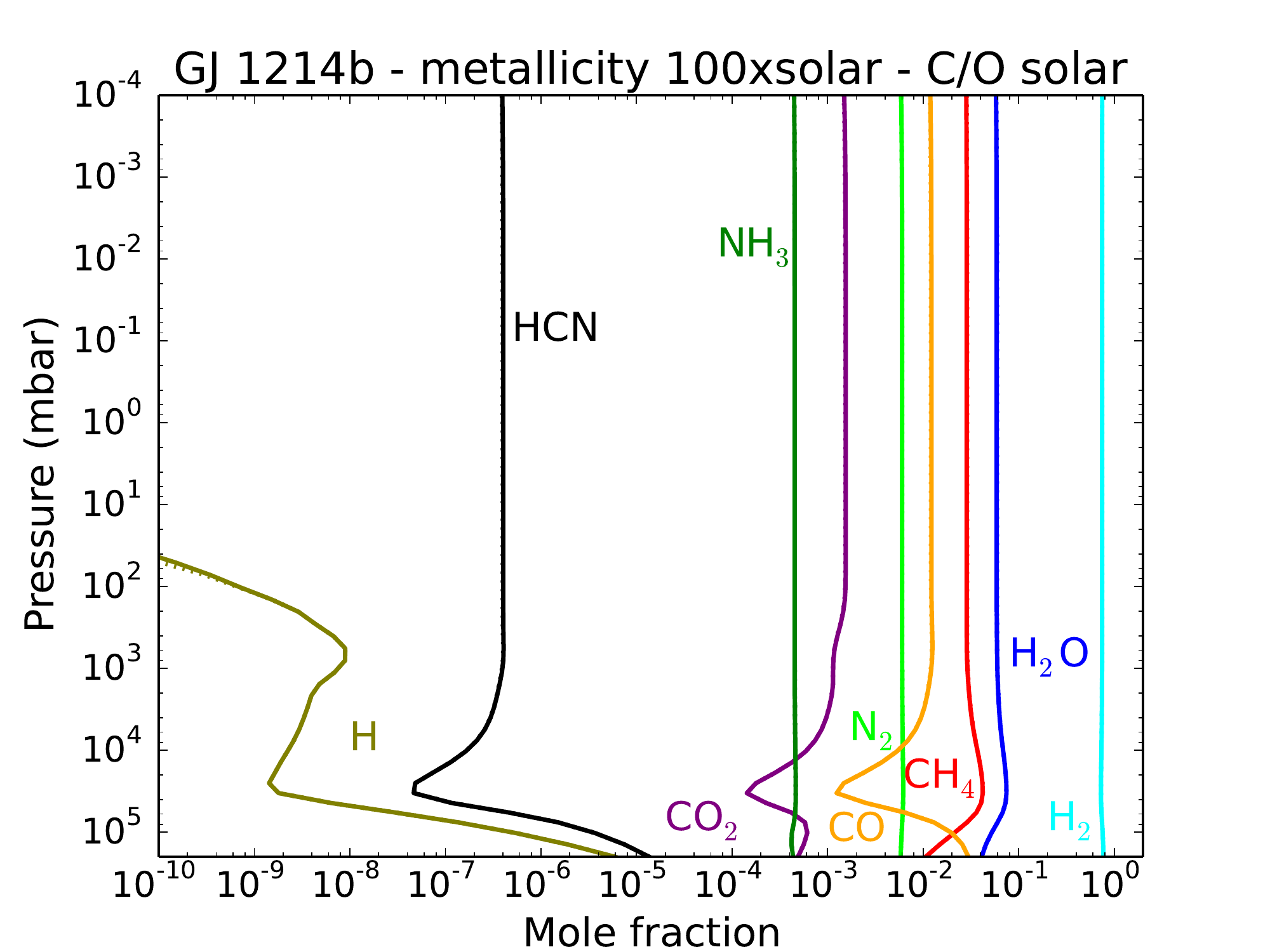}
\caption{Atmospheric chemical compositions of GJ 1214b obtained with the full scheme of \cite{Venot2012} (solid lines) and  the reduced scheme (dotted lines). solid and dotted lines are almost superimposed.}
\label{fig:GJ1214b}
\end{figure}

We have modelled another warm Neptune, GJ 1214b, with the average dayside profile calculated with another atmospheric model, the \texttt{Generic LMDZ GCM} \citep{charnay2015}. We have assumed an atmospheric metallicity of 100~$\times$~solar metallicity \citep{lodders2010} and used the corresponding parametrization of vertical mixing recommended by \cite{charnay2015}: $K_{zz}=3 \times 10^7 \times P^{-0.4}$cm$^2$s$^{-1}$, with $P$ in bar.
Even if the thermal profile of GJ~1214b is close to that of GJ~436b (see Fig. \ref{fig:PT}), the difference of temperature ($\sim$100 K) in the quenching area (10$^4$ mbar) leads to important differences in chemical composition between the two planets while having the same elemental abundances. Indeed, with a metallicity 100~$\times$~solar, the transition of C-bearing species between CO and CH$_4$ occurs at these temperature and pressure values ($\sim$1100 K and $\sim$10 bar). Whereas in GJ 436b's atmosphere, CO is quenched when more abundant than CH$_4$, in GJ 1214b, it is the opposite. Our model of GJ 1214b's atmosphere, in complement of that of GJ 436b with a metallicity 100 $\times$ solar, shows that the reduced chemical scheme is accurate in these two different atmospheric configurations (see Fig.~\ref{fig:GJ1214b}). In the pressure range of interest [0.1--1000] mbar, departures between the abundances obtained with the full and the reduced schemes are lower than 1\% for all species for which the reduced scheme is designed (see Table \ref{tab:variation_HD}). For this planet, CO and HCN are the two species that show the strongest variations in abundance between the two schemes.
Note that these differences of temperature and thus of chemical compositions result more from differences in the atmospheric models (\texttt{ATMO} and \texttt{Generic LMDZ GCM}) used to calculate the thermal profiles (e.g. internal temperature, opacities, treatment of alkali) rather than differences between the two planets which are quite similar in terms of irradiation. It has been shown by \cite{Baudino2017} that different assumptions lead to important variations in the results of atmospheric models. Our results show that the chemical composition is consequently also sensitive to these hypotheses.

\subsection{Hot Jupiters: HD 209458b and HD 189733b}\label{sec:HJ}
\begin{figure}[!htb]
\centering
\includegraphics[angle=0,width=\columnwidth]{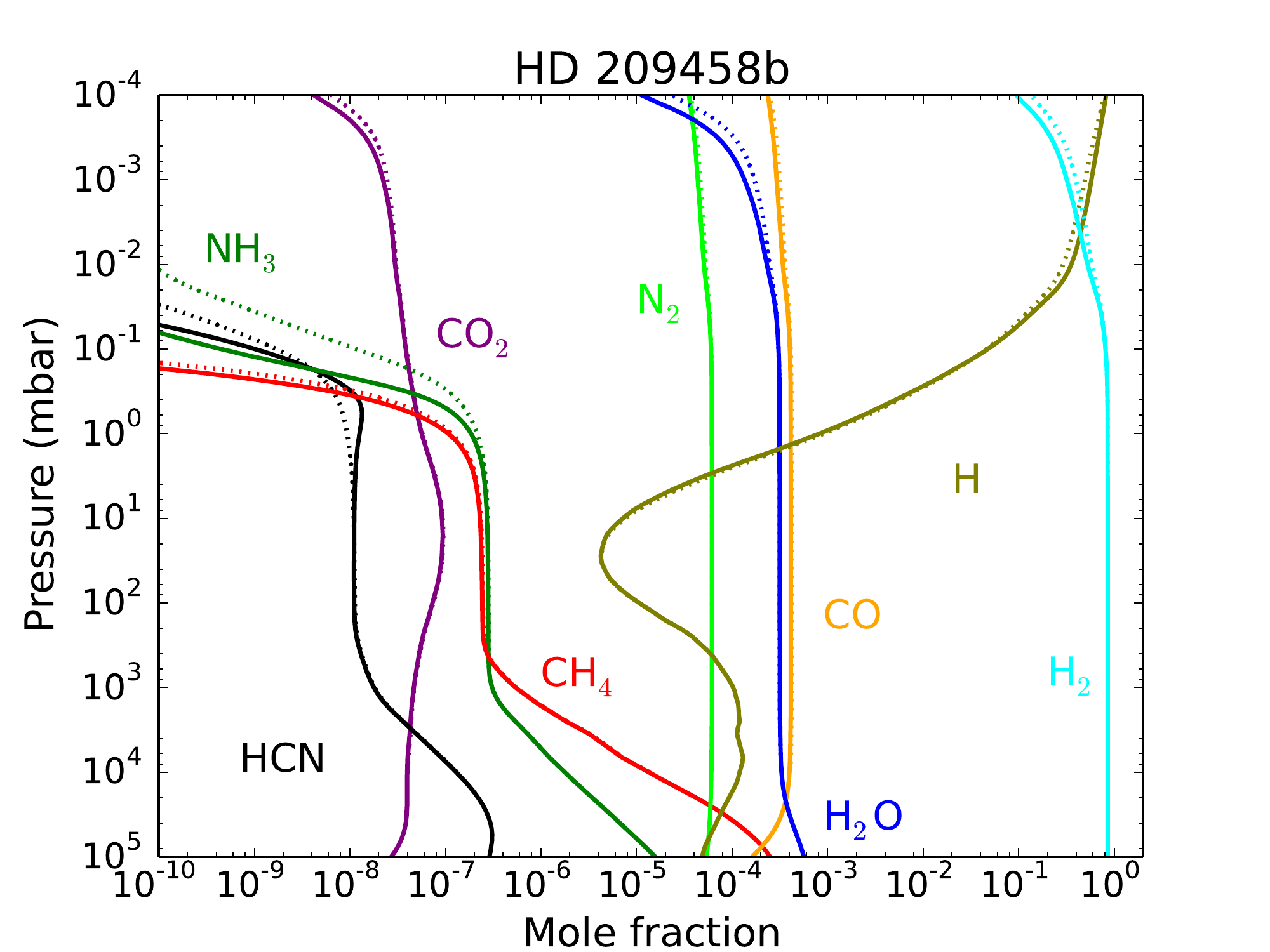}
\includegraphics[angle=0,width=\columnwidth]{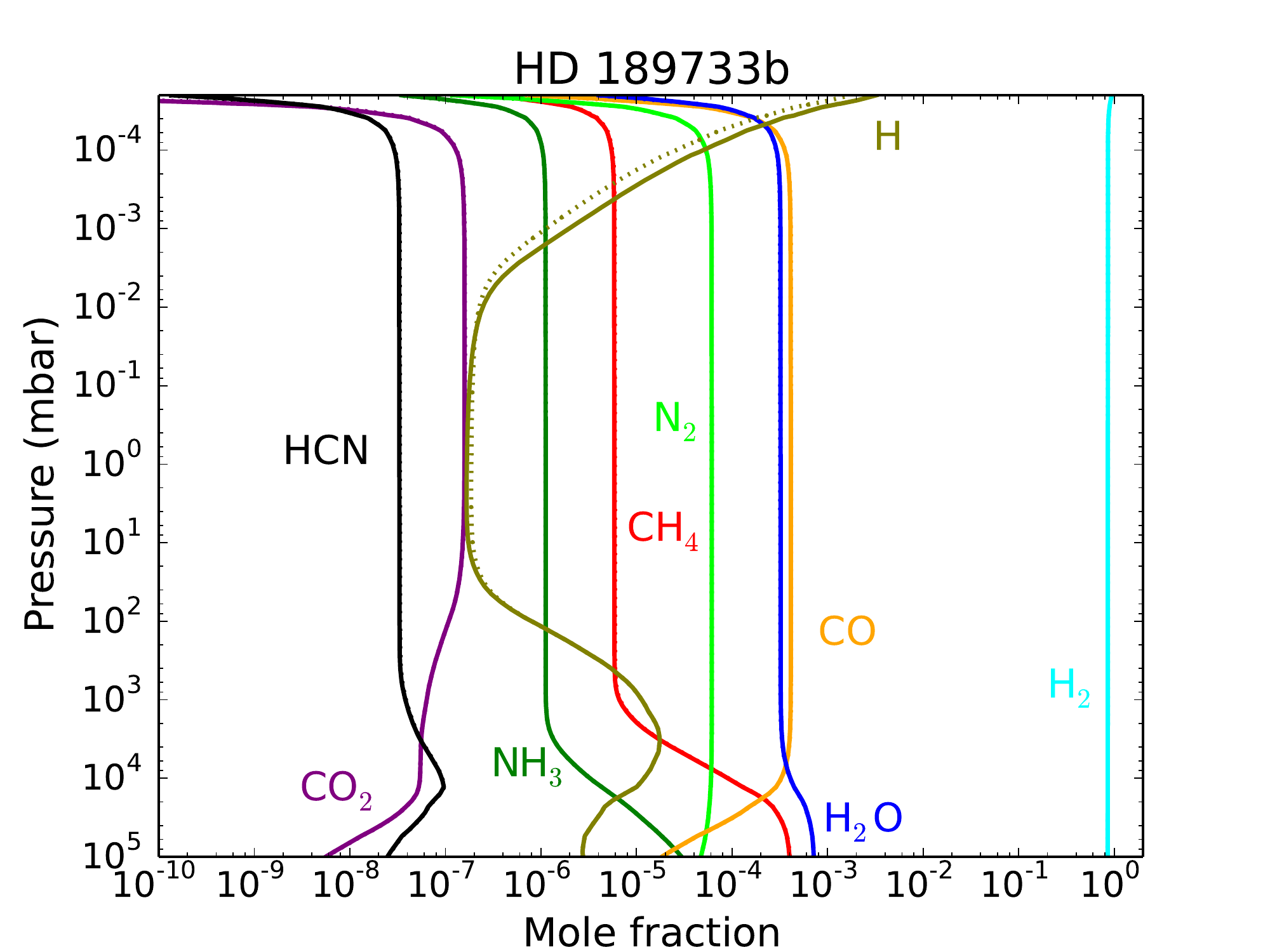}
\caption{Atmospheric chemical compositions of HD 209458b (top) and HD 189733b (bottom) obtained with the full scheme of \cite{Venot2012} (solid lines) and  the reduced scheme (dotted lines). Solid and dotted lines are almost superimposed.}
\label{fig:HD209_HD189}
\end{figure}
We have tested the validity of the reduced chemical scheme on the famous hot Jupiters HD 209458b and HD 189733b. We have used the thermal profiles and eddy diffusion coefficients profiles of \cite{moses2011} that were also used in \cite{Venot2012} with the full chemical scheme. We assume solar elemental abundances \citep{lodders2010} with a reduction of 20\% of oxygen, due to sequestration in the deep layers. The agreement between the results obtained with the two schemes is excellent (see Fig.~\ref{fig:HD209_HD189}). For the pressure range probed by observations, we listed in Table~\ref{tab:variation_HD} the maximum difference of abundances between the two chemical schemes for these two models. Contrary to the other models presented earlier, we observe for HD 209458b some relatively important deviations between the two schemes for pressures lower than 4 mbar for HCN (60\%), CH$_4$ (600\%), and NH$_3$ (2000\%). However, these deviations are acceptable as 1) they are at the limit of the pressure range probed by observations and 2) they concern species with a low abundance (<10$^{-6}$). Then, at even higher altitudes (from 10$^{-2}$ mbar), we observe deviations for CO$_2$ and H$_2$O (H and H$_2$ also but they are not in the list of species for which the reduced scheme is designed). However, these pressure levels are not probed by observations. For HD 189733b, the agreement between the two chemical schemes is higher. The variations of abundances for species of interest are lower than 1\% in the [0.1--1000] mbar region. The largest variation is due to HCN at 100 mbar.

%HD209458b : 31.91 minutes / 1.16 minutes (t=1d10s)
%HD189733b :37.09 minutes / 1.06 minutes (t=1d10s)

\begin{table*}[h]
\caption{For HD 209458b, HD 189733b, GJ 1214b models, maximum variations of abundances (in \%) for each species for which the reduced scheme is designed. The pressure level (@level in mbar) where the maximum variation is reached is indicated in parenthesis. These values are calculated within the regions probed by infrared observations ([0.1--1000] mbar).} \label{tab:variation_HD}
\centering
\begin{tabular}{l|l|l|l|l}
\hline \hline
Species & HD 209458b & HD 189733b & GJ 1214b\\
\hline
H$_2$O   &   1$\times$10$^{-1}$ (@1$\times$10$^{-1}$) &  6$\times$10$^{-1}$ (@3$\times$10$^{-1}$)  & 3$\times$10$^{-1}$ (@3$\times$10$^{1}$) \\
 \hline
 CH$_4$  &   6$\times$10$^{2}$ (@1$\times$10$^{-1}$)  &  1$\times$10$^{-1}$ (@3$\times$10$^{2}$) & 6$\times$10$^{-1}$ (@3$\times$10$^{2}$) \\
  \hline
CO  &   7$\times$10$^{-2}$ (@1$\times$10$^{-1}$)  & 5$\times$10$^{-1}$ (@6$\times$10$^{1}$)  & 1 (@2$\times$10$^{2}$) \\
 \hline
CO$_2$    &   1$\times$10$^{-2}$ (@3$\times$10$^{-1}$)    & 8$\times$10$^{-1}$ (@1$\times$10$^{-1}$) & 9$\times$10$^{-1}$ (@1$\times$10$^{-1}$) \\
 \hline
NH$_3$  &   2$\times$10$^{3}$ (@1$\times$10$^{-1}$) &  2$\times$10$^{-2}$ (@1$\times$10$^{-1}$) & 3$\times$10$^{-2}$ (@1$\times$10$^{-1}$)  \\
 \hline
HCN      &  6$\times$10$^{1}$ (@1$\times$10$^{-1}$)& 1 (@1$\times$10$^{2}$) & 1 (@2$\times$10$^{2}$) \\ 
\hline
\end{tabular}
\end{table*}

\subsection{Deep atmospheres of Uranus and Neptune}\label{sec:UN}
\begin{figure}[!htb]
\centering
\includegraphics[angle=0,width=\columnwidth]{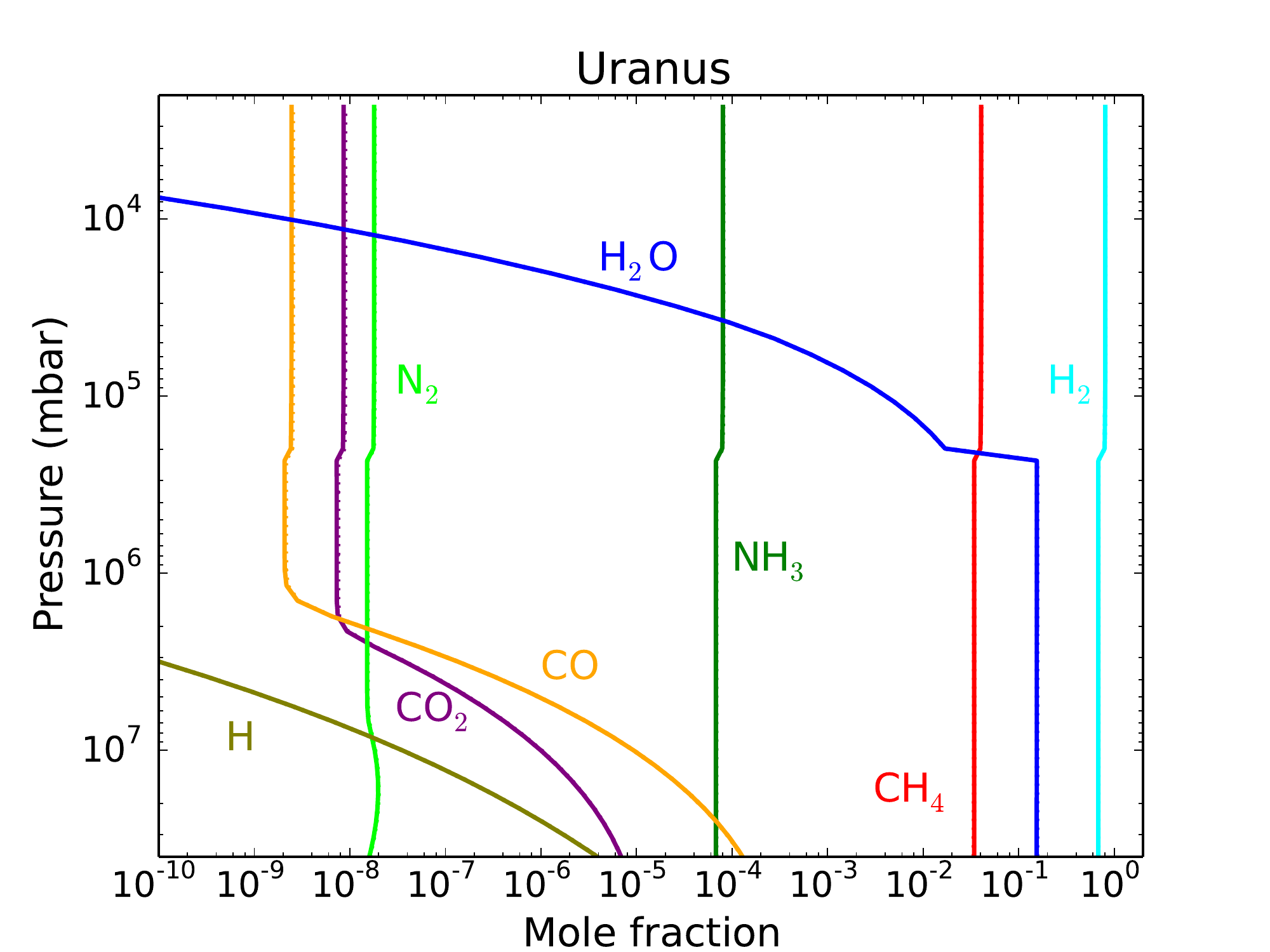}
\includegraphics[angle=0,width=\columnwidth]{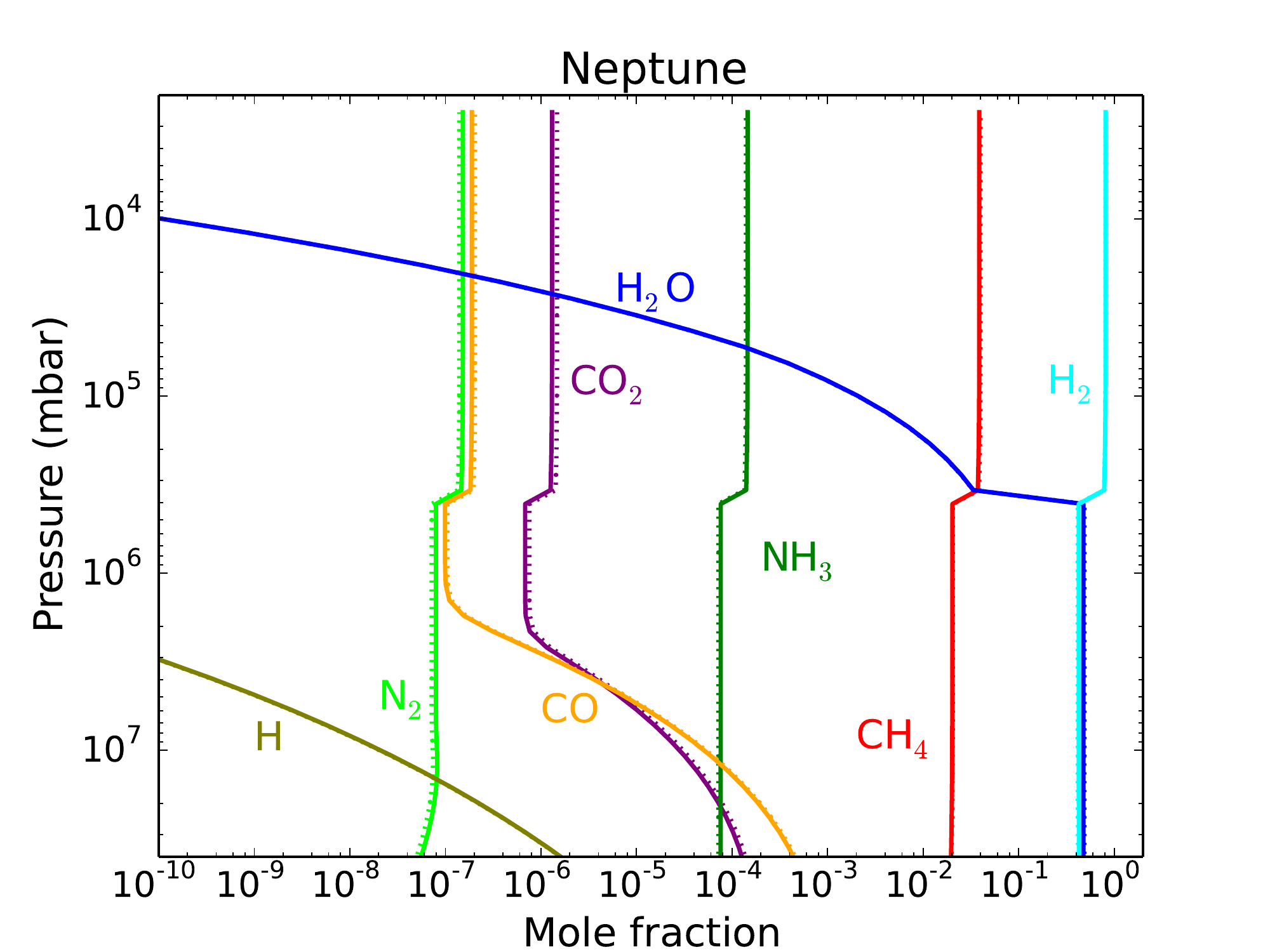}
\caption{Tropospheric chemical compositions of Uranus (top) and Neptune (bottom) computed with the full scheme of \cite{Venot2012} (solid lines) and  the reduced scheme (dotted lines). Solid and dotted lines are almost superimposed.}
\label{fig:ura_nept}
\end{figure}
We have applied our kinetic model to the deep atmospheres of Uranus and Neptune, as in \cite{cavalie2017}, using the `3-layer' thermal profiles calculated with the prescription of \cite{leconte2017}. Following \cite{cavalie2017}, we have set the eddy diffusion coefficient $K_{zz}$ to 10$^{8}$ cm$^2$.s$^{-1}$. Concerning the elemental abundances, we have assumed the following enrichment compared to the solar abundances \citep{lodders2010}: For both planets, we have assumed a N/H ratio of 0.5 times the solar value. For Uranus, we have assumed C/H and O/H ratios of 75 and 160 times the solar value, respectively.
For Neptune, we have assumed C/H and O/H ratios of 45 and 480 times the solar value, respectively. These values correspond to the nominal cases of \cite{cavalie2017}, allowing to reproduce the observational constraints available at 2 bar for CO and CH$_4$:  y$_{CH_4}$=0.04 for both planets and y$_{CO}$=2.0$\times$10$^{-7}$ at Neptune and y$_{CO}$<2.1$\times$10$^{-9}$ at Uranus.

The tropospheric compositions determined with the full and reduced schemes are presented Fig.~\ref{fig:ura_nept}. Chemical abundances obtained with both schemes are very close. We report in Table~\ref{tab:variation_GP_SD} the maximum variations of abundances between the two chemical schemes for the models of Uranus and Neptune in the pressure range [10$^6$--10$^7$] mbar. For the giant planets, we choose this range because it includes the quenching level. In the methodology we apply to the study of these planets, this level is the most important and decisive for the conclusions drawn concerning the elemental composition of the atmospheres \citep{Cavalie2014,cavalie2017}.
For the two planets, the variations of abundances between the two chemical scheme are low ($<$ 10\%) in the pressure range of interest. We remark that the maximum values are found for Neptune, which is not very surprising. Indeed, this atmosphere is the most enriched of this study (i.e. O/H and C/H are, respectively, 480 and 40 $\times$ the solar values), and thus the elemental conditions are the most distant from that of our nominal GJ 436b model used to develop the reduced chemical scheme.

\subsection{Atmosphere of a SD 1110-like brown dwarf}\label{sec:BD}

\begin{figure}[!htb]
\centering\includegraphics[angle=0,width=\columnwidth]{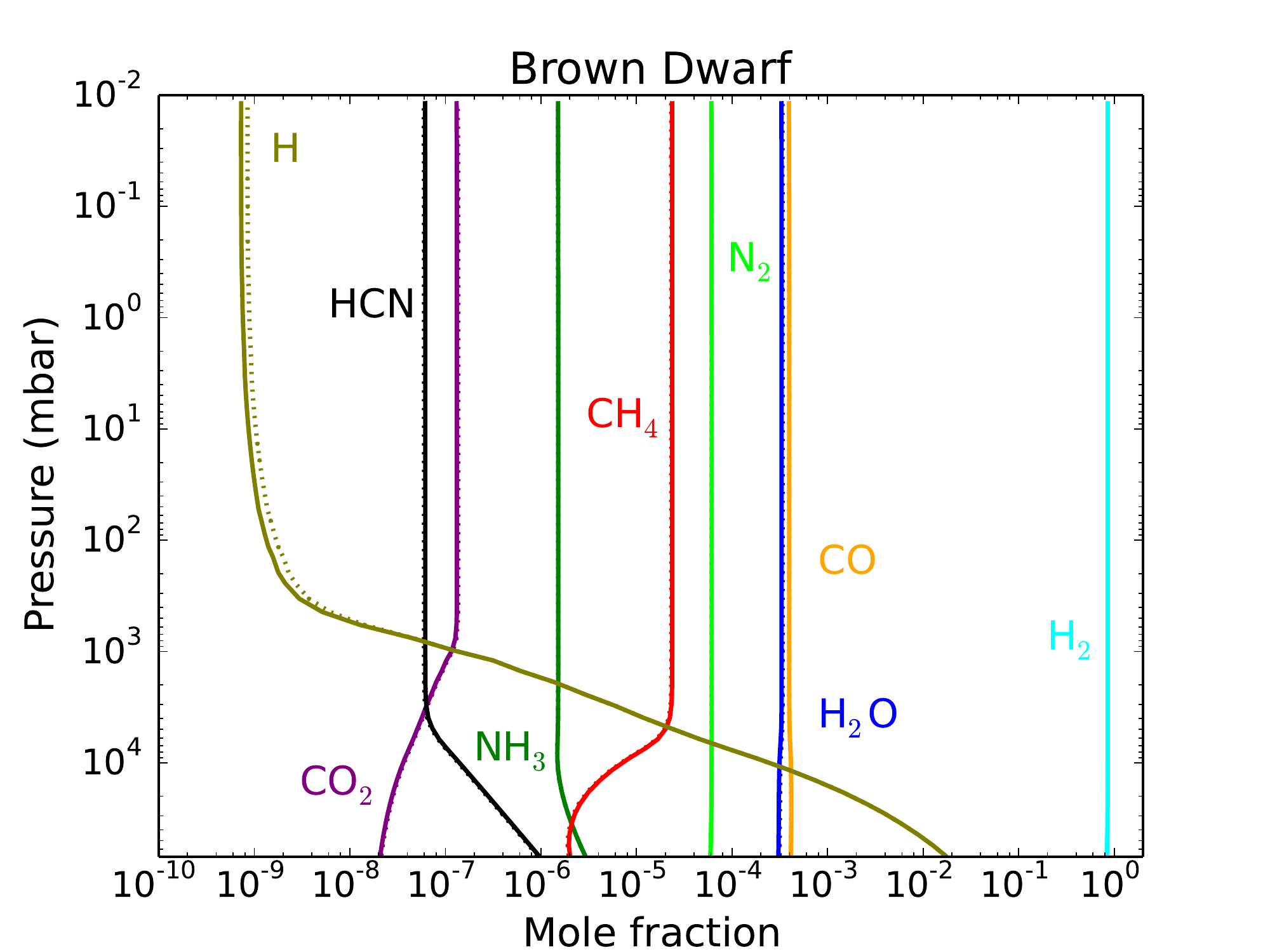}
\caption{Atmospheric chemical compositions of a SD 1110-like Brown Dwarf obtained with the full scheme of \cite{Venot2012} (solid lines) and  the reduced scheme (dotted lines). Solid and dotted lines are almost superimposed.}
\label{fig:BD}
\end{figure}

We have finally applied our kinetic model to study the chemical composition of a brown dwarf's atmosphere. With \texttt{ATMO} \citep{tremblin:2016aa,Drummond2016}, we have determined the thermal profile of a brown dwarf located at the L/T transition, with an effective temperature of 1000 K and log $g$=4.5. At the transition, L-type brown dwarfs are CO-dominated and become CH$_4$-dominated as they cool down towards the T-dwarf sequence. Non-equilibrium chemistry at the CO/CH$_4$ transition is thus an important process to correctly predict the transition as a function of effective temperature in the cooling sequence of brown dwarfs \citep{saumon2003}. The model used for this study is similar to the T5.5 brown dwarf SD 1110 \citep[see][]{stephens2009}. The thermal structure has been calculated assuming thermochemical equilibrium with solar elemental abundances \citep{lodders2010}. The abundance of condensates are explicitly included in this calculation, which corresponds to a reduction of $\sim$20\% of oxygen. For the kinetic model, we thus applied the same oxygen reduction and we have taken a constant vertical mixing of 10$^8$cm$^2$s$^{-1}$. Fig. \ref{fig:BD} shows that the chemical composition obtained with the reduced chemical scheme is similar to that obtained with the full scheme. For brown dwarfs, the pressure range that can be probed by infrared observations is slightly deeper than for exoplanets. Thus, we have focused our study here to the [10$^2$--10$^4$] mbar range \citep[see][]{morley2014}. In this range, the largest variation between abundances obtained with the two schemes is of 3\% and is due to HCN (at 2 bar) and CO$_2$ (at 100 mbar). H$_2$O and CH$_4$ present variations of about the same order of magnitude (2\% at 9 bar) (see Table~\ref{tab:variation_GP_SD}).

\begin{table*}[h]
\caption{For Uranus, Neptune, and SD 1110 models, maximum variations of abundances (in \%) for each species for which the reduced scheme is designed. The pressure level (@level in mbar) where the maximum variation is reached is indicated in parenthesis. These values are calculated within the region where quenching occurs for the Giant Planets ([10$^{6}$--10$^{7}$] mbar) and within the region probed by infrared observations ([10$^2$--10$^4$] mbar) for the Brown Dwarf.} \label{tab:variation_GP_SD}
\centering
%\small
\begin{tabular}{l|l|l|l}
\hline \hline
Species & Uranus & Neptune & SD 1110\\
\hline
% OK - actualise avec Red8
H$_2$O &   6$\times$10$^{-1}$ (@8$\times$10$^{6}$) & 2 (@9$\times$10$^{6}$) & 2 (@9$\times$10$^{3}$) \\
 \hline
CH$_4$     &    1$\times$10$^{-1}$ (@1$\times$10$^{6}$)& 1 (@9$\times$10$^{6}$) &  3 (@9$\times$10$^{3}$)  \\
  \hline
CO   & 3 (@1$\times$10$^{6}$) &  6 (@1$\times$10$^{6}$) & 2$\times$10$^{-1}$ (@9$\times$10$^{3}$) \\
 \hline
CO$_2$   &  3 (@1$\times$10$^{6}$) & 1$\times$10$^{1}$ (@1$\times$10$^{6}$) & 3 (@1$\times$10$^{2}$)\\
 \hline
NH$_3$      &   7$\times$10$^{-1}$ (@8$\times$10$^{6}$)  & 6 (@1$\times$10$^{6}$) & 9$\times$10$^{-2}$ (@1$\times$10$^{3}$)\\
 \hline
HCN      &  4$\times$10$^{-1}$ (@5$\times$10$^{6}$) & 5 (@1$\times$10$^{6}$) & 3 (@2$\times$10$^{3}$) \\ 
\hline
\end{tabular}
\end{table*}

\subsection{Out of validity: C-rich hot Jupiter}\label{sec:HD209_Crich}

\begin{figure}[!htb]
\centering
\includegraphics[angle=0,width=\columnwidth]{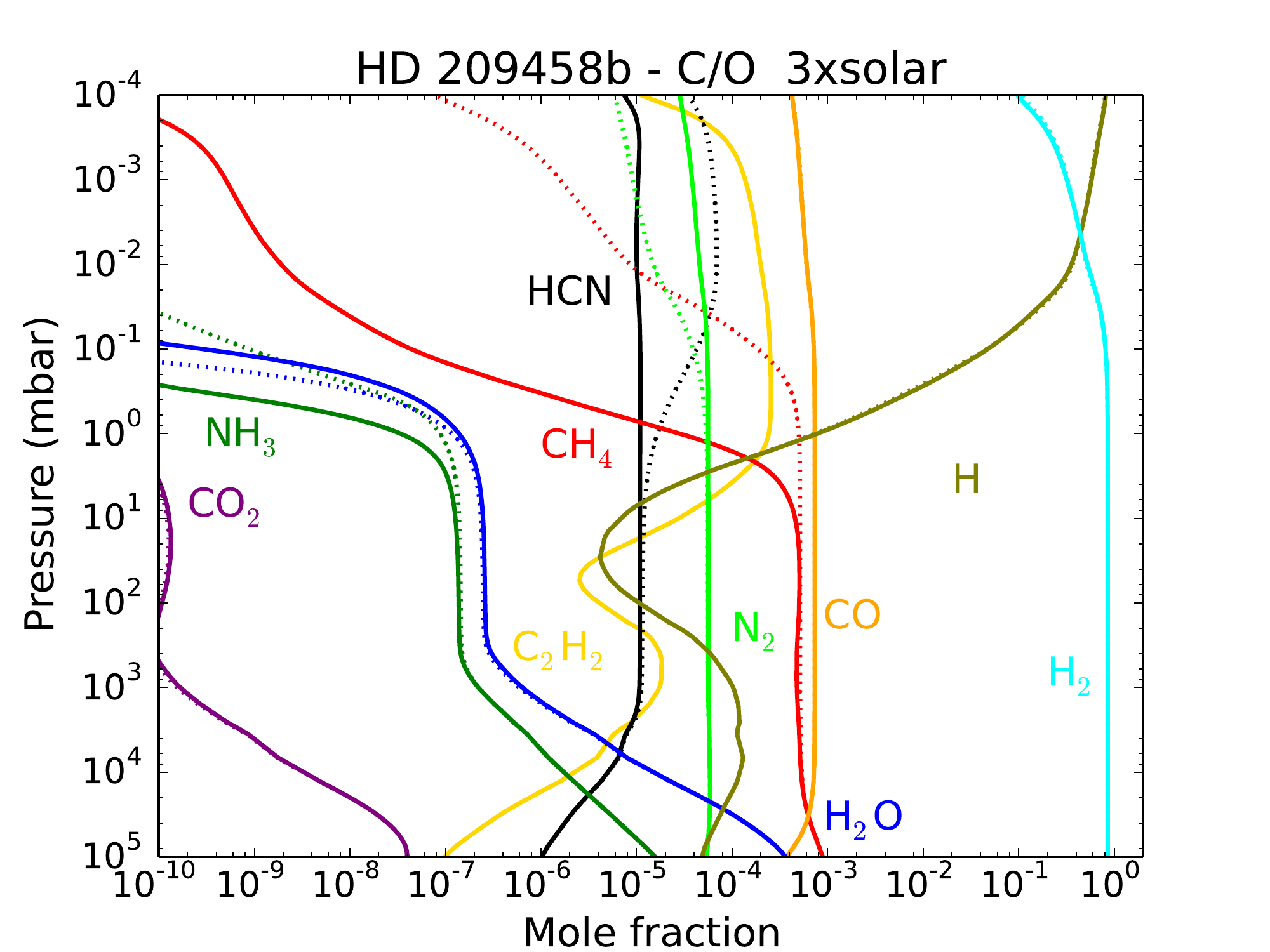}
\includegraphics[angle=0,width=\columnwidth]{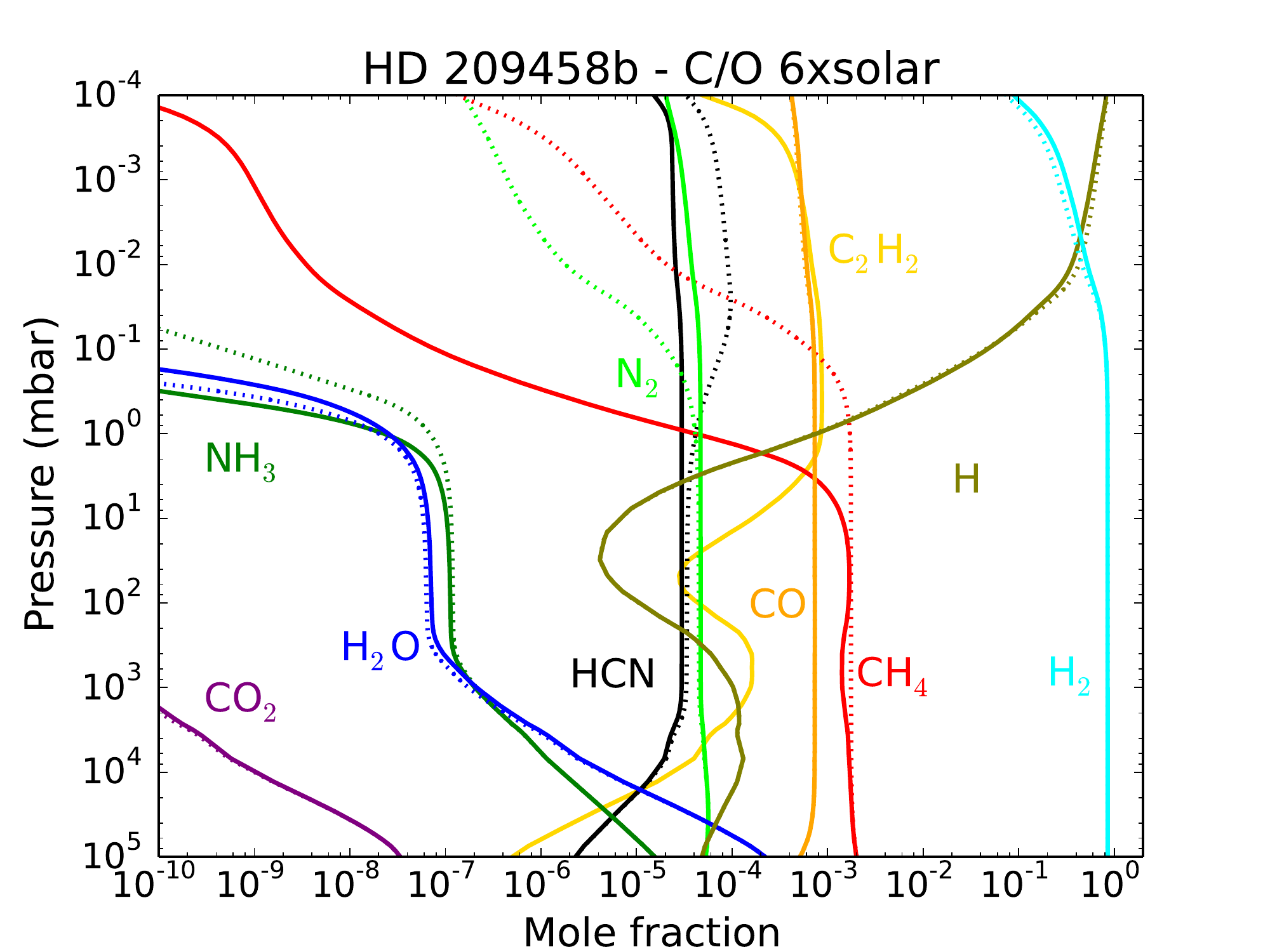}
\caption{Atmospheric chemical compositions of a HD209458b with a C/O ratios 3 $\times$ (top) and 6 $\times$ (bottom) solar obtained with the full scheme of \cite{Venot2012} (solid lines) and the reduced scheme (dotted lines). Deviations are easily visible at pressures lower than 10 mbar. The abundance of C$_2$H$_2$ has been determined with the full scheme only, as this species is not included in the reduced scheme.}
\label{fig:HD_Crich}
\end{figure}

As acetylene is not included in our reduced chemical scheme, the study of hot atmospheres rich in carbon might be problematic. To assess the extent of this error, we modelled HD~209458b, with our two chemical schemes, assuming two carbon-rich compositions: 3 $\times$ and 6 $\times$ the solar C/O ratio (and then removing 20\% of oxygen). One can see on Fig.~\ref{fig:HD_Crich} that deviations between the abundances obtained with the two schemes appear for pressures lower than $\sim$10 mbar. These differences are due to the important mixing ratio of C$_2$H$_2$; species that is included only in the full chemical scheme. The high amount of carbon carried by this species in the C$_0$-C$_2$ scheme is distributed in the other carboneous species in the reduced scheme, mainly CH$_4$, HCN, and CH$_3$. For species of interest, the variations of abundances obtained with the two schemes reach very important values, up to 8$\times$10$^{5}$ \% (see Table \ref{tab:variation_HD_Crich}). Such differences of abundances can have consequences on the synthetic spectra computed using these results, especially because of the absence of C$_2$H$_2$ as a source of opacity when using the reduced scheme's chemical composition. These results make us recommend not to use the reduced scheme to study hot carbon-rich atmosphere, or at least to be very cautious with the chemical composition predicted in the upper atmosphere (P<10 mbar).

\begin{table*}[h]
\caption{For HD 209458b with a C/O = 3 $\times$ and 6 $\times$ solar value, maximum variations of abundances (in \%) for each species for which the reduced scheme is designed. The pressure level (@level in mbar) where the maximum variation is reached is indicated in parenthesis. These values are calculated within the regions probed by infrared observations ([0.1--1000] mbar).} \label{tab:variation_HD_Crich}
\centering
%\small
\begin{tabular}{l|l|l}
\hline \hline
Species & C/O =3$\odot$ & C/O=6$\odot$\\
\hline
H$_2$O &   1$\times$10$^{2}$ (@1$\times$10$^{-1}$) & 9$\times$10$^{1}$ (@2$\times$10$^{-1}$)  \\
 \hline
CH$_4$     &    4$\times$10$^{5}$ (@1$\times$10$^{-1}$)& 8$\times$10$^{5}$ (@1$\times$10$^{-1}$)   \\
  \hline
CO   & 1$\times$10$^{-1}$ (@1$\times$10$^{-1}$) & 2$\times$10$^{-1}$ (@1$\times$10$^{-1}$)  \\
 \hline
CO$_2$   &  1$\times$10$^{2}$ (@1$\times$10$^{-1}$) & 9$\times$10$^{1}$ (@2$\times$10$^{-1}$) \\
 \hline
NH$_3$      &   1$\times$10$^{4}$ (@2$\times$10$^{-1}$)  & 1$\times$10$^{4}$ (@2$\times$10$^{-1}$) \\
 \hline
HCN      &  3$\times$10$^{2}$ (@1$\times$10$^{-1}$) & 1$\times$10$^{2}$ (@1$\times$10$^{-1}$) \\ 
\hline
\end{tabular}
\end{table*}

\section{Conclusion}\label{sec:concl}

We have developed a reduced chemical scheme for exoplanetary atmospheres from the C$_0$-C$_2$ chemical network of \cite{Venot2012}. This reduced scheme has been designed to reproduce the abundances of H$_2$O, CH$_4$, CO, CO$_2$, NH$_3$, and HCN and is available in the KIDA Database \citep{KIDA2012}. For these species of interest, we have validated this new scheme through a comparison with the abundances predicted by the full scheme. In many different cases (metallicities, C/O ratios, thermal profiles) the abundances are very similar, with only some percent of deviations (often less than 1\%) in the pressure range probed by infrared observations. The only limitation of this scheme concerns hot C-rich atmospheres. As the reduced chemical scheme does not include C$_2$H$_2$, the use of this scheme to study this kind of atmospheres (i.e. T$\gtrsim$ 1000K and C/O > 1) is not recommended. For our model of GJ436b, we have also performed an uncertainty propagation study on the full chemical scheme. The abundances obtained with the reduced scheme are included in the 1-$\sigma$ interval around the C$_0$-C$_2$ nominal abundance.

An even more drastic reduction of the scheme is possible, decreasing the desired level of accuracy in our methodology. However, such a reduction would lead to a chemical scheme giving deviations of abundances with the full scheme larger than what we obtained with the scheme presented in this paper. Also, the range of validity in terms of temperature of this very reduced scheme might be narrowed down, which could be problematic when used in 3D models of planets with a large day/night temperature gradient, such as the hot Jupiters WASP-43b, WASP-18b, WASP-103b, WASP-12b \citep{parmentier2017}.

The advantage of this reduced network is that it runs faster than the full chemical scheme from which it is extracted ($\sim$30 times faster). This gain in computational time is due to the reduction in the size of the system of differential equations (continuity equations) that must be solved to determine the steady state of the atmosphere. 
But the most important application we expect for this scheme is to be incorporated in 3D models. This would be an important step forward in view of a good interpretation of the future observations that will be provided by the next generation of telescopes (JWST, ARIEL) and for understanding the global chemical composition of exoplanets, brown dwarfs and the deep composition of solar system giant planets. For these latters, the recent Juno/MWR data \citep{Bolton2017}, presenting an unexpected NH$_3$ distribution below the condensation level, prove that robust 3D models accounting for chemistry and dynamics are highly needed to decipher their composition. These 3D models of the tropospheres of the solar system giant planets will be crucial to help select the latitude of any future entry probe \citep{arridge2014, mousis2014, mousis2016, mousis2018}.

\begin{acknowledgements}
The authors thanks the referee for his careful reading of the manuscript which allowed its improvement. O.V. and T.C. thanks the CNRS/INSU Programme National de Plan\'etologie (PNP) for funding support. PT acknowledges support from the European Research Council (grant no. 757858 -- ATMO). BD and EH acknowledge support from the STFC Consolidated Grant ST/R000395/1.
\end{acknowledgements}

\bibliographystyle{aa}
\bibliography{VENOT_ReducedScheme_arxiv}

\begin{thebibliography}{60}
\expandafter\ifx\csname natexlab\endcsname\relax\def\natexlab#1{#1}\fi

\bibitem[{{Ag{\'u}ndez} {et~al.}(2014){Ag{\'u}ndez}, {Parmentier}, {Venot},
  {Hersant}, \& {Selsis}}]{agu2014}
{Ag{\'u}ndez}, M., {Parmentier}, V., {Venot}, O., {Hersant}, F., \& {Selsis},
  F. 2014, Astronomy \& Astrophysics, 564, A73

\bibitem[{{Ag{\'u}ndez} {et~al.}(2012){Ag{\'u}ndez}, {Venot}, {Iro}, {Selsis},
  {Hersant}, {H{\'e}brard}, \& {Dobrijevic}}]{agu2012}
{Ag{\'u}ndez}, M., {Venot}, O., {Iro}, N., {et~al.} 2012, Astronomy \&
  Astrophysics, 548, A73

\bibitem[{{ANSYS, Inc.: San Diego}(2017)}]{chemkin}
{ANSYS, Inc.: San Diego}. 2017, Chemkin-Pro 18.2

\bibitem[{{Arridge} {et~al.}(2014){Arridge}, {Achilleos}, {Agarwal}, {Agnor},
  {Ambrosi}, {Andr{\'e}}, {Badman}, {Baines}, {Banfield}, {Barth{\'e}l{\'e}my},
  {Bisi}, {Blum}, {Bocanegra-Bahamon}, {Bonfond}, {Bracken}, {Brandt},
  {Briand}, {Briois}, {Brooks}, {Castillo-Rogez}, {Cavali{\'e}}, {Christophe},
  {Coates}, {Collinson}, {Cooper}, {Costa-Sitja}, {Courtin}, {Daglis}, {de
  Pater}, {Desai}, {Dirkx}, {Dougherty}, {Ebert}, {Filacchione}, {Fletcher},
  {Fortney}, {Gerth}, {Grassi}, {Grodent}, {Gr{\"u}n}, {Gustin}, {Hedman},
  {Helled}, {Henri}, {Hess}, {Hillier}, {Hofstadter}, {Holme}, {Horanyi},
  {Hospodarsky}, {Hsu}, {Irwin}, {Jackman}, {Karatekin}, {Kempf}, {Khalisi},
  {Konstantinidis}, {Kr{\"u}ger}, {Kurth}, {Labrianidis}, {Lainey}, {Lamy},
  {Laneuville}, {Lucchesi}, {Luntzer}, {MacArthur}, {Maier}, {Masters},
  {McKenna-Lawlor}, {Melin}, {Milillo}, {Moragas-Klostermeyer}, {Morschhauser},
  {Moses}, {Mousis}, {Nettelmann}, {Neubauer}, {Nordheim}, {Noyelles}, {Orton},
  {Owens}, {Peron}, {Plainaki}, {Postberg}, {Rambaux}, {Retherford}, {Reynaud},
  {Roussos}, {Russell}, {Rymer}, {Sallantin}, {S{\'a}nchez-Lavega}, {Santolik},
  {Saur}, {Sayanagi}, {Schenk}, {Schubert}, {Sergis}, {Sittler}, {Smith},
  {Spahn}, {Srama}, {Stallard}, {Sterken}, {Sternovsky}, {Tiscareno}, {Tobie},
  {Tosi}, {Trieloff}, {Turrini}, {Turtle}, {Vinatier}, {Wilson}, \&
  {Zarka}}]{arridge2014}
{Arridge}, C.~S., {Achilleos}, N., {Agarwal}, J., {et~al.} 2014, Planetary and
  Space Science, 104, 122

\bibitem[{{Baudino} {et~al.}(2017){Baudino}, {Molli{\`e}re}, {Venot},
  {Tremblin}, {B{\'e}zard}, \& {Lagage}}]{Baudino2017}
{Baudino}, J.-L., {Molli{\`e}re}, P., {Venot}, O., {et~al.} 2017, \apj, 850,
  150

\bibitem[{{Bean} {et~al.}(2018){Bean}, {Stevenson}, {Batalha},
  {Berta-Thompson}, {Kreidberg}, {Crouzet}, {Benneke}, {Line}, {Sing},
  {Wakeford}, {Knutson}, {Kempton}, {D{\'e}sert}, {Crossfield}, {Batalha}, {de
  Wit}, {Parmentier}, {Harrington}, {Moses}, {Lopez-Morales}, {Alam}, {Blecic},
  {Bruno}, {Carter}, {Chapman}, {Decin}, {Dragomir}, {Evans}, {Fortney},
  {Fraine}, {Gao}, {Garc{\'\i}a Mu{\~n}oz}, {Gibson}, {Goyal}, {Heng}, {Hu},
  {Kendrew}, {Kilpatrick}, {Krick}, {Lagage}, {Lendl}, {Louden}, {Madhusudhan},
  {Mandell}, {Mansfield}, {May}, {Morello}, {Morley}, {Nikolov}, {Redfield},
  {Roberts}, {Schlawin}, {Spake}, {Todorov}, {Tsiaras}, {Venot}, {Waalkes},
  {Wheatley}, {Zellem}, {Angerhausen}, {Barrado}, {Carone}, {Casewell},
  {Cubillos}, {Damiano}, {de Val-Borro}, {Drummond}, {Edwards}, {Endl},
  {Espinoza}, {France}, {Gizis}, {Greene}, {Henning}, {Hong}, {Ingalls}, {Iro},
  {Irwin}, {Kataria}, {Lahuis}, {Leconte}, {Lillo-Box}, {Lines}, {Lothringer},
  {Mancini}, {Marchis}, {Mayne}, {Palle}, {Rauscher}, {Roudier}, {Shkolnik},
  {Southworth}, {Swain}, {Taylor}, {Teske}, {Tinetti}, {Tremblin}, {Tucker},
  {van Boekel}, {Waldmann}, {Weaver}, \& {Zingales}}]{Bean2018}
{Bean}, J.~L., {Stevenson}, K.~B., {Batalha}, N.~M., {et~al.} 2018, ArXiv
  e-prints, arXiv:1803.04985

\bibitem[{{Bolton} {et~al.}(2017){Bolton}, {Lunine}, {Stevenson}, {Connerney},
  {Levin}, {Owen}, {Bagenal}, {Gautier}, {Ingersoll}, {Orton}, {Guillot},
  {Hubbard}, {Bloxham}, {Coradini}, {Stephens}, {Mokashi}, {Thorne}, \&
  {Thorpe}}]{Bolton2017}
{Bolton}, S.~J., {Lunine}, J., {Stevenson}, D., {et~al.} 2017, Science, 213, 5

\bibitem[{{Cavali{\'e}} {et~al.}(2014){Cavali{\'e}}, {Moreno}, {Lellouch},
  {Hartogh}, {Venot}, {Orton}, {Jarchow}, {Encrenaz}, {Selsis}, {Hersant}, \&
  {Fletcher}}]{Cavalie2014}
{Cavali{\'e}}, T., {Moreno}, R., {Lellouch}, E., {et~al.} 2014, \aap, 562, A33

\bibitem[{{Cavali{\'e}} {et~al.}(2017){Cavali{\'e}}, {Venot}, {Selsis},
  {Hersant}, {Hartogh}, \& {Leconte}}]{cavalie2017}
{Cavali{\'e}}, T., {Venot}, O., {Selsis}, F., {et~al.} 2017, \icarus, 291, 1

\bibitem[{{Charnay} {et~al.}(2015){Charnay}, {Meadows}, \&
  {Leconte}}]{charnay2015}
{Charnay}, B., {Meadows}, V., \& {Leconte}, J. 2015, \apj, 813, 15

\bibitem[{{Cooper} \& {Showman}(2006)}]{CS2006}
{Cooper}, C.~S. \& {Showman}, A.~P. 2006, The Astrophysical Journal, 649, 1048

\bibitem[{{de Pater} {et~al.}(2015){de Pater}, {Sromovsky}, {Fry}, {Hammel},
  {Baranec}, \& {Sayanagi}}]{dePater2015}
{de Pater}, I., {Sromovsky}, L.~A., {Fry}, P.~M., {et~al.} 2015, \icarus, 252,
  121

\bibitem[{{Dobrijevic} {et~al.}(2011){Dobrijevic}, {Cavali{\'e}}, \&
  {Billebaud}}]{Dobrijevic2011}
{Dobrijevic}, M., {Cavali{\'e}}, T., \& {Billebaud}, F. 2011, \icarus, 214, 275

\bibitem[{{Dobrijevic} {et~al.}(2003){Dobrijevic}, {Ollivier}, {Billebaud},
  {Brillet}, \& {Parisot}}]{dob2003}
{Dobrijevic}, M., {Ollivier}, J.~L., {Billebaud}, F., {Brillet}, J., \&
  {Parisot}, J.~P. 2003, \aap, 398, 335

\bibitem[{{Dobrijevic} \& {Parisot}(1998)}]{dob1998}
{Dobrijevic}, M. \& {Parisot}, J.~P. 1998, \planss, 46, 491

\bibitem[{{Drummond} {et~al.}(2018{\natexlab{a}}){Drummond}, {Mayne},
  {Manners}, {Carter}, {Boutle}, {Baraffe}, {H{\'e}brard}, {Tremblin}, {Sing},
  {Amundsen}, \& {Acreman}}]{Drummond2018}
{Drummond}, B., {Mayne}, N.~J., {Manners}, J., {et~al.} 2018{\natexlab{a}},
  \apjl, 855, L31

\bibitem[{{Drummond} {et~al.}(2018{\natexlab{b}}){Drummond},
  \setbox0=\hbox{Z}{Mayne}, {Manners}, {Baraffe}, {Goyal}, {Tremblin}, {Sing},
  \& {Kohary}}]{Drummond2018b}
{Drummond}, B., \setbox0=\hbox{Z}{Mayne}, N.~J., {Manners}, J., {et~al.}
  2018{\natexlab{b}}, \apj, 869, 28

\bibitem[{{Drummond} {et~al.}(2016){Drummond}, {Tremblin}, {Baraffe},
  {Amundsen}, {Mayne}, {Venot}, \& {Goyal}}]{Drummond2016}
{Drummond}, B., {Tremblin}, P., {Baraffe}, I., {et~al.} 2016, \aap, 594, A69

\bibitem[{{Fischer} {et~al.}(2011){Fischer}, {Kurth}, {Gurnett}, {Zarka},
  {Dyudina}, {Ingersoll}, {Ewald}, {Porco}, {Wesley}, {Go}, \&
  {Delcroix}}]{Fischer2011}
{Fischer}, G., {Kurth}, W.~S., {Gurnett}, D.~A., {et~al.} 2011, \nat, 475, 75

\bibitem[{{Gilli} {et~al.}(2017){Gilli}, {Lebonnois}, {Gonz{\'a}lez-Galindo},
  {L{\'o}pez-Valverde}, {Stolzenbach}, {Lef{\`e}vre}, {Chaufray}, \&
  {Lott}}]{gilli2017}
{Gilli}, G., {Lebonnois}, S., {Gonz{\'a}lez-Galindo}, F., {et~al.} 2017,
  \icarus, 281, 55

\bibitem[{Goldsmith {et~al.}(2012)Goldsmith, Magoon, \& Green}]{goldsmith2012}
Goldsmith, C.~F., Magoon, G.~R., \& Green, W.~H. 2012, The Journal of Physical
  Chemistry A, 116, 9033

\bibitem[{{H\'ebrard} {et~al.}(2015){H\'ebrard}, {Tomlin}, {Bounaceur}, \&
  {Battin-Leclerc}}]{hebrard2015}
{H\'ebrard}, E., {Tomlin}, A.~S., {Bounaceur}, R., \& {Battin-Leclerc}, F.
  2015, Proceedings of the Combustion Institute, 35, 607

\bibitem[{{Irwin} {et~al.}(2016){Irwin}, {Fletcher}, {Tice}, {Owen}, {Orton},
  {Teanby}, \& {Davis}}]{Irwin2016}
{Irwin}, P.~G.~J., {Fletcher}, L.~N., {Tice}, D., {et~al.} 2016, \icarus, 271,
  418

\bibitem[{Lebedev {et~al.}(2013)Lebedev, Okun, Chorkov, Tokar, \&
  Strelkova}]{lebedev2013}
Lebedev, A., Okun, M., Chorkov, V., Tokar, P., \& Strelkova, M. 2013, Journal
  of Mathematical Chemistry, 51, 73

\bibitem[{{Lebonnois} {et~al.}(2001){Lebonnois}, {Toublanc}, {Hourdin}, \&
  {Rannou}}]{lebonnois2001}
{Lebonnois}, S., {Toublanc}, D., {Hourdin}, F., \& {Rannou}, P. 2001, \icarus,
  152, 384

\bibitem[{{Leconte} {et~al.}(2017){Leconte}, {Selsis}, {Hersant}, \&
  {Guillot}}]{leconte2017}
{Leconte}, J., {Selsis}, F., {Hersant}, F., \& {Guillot}, T. 2017, \aap, 598,
  A98

\bibitem[{{Lef{\`e}vre} {et~al.}(2004){Lef{\`e}vre}, {Lebonnois}, {Montmessin},
  \& {Forget}}]{Lefevre2004}
{Lef{\`e}vre}, F., {Lebonnois}, S., {Montmessin}, F., \& {Forget}, F. 2004,
  Journal of Geophysical Research (Planets), 109, E07004

\bibitem[{Liang {et~al.}(2009)Liang, Stevens, \& Farrell}]{liang2009}
Liang, L., Stevens, J.~G., \& Farrell, J.~T. 2009, Proceedings of the
  Combustion Institute, 32, 527

\bibitem[{Lodders(2010)}]{lodders2010}
Lodders, K. 2010, in Principles and Perspectives in Cosmochemistry, ed.
  A.~Goswami \& B.~E. Reddy (Berlin, Heidelberg: Springer Berlin Heidelberg),
  379--417

\bibitem[{Lu \& Law(2005)}]{lu2005}
Lu, T. \& Law, C.~K. 2005, Proceedings of the Combustion Institute, 30, 1333

\bibitem[{{Mayne} {et~al.}(2017){Mayne}, {Debras}, {Baraffe}, {Thuburn},
  {Amundsen}, {Acreman}, {Smith}, {Browning}, {Manners}, \& {Wood}}]{mayne2017}
{Mayne}, N.~J., {Debras}, F., {Baraffe}, I., {et~al.} 2017, \aap, 604, A79

\bibitem[{McBride {et~al.}(1993)McBride, Gordon, \& Reno}]{mcbride1993nasa}
McBride, B., Gordon, S., \& Reno, M. 1993, NASA Technical Memorandum, 4513

\bibitem[{{Mendon{\c c}a} {et~al.}(2018){Mendon{\c c}a}, {Tsai}, {Malik},
  {Grimm}, \& {Heng}}]{Mendonca2018}
{Mendon{\c c}a}, J.~M., {Tsai}, S.-M., {Malik}, M., {Grimm}, S.~L., \& {Heng},
  K. 2018, ArXiv e-prints [\eprint[arXiv]{1808.00501}]

\bibitem[{{Morley} {et~al.}(2014){Morley}, {Marley}, {Fortney}, {Lupu},
  {Saumon}, {Greene}, \& {Lodders}}]{morley2014}
{Morley}, C.~V., {Marley}, M.~S., {Fortney}, J.~J., {et~al.} 2014, \apj, 787,
  78

\bibitem[{{Moses} {et~al.}(2013{\natexlab{a}}){Moses}, {Line}, {Visscher},
  {Richardson}, {Nettelmann}, {Fortney}, {Barman}, {Stevenson}, \&
  {Madhusudhan}}]{Moses2013}
{Moses}, J.~I., {Line}, M.~R., {Visscher}, C., {et~al.} 2013{\natexlab{a}},
  \apj, 777, 34

\bibitem[{{Moses} {et~al.}(2013{\natexlab{b}}){Moses}, {Madhusudhan},
  {Visscher}, \& {Freedman}}]{Moses2013CO}
{Moses}, J.~I., {Madhusudhan}, N., {Visscher}, C., \& {Freedman}, R.~S.
  2013{\natexlab{b}}, \apj, 763, 25

\bibitem[{{Moses} {et~al.}(2011){Moses}, {Visscher}, {Fortney}, {Showman},
  {Lewis}, {Griffith}, {Klippenstein}, {Shabram}, {Friedson}, {Marley}, \&
  {Freedman}}]{moses2011}
{Moses}, J.~I., {Visscher}, C., {Fortney}, J.~J., {et~al.} 2011, The
  Astrophysical Journal, 737, 15

\bibitem[{{Mousis} {et~al.}(2018){Mousis}, {Atkinson}, {Cavali{\'e}},
  {Fletcher}, {Amato}, {Aslam}, {Ferri}, {Renard}, {Spilker}, {Venkatapathy},
  {Wurz}, {Aplin}, {Coustenis}, {Deleuil}, {Dobrijevic}, {Fouchet}, {Guillot},
  {Hartogh}, {Hewagama}, {Hofstadter}, {Hue}, {Hueso}, {Lebreton}, {Lellouch},
  {Moses}, {Orton}, {Pearl}, {S{\'a}nchez-Lavega}, {Simon}, {Venot}, {Waite},
  {Achterberg}, {Atreya}, {Billebaud}, {Blanc}, {Borget}, {Brugger}, {Charnoz},
  {Chiavassa}, {Cottini}, {d'Hendecourt}, {Danger}, {Encrenaz}, {Gorius},
  {Jorda}, {Marty}, {Moreno}, {Morse}, {Nixon}, {Reh}, {Ronnet}, {Schmider},
  {Sheridan}, {Sotin}, {Vernazza}, \& {Villanueva}}]{mousis2018}
{Mousis}, O., {Atkinson}, D.~H., {Cavali{\'e}}, T., {et~al.} 2018, Planetary
  and Space Science, 155, 12

\bibitem[{{Mousis} {et~al.}(2016){Mousis}, {Atkinson}, {Spilker},
  {Venkatapathy}, {Poncy}, {Frampton}, {Coustenis}, {Reh}, {Lebreton},
  {Fletcher}, {Hueso}, {Amato}, {Colaprete}, {Ferri}, {Stam}, {Wurz}, {Atreya},
  {Aslam}, {Banfield}, {Calcutt}, {Fischer}, {Holland}, {Keller}, {Kessler},
  {Leese}, {Levacher}, {Morse}, {Mu{\~n}oz}, {Renard}, {Sheridan}, {Schmider},
  {Snik}, {Waite}, {Bird}, {Cavali{\'e}}, {Deleuil}, {Fortney}, {Gautier},
  {Guillot}, {Lunine}, {Marty}, {Nixon}, {Orton}, \&
  {S{\'a}nchez-Lavega}}]{mousis2016}
{Mousis}, O., {Atkinson}, D.~H., {Spilker}, T., {et~al.} 2016, Planetary and
  Space Science, 130, 80

\bibitem[{{Mousis} {et~al.}(2014){Mousis}, {Fletcher}, {Lebreton}, {Wurz},
  {Cavali{\'e}}, {Coustenis}, {Courtin}, {Gautier}, {Helled}, {Irwin}, {Morse},
  {Nettelmann}, {Marty}, {Rousselot}, {Venot}, {Atkinson}, {Waite}, {Reh},
  {Simon}, {Atreya}, {Andr{\'e}}, {Blanc}, {Daglis}, {Fischer}, {Geppert},
  {Guillot}, {Hedman}, {Hueso}, {Lellouch}, {Lunine}, {Murray}, {O`Donoghue},
  {Rengel}, {S{\'a}nchez-Lavega}, {Schmider}, {Spiga}, {Spilker}, {Petit},
  {Tiscareno}, {Ali-Dib}, {Altwegg}, {Bolton}, {Bouquet}, {Briois}, {Fouchet},
  {Guerlet}, {Kostiuk}, {Lebleu}, {Moreno}, {Orton}, \& {Poncy}}]{mousis2014}
{Mousis}, O., {Fletcher}, L.~N., {Lebreton}, J.~P., {et~al.} 2014, Planetary
  and Space Science, 104, 29

\bibitem[{{Nixon} {et~al.}(2010){Nixon}, {Achterberg}, {Romani}, {Allen},
  {Zhang}, {Teanby}, {Irwin}, \& {Flasar}}]{nixon2010}
{Nixon}, C.~A., {Achterberg}, R.~K., {Romani}, P.~N., {et~al.} 2010, Planetary
  and Space Science, 58, 1667

\bibitem[{{Parmentier} \& {Crossfield}(2017)}]{parmentier2017}
{Parmentier}, V. \& {Crossfield}, I. J.~M. 2017, {Exoplanet Phase Curves:
  Observations and Theory}, 116

\bibitem[{Pepiot-Desjardins \& Pitsch(2008)}]{pepiot2008}
Pepiot-Desjardins, P. \& Pitsch, H. 2008, Combustion and Flame, 154, 67

\bibitem[{Qiu {et~al.}(2016)Qiu, Cheng, Wang, Li, Li, Wang, \& Wu}]{qiu2016}
Qiu, L., Cheng, X., Wang, X., {et~al.} 2016, Energy \& Fuels, 30, 10875

\bibitem[{{Saumon} {et~al.}(2003){Saumon}, {Marley}, {Lodders}, \&
  {Freedman}}]{saumon2003}
{Saumon}, D., {Marley}, M.~S., {Lodders}, K., \& {Freedman}, R.~S. 2003, in IAU
  Symposium, Vol. 211, Brown Dwarfs, ed. E.~{Mart{\'{\i}}n}, 345

\bibitem[{{Stephens} {et~al.}(2009){Stephens}, {Leggett}, {Cushing}, {Marley},
  {Saumon}, {Geballe}, {Golimowski}, {Fan}, \& {Noll}}]{stephens2009}
{Stephens}, D.~C., {Leggett}, S.~K., {Cushing}, M.~C., {et~al.} 2009, \apj,
  702, 154

\bibitem[{Stolzenbach(2016)}]{stolzenbach2016}
Stolzenbach, A. 2016, Theses, {Universit{\'e} Pierre et Marie Curie - Paris VI}

\bibitem[{Stolzenbach {et~al.}(2014)Stolzenbach, Lef{\`e}vre, Lebonnois,
  M{\"a}{\"a}tt{\"a}nen, \& Bekki}]{stolzenbach2014}
Stolzenbach, A., Lef{\`e}vre, F., Lebonnois, S., M{\"a}{\"a}tt{\"a}nen, A., \&
  Bekki, S. 2014, in {EGU General Assembly 2014}, Vienna, Austria

\bibitem[{{Tinetti} {et~al.}(2018){Tinetti}, {Drossart}, {Eccleston},
  {Hartogh}, {Heske}, {Leconte}, {Micela}, {Ollivier}, {Pilbratt}, {Puig},
  {Turrini}, {Vandenbussche}, {Wolkenberg}, {Beaulieu}, {Buchave}, {Ferus},
  {Griffin}, {Guedel}, {Justtanont}, {Lagage}, {Machado}, {Malaguti}, {Min},
  {N{\o}rgaard-Nielsen}, {Rataj}, {Ray}, {Ribas}, {Swain}, {Szabo}, {Werner},
  {Barstow}, {Burleigh}, {Cho}, {du Foresto}, {Coustenis}, {Decin}, {Encrenaz},
  {Galand}, {Gillon}, {Helled}, {Morales}, {Mu{\~n}oz}, {Moneti}, {Pagano},
  {Pascale}, {Piccioni}, {Pinfield}, {Sarkar}, {Selsis}, {Tennyson}, {Triaud},
  {Venot}, {Waldmann}, {Waltham}, {Wright}, {Amiaux}, {Augu{\`e}res},
  {Berth{\'e}}, {Bezawada}, {Bishop}, {Bowles}, {Coffey}, {Colom{\'e}},
  {Crook}, {Crouzet}, {Da Peppo}, {Sanz}, {Focardi}, {Frericks}, {Hunt},
  {Kohley}, {Middleton}, {Morgante}, {Ottensamer}, {Pace}, {Pearson},
  {Stamper}, {Symonds}, {Rengel}, {Renotte}, {Ade}, {Affer}, {Alard}, {Allard},
  {Altieri}, {Andr{\'e}}, {Arena}, {Argyriou}, {Aylward}, {Baccani}, {Bakos},
  {Banaszkiewicz}, {Barlow}, {Batista}, {Bellucci}, {Benatti}, {Bernardi},
  {B{\'e}zard}, {Blecka}, {Bolmont}, {Bonfond}, {Bonito}, {Bonomo}, {Brucato},
  {Brun}, {Bryson}, {Bujwan}, {Casewell}, {Charnay}, {Pestellini}, {Chen},
  {Ciaravella}, {Claudi}, {Cl{\'e}dassou}, {Damasso}, {Damiano}, {Danielski},
  {Deroo}, {Di Giorgio}, {Dominik}, {Doublier}, {Doyle}, {Doyon}, {Drummond},
  {Duong}, {Eales}, {Edwards}, {Farina}, {Flaccomio}, {Fletcher}, {Forget},
  {Fossey}, {Fr{\"a}nz}, {Fujii}, {Garc{\'\i}a-Piquer}, {Gear}, {Geoffray},
  {G{\'e}rard}, {Gesa}, {Gomez}, {Graczyk}, {Griffith}, {Grodent}, {Guarcello},
  {Gustin}, {Hamano}, {Hargrave}, {Hello}, {Heng}, {Herrero}, {Hornstrup},
  {Hubert}, {Ida}, {Ikoma}, {Iro}, {Irwin}, {Jarchow}, {Jaubert}, {Jones},
  {Julien}, {Kameda}, {Kerschbaum}, {Kervella}, {Koskinen}, {Krijger}, {Krupp},
  {Lafarga}, {Landini}, {Lellouch}, {Leto}, {Luntzer}, {Rank-L{\"u}ftinger},
  {Maggio}, {Maldonado}, {Maillard}, {Mall}, {Marquette}, {Mathis}, {Maxted},
  {Matsuo}, {Medvedev}, {Miguel}, {Minier}, {Morello}, {Mura}, {Narita},
  {Nascimbeni}, {Nguyen Tong}, {Noce}, {Oliva}, {Palle}, {Palmer}, {Pancrazzi},
  {Papageorgiou}, {Parmentier}, {Perger}, {Petralia}, {Pezzuto},
  {Pierrehumbert}, {Pillitteri}, {Piotto}, {Pisano}, {Prisinzano}, {Radioti},
  {R{\'e}ess}, {Rezac}, {Rocchetto}, {Rosich}, {Sanna}, {Santerne}, {Savini},
  {Scandariato}, {Sicardy}, {Sierra}, {Sindoni}, {Skup}, {Snellen}, {Sobiecki},
  {Soret}, {Sozzetti}, {Stiepen}, {Strugarek}, {Taylor}, {Taylor}, {Terenzi},
  {Tessenyi}, {Tsiaras}, {Tucker}, {Valencia}, {Vasisht}, {Vazan}, {Vilardell},
  {Vinatier}, {Viti}, {Waters}, {Wawer}, {Wawrzaszek}, {Whitworth}, {Yung},
  {Yurchenko}, {Osorio}, {Zellem}, {Zingales}, \& {Zwart}}]{Tinetti2018}
{Tinetti}, G., {Drossart}, P., {Eccleston}, P., {et~al.} 2018, Experimental
  Astronomy, 53

\bibitem[{{Tremblin} {et~al.}(2016){Tremblin}, {Amundsen}, {Chabrier},
  {Baraffe}, {Drummond}, {Hinkley}, {Mourier}, \& {Venot}}]{tremblin:2016aa}
{Tremblin}, P., {Amundsen}, D.~S., {Chabrier}, G., {et~al.} 2016, \apjl, 817,
  L19

\bibitem[{{Tremblin} {et~al.}(2015){Tremblin}, {Amundsen}, {Mourier},
  {Baraffe}, {Chabrier}, {Drummond}, {Homeier}, \& {Venot}}]{tremblin:2015aa}
{Tremblin}, P., {Amundsen}, D.~S., {Mourier}, P., {et~al.} 2015, \apjl, 804,
  L17

\bibitem[{{Tsai} {et~al.}(2018){Tsai}, {Kitzmann}, {Lyons}, {Mendon{\c c}a},
  {Grimm}, \& {Heng}}]{Tsai2018}
{Tsai}, S.-M., {Kitzmann}, D., {Lyons}, J.~R., {et~al.} 2018, \apj, 862, 31

\bibitem[{{Tsai} {et~al.}(2017){Tsai}, {Lyons}, {Grosheintz}, {Rimmer},
  {Kitzmann}, \& {Heng}}]{Tsai2017}
{Tsai}, S.-M., {Lyons}, J.~R., {Grosheintz}, L., {et~al.} 2017, The
  Astrophysical Journal Supplement Series, 228, 20

\bibitem[{{Venot} {et~al.}(2014){Venot}, {Ag{\'u}ndez}, {Selsis}, {Tessenyi},
  \& {Iro}}]{Venot2014}
{Venot}, O., {Ag{\'u}ndez}, M., {Selsis}, F., {Tessenyi}, M., \& {Iro}, N.
  2014, \aap, 562, A51

\bibitem[{{Venot} {et~al.}(in prep.){Venot}, {Crouzet}, {Parmentier},
  {Tremblin}, {Gao}, {Powell}, {et~al.}}]{VenotWASP43b}
{Venot}, O., {Crouzet}, N., {Parmentier}, V., {et~al.} in prep., \apj

\bibitem[{{Venot} {et~al.}(2018){Venot}, {Drummond}, {Miguel}, {Waldmann},
  {Pascale}, \& {Zingales}}]{Venot2018ARIEL}
{Venot}, O., {Drummond}, B., {Miguel}, Y., {et~al.} 2018, Experimental
  Astronomy, 49

\bibitem[{{Venot} {et~al.}(2015){Venot}, {H{\'e}brard}, {Ag{\'u}ndez}, {Decin},
  \& {Bounaceur}}]{Venot2015}
{Venot}, O., {H{\'e}brard}, E., {Ag{\'u}ndez}, M., {Decin}, L., \& {Bounaceur},
  R. 2015, \aap, 577, A33

\bibitem[{{Venot} {et~al.}(2012){Venot}, {H{\'e}brard}, {Ag{\'u}ndez},
  {Dobrijevic}, {Selsis}, {Hersant}, {Iro}, \& {Bounaceur}}]{Venot2012}
{Venot}, O., {H{\'e}brard}, E., {Ag{\'u}ndez}, M., {et~al.} 2012, Astronomy \&
  Astrophysics, 546, A43

\bibitem[{{Visscher} {et~al.}(2010){Visscher}, {Moses}, \&
  {Saslow}}]{Visscher2010}
{Visscher}, C., {Moses}, J.~I., \& {Saslow}, S.~A. 2010, \icarus, 209, 602

\bibitem[{{Wakelam} {et~al.}(2012){Wakelam}, {Herbst}, {Loison}, {Smith},
  {Chandrasekaran}, {Pavone}, {Adams}, {Bacchus-Montabonel}, {Bergeat},
  {B{\'e}roff}, {Bierbaum}, {Chabot}, {Dalgarno}, {van Dishoeck}, {Faure},
  {Geppert}, {Gerlich}, {Galli}, {H{\'e}brard}, {Hersant}, {Hickson},
  {Honvault}, {Klippenstein}, {Le Picard}, {Nyman}, {Pernot}, {Schlemmer},
  {Selsis}, {Sims}, {Talbi}, {Tennyson}, {Troe}, {Wester}, \&
  {Wiesenfeld}}]{KIDA2012}
{Wakelam}, V., {Herbst}, E., {Loison}, J.-C., {et~al.} 2012, \apjs, 199, 21

\end{thebibliography}

\end{document}